\documentclass[oneside,english]{scrbook}

\usepackage[T1]{fontenc}
\usepackage[latin9]{inputenc}
\usepackage{geometry}
\usepackage{array}
\usepackage{refstyle}
\usepackage{float}
\usepackage{amsmath}
\usepackage{amsthm}
\usepackage{amssymb}
\usepackage{graphicx}

\makeatletter

\AtBeginDocument{\providecommand\eqref[1]{\ref{eq:#1}}}
\providecommand{\tabularnewline}{\\}
\RS@ifundefined{subsecref}
  {\newref{subsec}{name = \RSsectxt}}
  {}
\RS@ifundefined{thmref}
  {\def\RSthmtxt{theorem~}\newref{thm}{name = \RSthmtxt}}
  {}
\RS@ifundefined{lemref}
  {\def\RSlemtxt{lemma~}\newref{lem}{name = \RSlemtxt}}
  {}

\numberwithin{equation}{section}
\numberwithin{figure}{section}

\@ifundefined{date}{}{\date{}}
\usepackage{epigraph}
\usepackage{tikz}
\usetikzlibrary{automata, positioning}

\definecolor{myBlue}{rgb}{0,0.396,0.6}

\usepackage[colorlinks=true,linkcolor=myBlue,urlcolor=myBlue,citecolor=myBlue,anchorcolor=myBlue]{hyperref}
\usepackage[numbers,square,sort&compress]{natbib}

\usepackage{amsfonts}
\usepackage{physics}
\usepackage{chngcntr}
\usepackage{listings}
\usepackage{xcolor}
 
\definecolor{codegreen}{rgb}{0,0.6,0}
\definecolor{codegray}{rgb}{0.5,0.5,0.5}
\definecolor{codepurple}{rgb}{0.58,0,0.82}
\definecolor{backcolour}{rgb}{0.98,0.98,0.98}
 
\lstdefinestyle{mystyle}{
    backgroundcolor=\color{backcolour},   
    commentstyle=\color{codegreen},
    keywordstyle=\color{magenta},
    numberstyle=\tiny\color{codegray},
    stringstyle=\color{codepurple},
    basicstyle=\linespread{0.8}\footnotesize\ttfamily,
    breakatwhitespace=false,         
    breaklines=true,                 
    captionpos=b,                    
    keepspaces=true,                 
    numbersep=5pt,                  
    showspaces=false,                
    showstringspaces=false,
    showtabs=false,                  
    tabsize=2
}
 
\makeatother

\usepackage{listings}

\begin{document}
\title{Inference, prediction and optimization of non-pharmaceutical interventions
using compartment models: the PyRoss library}
\author{R.  Adhikari\footnote{ra413@cam.ac.uk}, Austen Bolitho, Fernando Caballero,  
\\
Michael E. Cates, Jakub Dolezal, Timothy Ekeh, Jules Guioth,
\\
Robert L. Jack, Julian Kappler, Lukas Kikuchi, Hideki Kobayashi, 
\\
Yuting I. Li, Joseph D. Peterson, Patrick Pietzonka, Benjamin Remez, \\
Paul B. Rohrbach, Rajesh Singh, and Günther Turk}
\date{\textbf{\normalsize{}University of Cambridge, United Kingdom}}

\maketitle

\textbf{PyRoss\footnote{\href{https://github.com/rajeshrinet/PyRoss}{https://github.com/rajeshrinet/PyRoss}}}
is an open-source Python library that offers an integrated platform for inference, prediction and optimisation of non-pharmaceutical interventions in age- and contact-structured epidemiological compartment models.
This report outlines the rationale and functionality of the PyRoss
library, with various illustrations and examples focusing on well-mixed, age-structured populations. The PyRoss library supports arbitrary
age-structured compartment models formulated stochastically (in terms
of master equations) or deterministically (as systems of differential
equations) and allows mid-run transitioning from one to the other.
By supporting additional compartmental subdivision \emph{ad} \emph{libitum},
PyRoss can emulate time-since-infection models and allows medical stages such as hospitalization or quarantine
to be modelled and forecast. The PyRoss library enables fitting to
epidemiological data, as available, using Bayesian parameter inference,
so that competing models can be weighed by their evidence. 
 PyRoss allows fully Bayesian forecasts of the impact of idealized
non-pharmaceutical interventions (NPIs) by convolving uncertainties arising from epidemiological data,
model choice, parameters, and intrinsic stochasticity. Algorithms
to optimize time-dependent NPI scenarios against user-defined cost
functions are included. PyRoss's current age-structured compartment
framework for well-mixed populations will in future reports be extended
to include compartments structured by location, occupation, use of
travel networks and other attributes relevant to assessing disease
spread and the impact of NPIs. We argue that such compartment
models, by allowing social data of arbitrary granularity to be combined
with Bayesian parameter estimation for poorly-known disease variables,
could enable more powerful and robust prediction than other approaches to detailed epidemic modelling. We invite
others to use the PyRoss library for research to address today's COVID-19 crisis, and to plan for future
pandemics.

\tableofcontents{}

\newpage{}

\chapter{Introduction\label{chapter:1}}

This report introduces PyRoss, a Python library for inference, forecasting
and optimisation of non-pharmaceutical interventions (NPIs), using compartment
models of epidemics. These models are very widely used in epidemiology
\cite{anderson1992infectious,keeling2011modeling,bailey1975mathematical},
including simple examples such as SIR, but also more complicated variants
\cite{anderson1980spread,wearing2005appropriate,krylova2013effects,lloyd2001realistic},
some of which are discussed further below. The central modelling
philosophy is to group individuals into compartments, which correspond
to disease states (such as susceptible / infectious) and may be
further divided by age, and by objective medical states (for example,
seropositive, hospitalized, in ICU, on ventilator). This leads to
a broad class of models that are intermediate in detail between simple
compartment models (such as SIR \cite{anderson1992infectious}) and
agent-based or similar models where the population is disaggregated
into synthetic individuals \cite{barbosa2018human}. 

The functionality of PyRoss not only includes simulation of such compartment
models, but also automates inference of their parameters from data,
and construction of Bayesian forecasts. In addition, it offers framework
for modelling NPIs, and optimisation of their parameters (for example,
the length of a lockdown) against user-defined cost functions. While the simulation aspects of PyRoss may be
comparable to those available elsewhere (such as epiModel \cite{jenness2018epimodel}), its integration of inference and optimisation engines create enhanced functionality within a unified and relatively transparent open-source coding environment.

\section{Rationale for compartment models}

PyRoss is built around Markovian compartment models, in which a system's
future evolution depends only on the current occupancies of the compartments.
This leads to several useful simplifications -{}- the average dynamics
of a large population can be obtained by solving a system of ordinary
differential equations, and the stochastic dynamics of a finite population
can be simulated by a simple (Markovian) stochastic process. On the
other hand, such models are sometimes criticised \cite{keeling2011modeling}
because they result in exponential distributions of residence times
in each compartment, which is often inconsistent with realistic disease
progression. To address this, time-since-infection (TSI) models may
be used \cite{kermack1927contribution,reddingius1971notes,sattenspiel1995structured,hoppensteadt1974age,diekmann2010construction,thieme1993may,keeling2011modeling}.

Nevertheless, the behaviour of TSI models can be captured by compartment
models, at the modest cost of introducing additional compartments. This
is called the method of stages \cite{anderson1980spread,wearing2005appropriate,krylova2013effects,lloyd2001realistic}.
In stochastic modelling, it amounts to replacing a non-Markovian
model by a Markovian one on a larger space (which here is a
model with extra compartments). PyRoss supports an unlimited number of
compartments, which achieves high generality while retaining the conceptual
and computational advantages of compartment models. Subdivision of a compartment into $k$ stages, even for modest values such as $k=5,$ already allow considerable flexibility
in capturing non-Markovian distributions of residence times within different
disease states. Larger values of $k$ can of course be used, if required.

Lifting any restriction on the number of compartments has  advantages beyond the TSI issue just described. For example, the medical data
used to infer parameters is often provided in the form of
compartment populations, such as ICU occupancy or serological test
data (perhaps segregated by age). The observed population of these added compartments
can then inform the parameters of the model, improving
predictions of the future prevalence of these (and other) outcomes. This is valuable when,
as with COVID-19, avoiding saturation of ICU or ventilator provision
is a policy objective.

One standard use of compartment
models is to address age-structured populations. Crucially, this
resolves the assortative character of social contacts through which
disease transmission occurs. By compartmenting into age bands, representing
the contact rates between each pair of bands in terms of a contact
matrix, and partitioning this matrix further into home- based, school-based,
workplace-based and `other' components (on which NPIs have separately estimable
effects) one can, within a well-mixed population model, prediction
epidemic evolution and NPIs outcomes in countries with different demographic
and societal structures. Indeed the first use of PyRoss was to make
such predictions for COVID-19 in India \cite{Singh2020} where three-generation
households are widespread. More generally, resolving the age structure
of an epidemic is clearly essential in modelling diseases whose transmission
and/or morbidity shows a strong age-dependence.

\section{NPIs and compartment structure}

NPIs are modelled in PyRoss using either user-specified or inference-based
modifications to the contact matrices. For instance the closure of
schools reduces the matrix elements between school-going age groups
by the school-based contact contribution. It is clear in this context
that subdivision to resolve the contact matrix more finely (for instance,
splitting work environments by occupation or sector, and adding retail,
catering and entertainment contributions) would improve predictive
power. Such models could address the fact that NPIs such as social
distancing will have different efficacies (and also different costs,
see below) in each context. Although PyRoss currently implements only
home, school and work compartments, finer subdivision requires no
fundamental change to its structure.

A further extension of PyRoss adds a compartment-based treatment of
spatial structure whereby locally well-mixed populations in different
neighbourhoods exchange individuals by means of additional, reversibly
populated compartments describing a ``commuterverse''. The resulting
codebase, PyRossGeo \textbf{\footnote{\href{https://github.com/lukastk/PyRossGeo}{https://github.com/lukastk/PyRossGeo}}},
which has been implemented for a model of Greater London, will be
the subject of a separate report. As described there, this approach
to spatial modelling could offer advantages over agent-based and other
approaches, particularly in mid-epidemic where disease dynamics may
become effectively deterministic even at local scale (at least in
a city as densely populated as London).

\section{Limitations} Of course, compartment models are not universally applicable. One
drawback emerges on considering strong lockdown scenarios in which
transmission between households becomes extremely low. This violates
the concept of a well-mixed population; disease cannot spread no matter
how high the intra-household contact rate. However, it seems plausible
that by transferring attention from individuals to households (each
described as a set of several individuals that becomes exposed when
any one of them is infected) a compartment model for this situation
could also be developed. Challenges also arise in modeling track-and-test,
quarantine and other individually resolved NPIs, but we suspect some
of these challenges can also be overcome by careful compartment design.

We surmise that the full potential of compartment models
in epidemic modelling remains unrealized. To help achieve this potential,
the PyRoss library currently supports general age-structured compartment models
formulated stochastically (in terms of so-called `chemical master
equations' or CMEs) or deterministically (as systems of differential
equations). The library can automatically switch between stochastic
and deterministic descriptions such that the more expensive stochastic
sampling is used when compartment numbers are small, as they are in
the early and late stages of an epidemic, while the less expensive
deterministic sampling is used when compartment numbers are large,
as they are near the peak of the epidemic, at least within well-mixed
models. This allows for accurate and efficient sampling of entire
epidemiological trajectories taking into account the intrinsic stochasticity
of the transmission process.

Perhaps the greatest limitation of compartment models is the tendency toward proliferation within such models of large numbers of parameters to describe the transition rates between compartments, many of which cannot be directly measured. A disciplined methodology for parameter estimation is then essential.

\section{Bayesian parameter inference and forecasting}

Accordingly, PyRoss supports fitting to epidemiological (and other)
data using Bayesian parameter inference and model selection. We use
the Gauss-Markov limit of the discrete-state continuous-time Markov
process described by the relevant chemical master equation for this
purpose. Then, the standard machinery of Gaussian process regression
can be used to infer parameters. The kernels of the Gaussian process
are adapted, via the Gauss-Markov limit, to the models being fitted
and these adapted kernels are obtained from the solution of systems
of ordinary differential equations. The numerous advantages of Gaussian
process regression, in particular tractable inference and the ability
to admit latent variables, accrue automatically. The Bayesian {\em
model evidence}, necessary for model comparison, model selection
and model averaging, alongside the {\em Fisher information matrix},
necessary for assessing parameter sensitivities, are thereby obtained
in PyRoss without the need for computationally expensive sampling
methods such as Markov Chain Monte Carlo.

Within PyRoss the future course of an epidemic can be predicted, with
Bayesian confidence estimation, from models fitted to data gathered
up to the present. These forecasts take into the account uncertainties
in data, the epidemiological parameters, choice of models, and the
intrinsic stochasticity of the transmission. The Bayesian methodology
is known to provide a principled way of managing uncertainty \cite{mackay2003information,mackay1992bayesian},
which is inherent in epidemic modelling and prediction.

Bayesian methods also defend against over-fitting, which is a risk
in all models with large numbers of parameters. For instance, the CovidSim
model of \cite{ferguson2020impact} has, we estimate, over 600 parameters,
many of which are not readily accounted for even after geographical,
social and other data inputs are identified. Over-fitted models, containing parameters unconstrained by data, are
liable to be precise but inaccurate forecasters, particularly under
data-poor conditions as is currently the case for COVID-19. In contrast,
relatively simple compartment models can be adequate forecasters under these same
conditions. An appendix to this report
gives a tutorial discussion of
the Bayesian approach to this topic for the benefit of those unfamiliar
with it.

\section{Optimising the outcomes of NPIs}

NPIs such as lockdown or social distancing each carry a cost (including,
but not limited to, economic, medical, social and ethical costs).
This can differ substantially between interventions. For measures
that treat different age groups differently, age-structured models
can be used to assess the their differential impact. Similar remarks
apply to other types of compartmenting in more general models, for
instance by geographic, social or industrial sector, as we plan to consider
in subsequent reports.

By representing NPIs through their effects on contact matrices, PyRoss
can provide Bayesian forecasts of the impact of specific interventions
(albeit limited in the currently implemented examples to the age-structured
case). Intervention strategies that extremise imputed costs can then
be found without leaving the platform.

The optimisation methodology provided within PyRoss a principled
way of minimising harm in sustained application of NPIs. The exact
definition of `harm' is of course controversial and subjective. To
create a well posed optimization problem, the first step is formulation
of a cost function which encodes mathematically how one type of harm
(such as fatalities) is weighed against another (such as unemployment).
For obvious reasons, PyRoss leaves the choice of cost function to
the user, although a schematic example of the approach is described later in this
report.

\section{Provenance}

PyRoss is named after Sir Ronald Ross, doctor, mathematician and poet.
In 1898 he made \textquotedbl the great discovery\textquotedbl{} in
his laboratory in Calcutta \textquotedbl that malaria is conveyed
by the bite of a mosquito\textquotedbl . Ross won the Nobel Prize
for Physiology or Medicine in 1902 and helped lay the foundations
of the mathematical modelling of infectious diseases.

As previously mentioned, PyRoss was initially developed to study the
age-structured impact of social distancing on the COVID-19 epidemic
in India \cite{Singh2020}. Currently it is being developed by the
authors of this report who form a task team of the RAMP initiative
(Rapid Assistance in Modelling the Pandemic) coordinated by the Royal
Society. The library can be used directly to study the transmission,
and mitigation through NPIs, of infectious diseases for any well mixed
demographic where age and social contact structures are available.
As described already, extensions to more general compartment models
are either nearing completion (PyRossGeo), or readily envisaged within
the same coding structure. More will be added as resources allow.

Compartment models in epidemiology have a long history and while a
number of references are cited below, we have not had time to survey
the literature as thoroughly as we would like in preparing this report.
Some parts may therefore cover territory familiar to experts, but
our aim here is to present PyRoss to a wider scientific community
as an open-source, well coded Python library that combines the flexibility
and generality of age-stratified (and prospectively more general)
compartment models with appropriate inference and optimization tools.
We hope this resource might prove equally valuable to seasoned epidemic
modellers and to the many now entering the field for the first time.

\section{Remaining chapters}

The remainder of this report is structured as follows. In Chapter
2 we provide an overview of metapopulation models of infectious diseases,
focussing on age and contact structures. We then sketch a pipeline
where models and epidemiological data can be combined, forecasts can
be made, and the impact of interventions can be studied in a fully
Bayesian manner. In Chapter 3 we outline the theory of inference,
prediction, and optimal control of the abstract Markovian epidemic
where the pipeline of the preceding chapter can be feasibly realised.
In Chapter 4 we provide a bestiary of epidemiological compartment
models that illustrate but by no means exhaust PyRoss's capabilities for model construction.
We discuss compartment subdivision
to allow non-exponentially distributed infectious periods and also
to uncouple combinations of rate parameters that would otherwise not
be independent. Although the specific examples we discuss have been
hard-optimized for the PyRoss library, we also describe how user-defined
models can be implemented with only mildly reduced efficiency while retaining full use of the surrounding inference and optimisation tools. In
Chapter 5 we present several fully coded worked examples in PyRoss, touching upon
sampling, inference, prediction, control and optimal control. As in
the preceding chapter, our purpose is illustrative and focuses on
simple rather than fully realistic cases.
We conclude in Chapter 6 with a brief summary. 

\chapter{Inference, Prediction and Optimized Intervention for Compartment Models}

\begin{figure*}
\centering\includegraphics[width=0.9\textwidth]{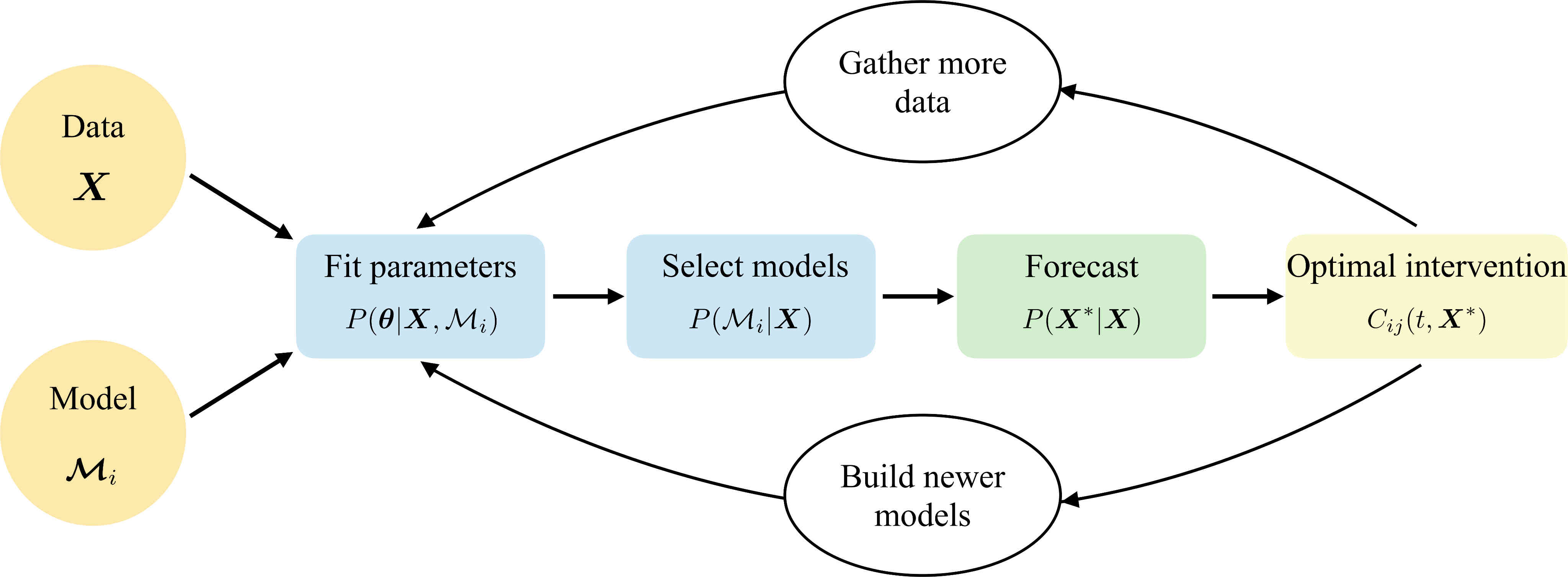}\caption{\textbf{Inference, prediction and intervention framework in PyRoss.}
For explanations of mathematical symbols see Chapter \ref{chap:markov}.}
\label{IPIfig}\label{fig:pipeline}
\end{figure*}
We now specify in more detail the types of model PyRoss currently
implements and outline how these form part of an integrated inference,
prediction and optimization pipeline, shown schematically in Fig.~\ref{fig:pipeline}.

\section{Compartment models}

PyRoss is designed to simulate structured epidemiological compartment
models. The basic variable in this class of models is a metapopulation
labeled by its epidemiological state (susceptible, infectious, recovered/removed,
etc.,) and additional attributes like age, gender, geographic location
and so on \cite{anderson1992infectious,keeling2011modeling,towers2012social,ferguson2006strategies,hethcote2000mathematics}.
These additional attributes define the structure of the model. The user can specify more disease-state compartments than is traditional,
and examples described later in this report include not only susceptible
($S$), infectious ($I$), exposed ($E$), quarantined ($Q$) and
recovered/removed ($R$) states but also, for example, subdivisions
of I into asymptomatic and symptomatic, with the latter further divided
into hospitalized and in intensive care. Additionally, the infectious
class or classes can be subdivided into $k$ time stages \cite{anderson1980spread,wearing2005appropriate,krylova2013effects,lloyd2001realistic}
to approximate fixed overall residence times as described in Chapter
\ref{chapter:1}. Similar latent compartments can also be used to
overcome other constraints on the disease dynamics that arise from
a few-compartment approach.

The compartment subdivision scheme exploits
PyRoss's efficient representation of transitions between compartments as Markovian
jump processes of given rate. This means that the residence time in
any (sub-)compartment is exponentially distributed unless the jump
rates themselves vary in time (as they do if an NPI regime is changed).
The time evolution of the compartment occupancy variables are accordingly
described by chemical master equations with time-dependent rates and,
when compartmental fluctuations are small so that deterministic dynamics
arises, by ordinary differential equations.

\section{Age structure and contact matrices}

\begin{figure*}[t]
\centering\includegraphics[width=0.98\textwidth]{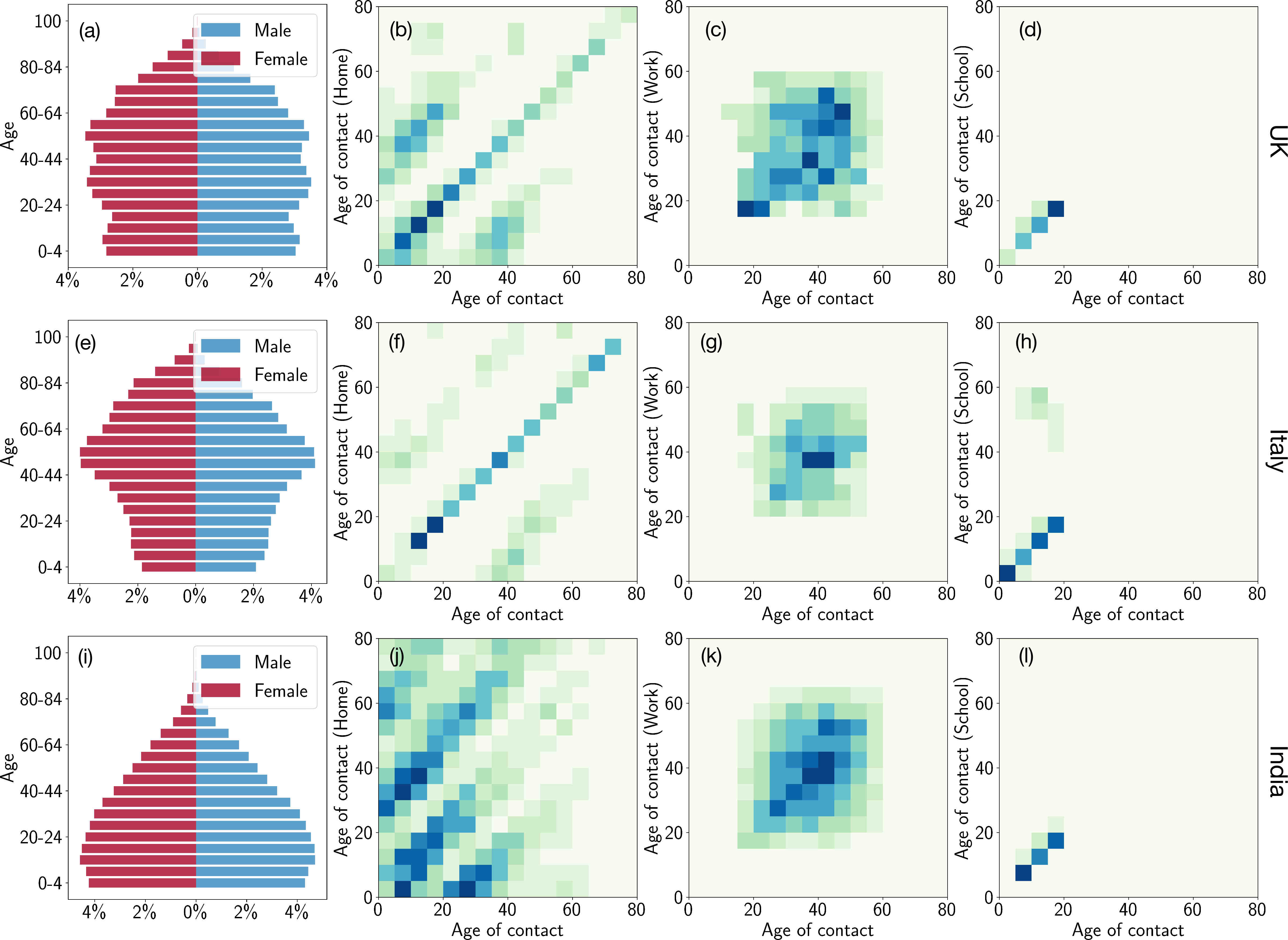}\caption{\textbf{Age and contact structures of the populations of UK (top),
Italy and India (bottom).} The first column shows population pyramids
by age and gender. The second third and fourth columns show the contact
structures in households, workplaces and schools respectively. The
darker colours represent greater contacts. Taken from \cite{prem2017projecting}.
\label{fig:age-contact}}
\end{figure*}

An infectious disease is spread by social contacts which are typically
assortative in age \cite{anderson1992infectious,glasser2012mixing}.
Therefore, it is important to account for the country-specific age
and social contact structures when modelling the spread of infection.
Also, in diseases where morbidity and mortality are strongly age-dependent,
as with COVID-19, it is important not only to know how many people
are likely to be infected but also how they are distributed in age.
For example, India is one of the few countries in the world with a
high prevalence of three-generation households. This poses a large
risk to the elderly, as they are more likely to catch the disease
from second- and third-generation members of the household with greater
social contacts. PyRoss has already provided the first age-structured
forecasts of morbidity and mortality of the COVID-19 epidemic in India
\cite{Singh2020} and can do so for any community in which the appropriate
age stratified contact structure is available, including the 152 countries
reported in \cite{prem2017projecting}; see Fig.~\ref{fig:age-contact}
for examples.

The resulting differences between countries can be substantial. For
instance, with a simple SIR model \cite{ross1916application,keeling2011modeling,anderson1992infectious,bailey1975mathematical},
the difference between age and contact structures in UK and India
leads to a basic reproduction ratio $\mathcal{R}_{0}$ that is more
than 50 percent higher in the latter case; see Table~\ref{two-one}.
This of course means that identical NPIs could suppress the epidemic
in one country and fail completely to do so in another. (For similar
reasons it could fail in cities but succeed in the countryside, for
example.)

\begin{table}
\global\long\def\arraystretch{1.5}%
 \centering %
\begin{tabular}{|>{\centering}p{2cm}|>{\centering}p{6cm}|}
\hline 
Country & Basic reproductive ratio\tabularnewline
\hline 
\hline 
UK & $\mathcal{R}_{0}=82\beta$\tabularnewline
\hline 
Italy & $\mathcal{R}_{0}=119\beta$\tabularnewline
\hline 
India & $\mathcal{R}_{0}=136\beta$\tabularnewline
\hline 
\end{tabular}\caption{Country-specific basic reproductive ratio of the age-structured SIR
model for fixed probability of infection on contact $\beta$ and recovery
rate $\gamma=1/7$.}
\label{two-one}
\end{table}

\section{Modelling and optimizing NPIs}

The purpose of many NPIs, such as lockdown and social distancing,
is to reduce either the frequency of social contacts or the transmission
rates at each contact. Absorbing the latter into the former, such
NPIs simply alter the contact matrix. Age-structured models can therefore
address age-structured NPIs; the examples in this report are limited
to such cases, retaining for simplicity the simplest three-way breakdown
of contact spaces (Fig.~\ref{fig:age-contact}). As mentioned in
Chapter \ref{chapter:1}, further compartmenting in principle allows
resolution of NPIs that affect specific subsets of social contacts,
such as closing restaurants or public transport systems, without raising
major issues of principle.

An obvious example is schools closure, an NPI for which the schools
channel of the contact matrix is set to zero -- or perhaps some nonzero
multiplier of its normal value to describe partial opening. Another
simple case is a phased unlock stratified by age-bands. A model with
additional structure, such as workplaces stratified by occupation
or sector, should allow the outcome of more nuanced NPIs to be predicted
in the same way. 

In PyRoss, we partition contacts into spheres of home, workplace,
school and all other categories, such that the contact matrix can
be written as
\begin{align}
C_{ij}(t) & =a^{H}(t)C_{ij}^{H}+a^{W}(t)C_{ij}^{W}+a^{S}(t)C_{ij}^{S}+a^{O}(t)C_{ij}^{O}.\label{eq:CM}
\end{align}
The social contact matrix $C_{ij}$ denotes the average number of
contacts made per day by an individual in class $i$ with an individual
in class $j$. Clearly, the total number of contacts between group
$i$ to group $j$ must equal the total number of contacts from group
$j$ to group $i$, and thus, for populations of fixed size the contact
matrices obey the reciprocity relation $N_{i}C_{ij}=N_{j}C_{ji}$
\cite{busenberg1991general,wallinga2006using}, where $N_{i}$ is
the population in age-group $i$. 

In general, PyRoss expects
NPIs to be represented by time-dependent contact matrices. Its inference
platform for learning from data is structured accordingly to infer
time dependent parameters where appropriate. For example, sufficient
age-stratified disease data spanning an initially unlocked and subsequently
locked-down state, should allow Bayesian prediction of the effects
not just of full unlock but of
a partial unlock comprising school closure plus home sequestration
of people over 70 (say). There are clearly limitations to this, in
that the effect of a completely novel intervention cannot be predicted
without prior knowledge of its likely effects on contact matrices.
However such priors might be provided by expert judgement. Then, upon
actually starting the intervention, incoming data can refine and update
model parameters to give increasingly confident prediction of its
future effects.

For policy purposes it is obviously desirable to allow objective comparison
of alternative NPIs. PyRoss supports user-defined cost functions to
allow this. Intervention strategies that extremise the chosen cost
function can then be found using optimization tools which are built
into PyRoss as detailed in Chapter~\ref{chap:markov}.

\chapter{Techniques\label{chap:markov}}

In this chapter, we describe age-structured epidemiological compartment
models formulated as discrete-state continuous-time Markov processes.
We present the chemical master equation (CME) that describes such
processes and then provide the diffusion and Gauss-Markov approximations
of the process. The Gauss-Markov approximation is used for Bayesian
parameter inference and to compute the model evidence. Bayesian posterior
predictive distributions are used to provide forecasts given epidemiological
data. Such forecasts convolve uncertainties arising from epidemiological
data, parameters, models, and intrinsic stochasticity. Non-pharmaceutical
interventions (NPI) are imposed by altering the contact structures
of the models. Incorporating uncertainties, a Bayesian forecast of
the effect of such interventions can be obtained. The protocol of
the NPI - defined for instance by points of triggering and its duration
- can be optimized by extremizing a supplied cost function. This sequence
of inference, prediction and intervention can be applied iteratively
with the arrival of newer data, as shown pictorially in Fig.(\ref{fig:pipeline}),
leading to improved epidemiological compartment models. 

\section{Discrete-state continuous-time Markov process \label{sec:Discrete-state-continuous-time-m}}

We consider a structured metapopulation 
\begin{equation}
\boldsymbol{n}=(n_{1},\ldots n_{L\times M})\label{eq:notation}
\end{equation}
consisting of $L$ classes of epidemiological states and $M$ age-compartments.
The $\xi$-th transition between compartments can be written down
in its most general form as

\begin{equation}
\xi\text{-th transition step:}\quad\boldsymbol{n}\xrightarrow{w_{\xi}}\boldsymbol{n}+\boldsymbol{r}_{\xi}\label{eq:population_change}
\end{equation}
where $\boldsymbol{r}_{\xi}$ is the vector of change and $w_{\xi}$
is the rate for the transition $\xi$. This gives the chemical master
equation (CME) for the evolution of joint distribution over states
$P(\boldsymbol{n},t)$ : 
\begin{equation}
\partial_{t}P(\boldsymbol{n},t)=\sum_{\xi}\left[w_{\xi}(t,\boldsymbol{\theta},\boldsymbol{n}-\boldsymbol{r}_{\xi})P(\boldsymbol{n}-\boldsymbol{r}_{\xi},t)-w_{\xi}(t,\boldsymbol{\theta},\boldsymbol{n})P(\boldsymbol{n},t)\right].\label{eq:Master}
\end{equation}
Here $\boldsymbol{\theta}=(\theta_{1},\ldots,\theta_{k})$ is the
set of parameters for transitions between the states. The CME describes
a discrete-state continuous-time Markov process on the positive integers.
The CME resists solution for all but the simplest form of the transition
rates, which can be at most linear. Since the transmission of contagion
necessarily involves the contact of, at least, pairs the rates in
epidemiological models cannot be linear. Numerical sampling, or, analytical
approximations become necessary to extract model behavior from the
CME. We describe the analytical approximations in the next section
and numerical sampling in Section (\ref{sec:Numerical-methods}).

\section{Approximations}

The family of approximations we consider here replaces the discrete-state
continuous-time Markov process by a continuous-time continuous-state
Markov process. In this diffusion limit, the transitions $\boldsymbol{n}\xrightarrow{}\boldsymbol{n}+\boldsymbol{r}_{\xi}$
in the discrete state space $\boldsymbol{n}$ are replaced by transitions
$\boldsymbol{x}\xrightarrow{}\boldsymbol{x}+d\boldsymbol{r}$, in
a continuous state space $\boldsymbol{x}$ with continuous increments
$d\boldsymbol{r}$ and appropriately chosen rates. The continuous
state $x$ is usually the discrete state rescaled by the size of the
population, $\boldsymbol{x}=\boldsymbol{n}/N$. This makes it apparent
that the diffusion approximation is appropriate when the population
size $N$ is large. Here $N=\sum_{i=1}^{M}N_{i}$ and $N_{i}$ is
total population in the age-group $i=1,2,\dots M$. 

\subsection{Diffusion limit and the $\Omega$-expansion\label{subsec:diffusion limit}}

Formally, the diffusion limit is obtained by truncating the Kramers-Moyal
expansion of the CME to second order. The theorem due to Pawula \cite{pawula1967approximation}
constrains this truncation to be the only one that yields positive-definite
probability distributions. The result is the so-called \emph{chemical
Fokker-Planck equation }(CFPE) with drift and diffusion coefficients
that are jump moments of the transition rates of the CME \cite{gardiner1985handbook,vanKampen1992stochastic}:

\begin{equation}
\partial_{t}P(\boldsymbol{x},t)=L(t\boldsymbol{,\theta,x})P(\boldsymbol{x},t),\quad L=-\frac{\partial}{\partial x_{i}}A_{i}(t,\boldsymbol{\theta},\boldsymbol{x})+\tfrac{1}{2}\frac{\partial^{2}}{\partial x_{i}\partial x_{j}}B_{ij}(t,\boldsymbol{\theta},\boldsymbol{x}),\label{eq:CFPE}
\end{equation}
where
\begin{align}
\boldsymbol{A}(t,\boldsymbol{\theta},\boldsymbol{x}) & =\sum_{\xi}\boldsymbol{r}_{\xi}w_{\xi}(t,\boldsymbol{\theta},\boldsymbol{x}),\label{eq:First}\\
\boldsymbol{B}(t,\boldsymbol{\theta},\boldsymbol{x}) & =\sum_{\xi}\boldsymbol{r}_{\xi}\boldsymbol{r}_{\xi}w_{\xi}(t,\boldsymbol{\theta},\boldsymbol{x}).\label{eq:Second}
\end{align}
The equivalent Itô stochastic differential equation (SDE) is, 
\[
d\boldsymbol{x}=\boldsymbol{A}(t,\boldsymbol{\theta},\boldsymbol{x})dt+\boldsymbol{\sigma}(t,\boldsymbol{\theta},\boldsymbol{x})\cdot d\boldsymbol{W}
\]
 where $\boldsymbol{\sigma}$ is such that $\boldsymbol{\sigma}\boldsymbol{\sigma}^{T}=\boldsymbol{B}$
and $\boldsymbol{W}$ is a $L\times M$ dimensional Wiener process
with zero mean and unit variance.

The diffusion approximation yields an Ito process with configuration-dependent
noise. A further approximation due to Van Kampen decomposes the diffusion
process into a mean process without noise and a fluctuation which
is described by a time-dependent Ornstein-Uhlenbeck process. The resulting
process is both Markovian and Gaussian. In this linear noise approximation
(LNA), the state is expressed as,

\begin{equation}
\boldsymbol{x}(t)=\boldsymbol{x}^{0}(t)+\frac{1}{\sqrt{\Omega}}\boldsymbol{x}^{1}(t) \label{eq:ansatz}
\end{equation}
where $\Omega=N$ is the system size. The two components satisfy
\begin{equation}
d\boldsymbol{x}^{0}(t)=\boldsymbol{A}(t,\boldsymbol{\theta},\boldsymbol{x}^{0})dt,
\end{equation}

\begin{equation}
d\boldsymbol{x^{1}}=\boldsymbol{J}(t,\boldsymbol{\theta},\boldsymbol{x}^{0})\cdot\boldsymbol{x^{1}}dt+\boldsymbol{\sigma}(t,\boldsymbol{\theta},\boldsymbol{x}^{0})\cdot d\boldsymbol{W},
\end{equation}
where 
\[
J_{ij}(t,\boldsymbol{\theta},\boldsymbol{x}^{0})=\partial_{j}A_{i}(t,\boldsymbol{\theta},\boldsymbol{x})\Big|{}_{\boldsymbol{x}=\boldsymbol{x}^{0}}
\]
is the time-dependent Jacobian evaluated at $\boldsymbol{x}^{0}(t)$.
The lowest order term $\boldsymbol{x}^{0}$ describes the mean evolution
whereas the next order term $\boldsymbol{x}^{1}$ characterises the
Gaussian fluctuation around the deterministic trajectory. The fluctuation
$\boldsymbol{x}^{1}$ is a non-stationary Ornstein Uhlenbeck process
whose mean and variance can be obtained for any point on the mean
trajectory. 

As a note of caution, there are some subtleties associated with whether
the high order terms can grow significantly over time. This is particularly
relevant to us as the growth is exponential at the early stage of
an epidemic, where the approximation is not likely to be good. 

\subsection{Deterministic limit \label{subsec:Deterministic-limit}}

From the $\Omega$-expansion, we obtain the deterministic limit for
free as the evolution of the mean $\bar{\boldsymbol{x}}=\boldsymbol{x}^{0}+\expval{\boldsymbol{x}^{1}}$.
It is simple to show that the mean obeys the same equation as $\boldsymbol{x}^{0}$, 

\[
\frac{d\bar{\boldsymbol{x}}}{dt}=\boldsymbol{A}(t,\boldsymbol{\theta},\boldsymbol{\bar{x}})
\]
In \lstinline!pyross.deterministic!, the rescaling of $\beta$ is
built in to suit simulations with both intensive and extensive variables. 

\section{Inference\label{sec:Inference}}

In this section, we describe inference of parameters $\boldsymbol{\theta}=(\theta_{1},\ldots,\theta_{k})$
performed by the \lstinline!pyross.inference! module. The data consists
of the time series
\begin{alignat}{1}
\boldsymbol{X} & =\{x_{i}(t_{\nu})\,\,|\,\,i=1,\ldots M\times L;\nu=1,\ldots N_{t}\}\label{eq:tseries}
\end{alignat}
of the values of the components $x_{i}(t_{\nu})$ at the times $t_{\nu}$.
The module is capable of inferring the parameters with full and partial
information; in the latter case, the latent variables are also inferred.
The module determines Bayesian credible intervals for the inferred
parameters, which can be supplied to \lstinline!pyross.forecast!
to perform Bayesian forecasting with both parameter uncertainty and
inherent stochasticity. In the future, mode selection will be added
to \lstinline!pyross.inference! to measure how well the various epidemiological
models fit the data. 

\subsection{Non-stationary Gauss-Markov process}

Consider the time interval $(t_{\nu},t_{\mu}$) with initial condition
$\boldsymbol{x}(t_{\nu})=\boldsymbol{x'}$. The outcome of the $\Omega$-expansion
is a Gauss-Markov process, implying that the conditional probability
$P_{1|1}(\boldsymbol{x},t_{\mu}|\boldsymbol{x}',t_{\nu})$ follows
a Gaussian distribution determined uniquely by its conditional mean
$\bar{\boldsymbol{x}}(t_{\mu})=\expval{\boldsymbol{x}(t_{\mu})|\boldsymbol{x}'}$
and conditional variance, denoted as $\boldsymbol{\Sigma}$ \cite{Komorowski09}. 

\[
\boldsymbol{x},t_{\mu}|\boldsymbol{x}',t_{\nu}\:\sim\mathcal{N}(\bar{\boldsymbol{x}}(t_{\mu}),\boldsymbol{\Sigma}(t_{\mu}))
\]
Recall from section \ref{subsec:Deterministic-limit} that the mean
obeys the deterministic equation, 

\[
\frac{d\bar{\boldsymbol{x}}}{dt}=\boldsymbol{A}(t,\boldsymbol{\theta},\boldsymbol{\bar{x}})
\]
with the initial condition $\bar{\boldsymbol{x}}(t_{\nu})=\boldsymbol{x}'$,
which can be solved numerically with \lstinline!pyross.deterministic!.
The conditional variance, defined as, 

\[
\boldsymbol{\Sigma}(t)=\expval{\boldsymbol{x}(t)\boldsymbol{x}^{T}(t)|\boldsymbol{x}',t_{\nu}}-\bar{\boldsymbol{x}}(t)\bar{\boldsymbol{x}}(t)^{T}
\]
only has contributions from the fluctuating term $\boldsymbol{x}^{1}$
since $\boldsymbol{x}^{0}$ evolves deterministically,

\[
\boldsymbol{\Sigma}(t)=\frac{1}{\Omega}\left[\expval{\boldsymbol{x}^{1}(t)(\boldsymbol{x}^{1})^{T}(t)|0,t_{\nu}}\right]
\]
where we note that the mean of $\boldsymbol{x}^{1}$ remains zero
as long as we choose $\boldsymbol{x}^{1}(t_{\nu})=0$ (which can always
be done as the splitting of the initial condition $\boldsymbol{x}$'
into $\boldsymbol{x}^{0}$ and $\boldsymbol{x}^{1}$ is arbitrary).
It is a well established result that the covariance of an Ornstein-Uhlenbeck
process evolves according to the time-dependent differential Lyapunov
equation \cite{vanKampen1992stochastic,gardiner1985handbook}, 

\begin{equation}
\dot{\boldsymbol{\Sigma}}(t)=\boldsymbol{J}(\bar{\boldsymbol{x}}(t),t)\boldsymbol{\Sigma}(t)+\boldsymbol{\Sigma}(t)\boldsymbol{J}^{T}(\bar{\boldsymbol{x}}(t),t)+\boldsymbol{B}(\bar{\boldsymbol{x}}(t),t)/\Omega\label{eq:Lyapunov}
\end{equation}
For our models of interest, the Lyapunov equation cannot be solved
analytically and is computed numerically in \lstinline!pyross.inference!.
For improved accuracy, $\boldsymbol{J}$ and $\boldsymbol{B}$, both
of which are functions of $\bar{\boldsymbol{x}}(t)$, are computed
by interpolating the numerical solution of the deterministic equation
using Chebyshev polynomials. Once both the mean and the variance are
computed, the conditional probability $P_{1|1}(\boldsymbol{x},t_{\nu+1}|\boldsymbol{x}(t_{\nu}),t_{\nu})$
is given by 

\[
\ln P_{1|1}(\boldsymbol{x},t_{\mu}|\boldsymbol{x}',t_{\nu})=-\left(\frac{1}{2}\boldsymbol{\Delta}_{\mu}^{T}\boldsymbol{\Sigma}_{\mu}^{-1}\boldsymbol{\Delta}_{\mu}+\frac{1}{2}\ln\det2\pi\boldsymbol{\Sigma}_{\mu}\right)
\]
where $\Delta_{\mu}=\boldsymbol{x}-\bar{\boldsymbol{x}}(t_{\mu})$
is the deviation of the endpoint $\boldsymbol{x}$ from the deterministic
trajectory. 

\subsection{Parameter inference}

The conditional probability can be computed as above for each time
step, the posterior, using flat priors enforcing positive parameters,
is given as the sum of the conditional probabilities, 

\begin{equation}
\ln P(\boldsymbol{\theta}|\boldsymbol{X})=-\sum_{\mu=1}^{N_{t}}\left(\frac{1}{2}\boldsymbol{\Delta}_{\mu}^{T}\boldsymbol{\Sigma}_{\mu}^{-1}\boldsymbol{\Delta}_{\mu}+\frac{1}{2}\ln\det2\pi\boldsymbol{\Sigma}_{\mu}\right).\label{eq:detPost}
\end{equation}
Here $\boldsymbol{\Delta}_{\mu}=\boldsymbol{x}(t_{\mu})-\bar{\boldsymbol{x}}(t_{\mu})$.
This is then used to numerically obtain the MAP estimates for the
parameters. Near the onset of an epidemic, the higher order fluctuations
can grow exponentially and significantly deviate from the mean trajectory,
to resolve this issue, one needs to sample more frequently where the
growth rate (effectively $R_{0}$) is large. 

\subsection{Latent variables}

With limited information, we may only be able to observe a few components
$\boldsymbol{X}_{\mathrm{Reduced}}=\{x_{i}(t_{\nu})\,\,|\,\,i=1,...M_{\mathrm{Reduced}}\times L_{\mathrm{Reduced}};\nu=1,\ldots N_{t}\}$,
e.g. the number of confirmed cases and deaths. 

The first step towards inference for latent variables is to express
the probability of an full time series for a given set of parameters.
One would then need to trace out unobserved degrees of freedom. However,
this is hard to do in practice, because the joint distribution may
be highly non-Gaussian. This is the case even though the propagator
itself is Gaussian, because $\boldsymbol{\varSigma}$ and $\bar{\boldsymbol{x}}$
may depend on $\boldsymbol{x}$. Nonetheless, Gaussian inference is
possible for latent variables if the system-size expansion remains
approximately valid for time intervals as long as the whole observed
time series. Then, all typical trajectories remain close to the mean
of the time series, evolving according to the ODE 
\begin{equation}
\partial_{t}\boldsymbol{\bar{\boldsymbol{x}}}=\boldsymbol{A}(\bar{\boldsymbol{x}}),
\end{equation}
with the drift vector $\boldsymbol{A}$. One has to be careful, though,
because the Jacobian of $\boldsymbol{A}$ typically has positive eigenvalues
which results in exponential growth of any deviation from the mean.
We need a sufficiently large system size $N$. 

We now write $\boldsymbol{x}(t)=\bar{\boldsymbol{x}}(t)+\boldsymbol{\Delta}(t)$.
The variable $\boldsymbol{\Delta}=\boldsymbol{x}^{1}/\sqrt{\Omega}$
and therefore also evolves according to an Ornstein Uhlenbeck process.
The conditional probability distribution for $\boldsymbol{\Delta}$
at time $t_{\mu}$ given its value at time $t_{\nu}$ is 
\begin{equation}
\boldsymbol{\Delta}_{\mu},t_{\mu}|\boldsymbol{\Delta}_{\nu},t_{\nu}\sim\mathcal{N}(\boldsymbol{U}_{\mu}\boldsymbol{\Delta}_{\nu},\boldsymbol{\Sigma}_{\mu})
\end{equation}
where the matrix $\boldsymbol{U}_{\mu}=\boldsymbol{U}(t_{\mu})$ is
the time evolution operator, defined by the equation $\partial_{t}\boldsymbol{U}(t)=\boldsymbol{J}(\bar{\boldsymbol{x}}(t))\boldsymbol{U}(t)$.
It yields the mean of $\boldsymbol{\Delta}(t_{\mu})$ for given $\boldsymbol{\Delta}(t_{\nu})$.
$\boldsymbol{\Sigma}_{\mu}$ is calculated from Eq.\ref{eq:Lyapunov},
but $\bar{x}$ is different, since we couldn't update it with a recent
initial condition. For a Gaussian initial distribution with covariance
$\boldsymbol{\Sigma}_{0}$, the conditional probabilities are concatenated
to yield the joint distribution. The overall covariance matrix for
the vectors $(\boldsymbol{\Delta}(t_{1}),\ldots,\boldsymbol{\Delta}(t_{N_{t}}))^{T}$
is given by the inverse of the following tridiagonal block matrix:
\begin{equation}
\boldsymbol{\varSigma}^{-1}=\left(\begin{matrix}\\
\boldsymbol{\Sigma}_{1}^{-1}+\boldsymbol{U}_{2}^{T}\boldsymbol{\Sigma}_{2}^{-1}\boldsymbol{U}_{2} & -{\normalcolor \boldsymbol{U}}_{2}^{T}\boldsymbol{\Sigma}_{2}^{-1} & 0 & \dots\\
\\
-\boldsymbol{\Sigma}_{2}^{-1}\boldsymbol{U}_{2} & \boldsymbol{\Sigma}_{2}^{-1}+\boldsymbol{U}_{3}^{T}\boldsymbol{\Sigma}_{3}^{-1}\boldsymbol{U}_{3} & -\boldsymbol{U}_{2}^{T}\boldsymbol{\Sigma}_{2}^{-1} & 0 & \dots\\
\\
0 & -\boldsymbol{\Sigma}_{3}^{-1}\boldsymbol{U}_{3} & \ddots & \ddots\\
\\
\vdots & 0 & \ddots & \ddots
\end{matrix}\right).\label{eq:bigmatrix}
\end{equation}
Note that after the inversion $\boldsymbol{\varSigma}$ typically
has nonzero entries everywhere. With latent observable, we simply
eliminate all rows and columns associated with (sub-)indices other
than those known. The joint distribution of the remaining then follows
as a multivariate Gaussian with the reduced covariance and the mean
for the observed variables. In the \lstinline!pyross.inference! module,
the initial conditions for $\boldsymbol{x}$ are also treated as parameters
to be inferred, as they are unknown for the unobserved variables.

\subsection{Model evidence\label{subsec:Model-evidence}}

Bayesian credible intervals (BCI), or standard deviation of the maximum
a posteriori (MAP) estimates, can be understood by writing $\boldsymbol{H}=-\boldsymbol{\nabla\nabla}\ln P(\boldsymbol{\theta}|\boldsymbol{X},\mathcal{M}_{i})|_{\boldsymbol{\theta}^{*}}$
for the Hessian, and Taylor expanding the log posterior of the parameters
$\boldsymbol{\theta}$, given the data $\boldsymbol{X}$ and the model
$\mathcal{M}_{i}$ with $\Delta\boldsymbol{\theta}=\boldsymbol{\theta}-\boldsymbol{\theta}^{*}$,
where $\boldsymbol{\theta}^{*}$ are the MAPs \cite{mackay2003information,singh2018fast},
\begin{equation}
P(\boldsymbol{\theta}|\boldsymbol{X},\mathcal{M}_{i})\approx P(\boldsymbol{\theta}^{*}|\boldsymbol{X},\mathcal{M}_{i})\exp\left(-\frac{1}{2}\Delta\boldsymbol{\theta}^{T}\boldsymbol{H}\Delta\boldsymbol{\theta}\right).\label{eq:GPost}
\end{equation}
This is a local Gaussian approximation to the posterior around its
mode with covariance matrix $\boldsymbol{H}^{-1}$. The square root
of the diagonal elements of this covariance matrix gives the BCIs. 

In case the functional form of the posterior distribution is known,
numerical errors associated with finite differences methods can be
circumvent by using automatic differentiation (eg 'autograd \cite{maclaurin2015autograd}').
This allows for exact and rapid computation of the Hessian matrix,
neither over- nor underestimating the uncertainty of our estimates.

Mostly branded 'the second level of inference' after the first level
of obtaining the MAP estimates, the arguably most important feature
of Bayesian inference is model selection. The posterior probability
for each model is by Bayes' theorem 
\begin{equation}
P(\mathcal{M}_{i}|\boldsymbol{X})\propto P(\boldsymbol{X}|\mathcal{M}_{i})P(\mathcal{M}_{i}).\label{eq:ModPost}
\end{equation}

In the following we assume a flat model prior $P(M_{i})$, ie we a
priori have no reason to prefer one model over an other. This leaves
us with the model evidence $P(\boldsymbol{X}|M_{i})$, which at the
same time is the normalization constant in Bayes' theorem for the
posterior of the parameters $\boldsymbol{\theta}$ given the data
\begin{equation}
P(\boldsymbol{\theta}|\boldsymbol{X},\mathcal{M}_{i})=\frac{P(\boldsymbol{X}|\boldsymbol{\theta},\mathcal{M}_{i})P(\boldsymbol{\theta}|\mathcal{M}_{i})}{P(\boldsymbol{X}|\mathcal{M}_{i})}.\label{eq:ParamPost}
\end{equation}

In order to evaluate the evidence, we have to marginalize the likelihood
function over the often high-dimensional parameter space
\begin{equation}
P(\boldsymbol{X}|\mathcal{M}_{i})=\int P(\boldsymbol{X}|\boldsymbol{\theta},\mathcal{M}_{i})P(\boldsymbol{\theta}|\mathcal{M}_{i})d\boldsymbol{\theta}.\label{eq:Evd}
\end{equation}

However, if the posterior in \eqref{ParamPost} is reasonably well
approximated by a Gaussian, which is expected to be increasingly accurate
the more data we use, we can use Laplace's method to approximate the
evidence by 
\begin{equation}
P(\boldsymbol{X}|\mathcal{M}_{i})\approx\underbrace{P(\boldsymbol{X}|\boldsymbol{\theta}^{*},\mathcal{M}_{i})}_{Best\,fit\,likelihood}\underbrace{P(\boldsymbol{\theta}^{*}|\mathcal{M}_{i})(2\pi)^{k/2}\det\{\boldsymbol{H}\}^{-1/2}}_{Occam\,factor},\label{eq:Gevd}
\end{equation}
the height of the peak of the integrand in \eqref{Evd} times its
width. The Occam factor automatically penalizes over-fitting \cite{mackay2003information}.
By comparing the contributions of best fit likelihood and Occam factor
one can at least qualitatively decide whether this approximation is
expected to be sufficient. The evaluation is of course orders of magnitude
faster than computing the integral in \eqref{Evd}.

\subsection{Nested sampling}

Nested sampling (eg via 'nestle' package \cite{nestle}) is an algorithm
invented by John Skilling \cite{skilling2006nested}, dealing with
those high-dimensional integrals in the evidence calculation \eqref{Evd}
by reducing them to one-dimensional integrals over unit range. We
have found this to be rather slow for most of our models, but since
the evidence has to be calculated only once per model, it might still
be feasible to use nested sampling. Especially, when not enough data
is available and the Gaussian approximation \eqref{Gevd} fails to
be sufficient, this should be used. The method is slow when running
on a single core for the inference on the manifold as described in
sec \ref{sec:Inference}. However, we have found that for the tangent
space inference described in section \ref{sec:Tangent-space-inference},
nestle is actually reasonably fas. This is simply because of the much
faster log-posterior evaluations. With parallel processing of the
nested sampling, log-posterior evaluation takes few seconds on a 8
core machine for $M=2$ and $N=5\times10^{4}$. 

\section{Tangent space inference\label{sec:Tangent-space-inference}}
As mentioned before, a self-consistent method of taking the diffusion limit of equation
\eqref{Master} is van Kampen's system size expansion \cite{vanKampen1992stochastic}.
With the population densities $\boldsymbol{x}\equiv\boldsymbol{n}/\Omega$,
where $\Omega$ is the system size - the total population size, the
a posteriori justified ansatz of the $\Omega$-expansion is 
where $\boldsymbol{x}_{\nu}=\boldsymbol{x}(t_{\nu})$. 
The mean in \ref{eq:ansatz} evolves according to the macroscopic rate equations
(MRE) 
\begin{equation}
\dot{\bar{\boldsymbol{x}}}_{\nu}=\boldsymbol{A}(t_{\nu},\bar{\boldsymbol{x}}_{\nu},\boldsymbol{\theta}),\label{eq:MRE}
\end{equation}
and can thus be found by solving this system of ODEs. The CFPE is
equivalent to the Ito stochastic differential equation describing
the dynamics of $\boldsymbol{u}$, the so called \emph{chemical Langevin
equation }(CLE),
\begin{equation}
d\boldsymbol{u}_{\nu}=\boldsymbol{J}(t_{\nu},\bar{\boldsymbol{x}}_{\nu},\boldsymbol{\theta})\cdot\boldsymbol{u}_{\nu}dt+\boldsymbol{\sigma}(t_{\nu},\bar{\boldsymbol{x}}_{\nu},\boldsymbol{\theta})\cdot d\boldsymbol{W},\quad\boldsymbol{\sigma}\boldsymbol{\sigma}^{T}=\boldsymbol{B}(t_{\nu},\bar{\boldsymbol{x}}_{\nu},\boldsymbol{\theta}),\label{eq:CLE}
\end{equation}
with $J_{ab}=\partial_{b}A_{a}$, and $\boldsymbol{W}$ being a multi-dimensional
Wiener process. Thus, in this approximation the population density
evolves according to the stochastic process 
\begin{equation}
d\boldsymbol{x}_{\nu}=\boldsymbol{A}(t_{\nu},\bar{\boldsymbol{x}}_{\nu},\boldsymbol{\theta})dt+\Omega^{-\frac{1}{2}}d\boldsymbol{u}_{\nu},\label{eq:SDE}
\end{equation}
driven by a small (suppressed by a factor of $\Omega^{-\frac{1}{2}}$)
Ornstein-Uhlenbeck noise. Formally re-writing equation \eqref{SDE}
we get
\begin{align}
\dot{\boldsymbol{x}}_{\nu} & =\dot{\bar{\boldsymbol{x}}}_{\nu}+\Omega^{-\frac{1}{2}}\dot{\boldsymbol{u}}_{\nu}\nonumber \\
 & =\boldsymbol{A}(t_{\nu},\bar{\boldsymbol{x}}_{\nu},\boldsymbol{\theta})+\Omega^{-\frac{1}{2}}\left(\boldsymbol{J}(t_{\nu},\bar{\boldsymbol{x}}_{\nu},\boldsymbol{\theta})\cdot\boldsymbol{u}_{\nu}+\boldsymbol{\epsilon}_{\nu}\right)\nonumber \\
 & =\boldsymbol{A}(t_{\nu},\bar{\boldsymbol{x}}_{\nu},\boldsymbol{\theta})+\boldsymbol{J}(t_{\nu},\bar{\boldsymbol{x}}_{\nu},\boldsymbol{\theta})\cdot\left(\boldsymbol{x}_{\nu}-\bar{\boldsymbol{x}}_{\nu}\right)+\boldsymbol{\eta}_{\nu},\label{eq:formal}
\end{align}
with the Gaussian white noises $\boldsymbol{\epsilon}_{\nu}\sim\mathcal{N}\left(\boldsymbol{0},\boldsymbol{B}(t_{\nu},\bar{\boldsymbol{x}}_{\nu},\boldsymbol{\theta})\right)$
and $\boldsymbol{\eta}_{\nu}\sim\mathcal{N}\left(\boldsymbol{0},\frac{1}{\Omega}\boldsymbol{B}(t_{\nu},\bar{\boldsymbol{x}}_{\nu},\boldsymbol{\theta})\right)$.
Thus, with the change of measure 
\begin{equation}
P(\dot{\boldsymbol{x}}_{\nu})d\dot{\boldsymbol{x}}=P(\boldsymbol{\eta}_{\nu})d\boldsymbol{\eta}_{\nu}\label{eq:measure}
\end{equation}
and the Jacobian $\partial\eta_{\nu}^{i}/\partial x_{\nu}^{j}=\delta^{ij}$
we obtain a multivariate normal distribution for the likelihood of
$\dot{\boldsymbol{x}}$, given the parameters $\boldsymbol{\theta}$
\begin{equation}
P(\dot{\boldsymbol{x}}|\boldsymbol{\theta})=\prod_{\nu=1}^{T}\mathcal{N}\left(\langle\dot{\boldsymbol{x}}_{\nu}\rangle,\frac{1}{\Omega}\boldsymbol{B}(t_{\nu},\bar{\boldsymbol{x}}_{\nu},\boldsymbol{\theta})\right),\label{eq:Like}
\end{equation}
with the mean velocity defined by 
\begin{equation}
\langle\dot{\boldsymbol{x}}_{\nu}\rangle\equiv\boldsymbol{A}(t_{\nu},\bar{\boldsymbol{x}}_{\nu},\boldsymbol{\theta})+\boldsymbol{J}(t_{\nu},\bar{\boldsymbol{x}}_{\nu},\boldsymbol{\theta})\cdot\left(\boldsymbol{x}_{\nu}-\bar{\boldsymbol{x}}_{\nu}\right).\label{eq:mean}
\end{equation}
Assuming improper informative priors, i.e,, flat priors enforcing
positive parameters, the log-posterior distribution is
\begin{equation}
\ln P(\boldsymbol{\theta}|\dot{\boldsymbol{x}})=-\sum_{\nu=1}^{T}\left[\frac{\Omega}{2}\left(\dot{\boldsymbol{x}}_{\nu}-\langle\dot{\boldsymbol{x}}_{\nu}\rangle\right)^{T}\cdot\boldsymbol{B}^{-1}\cdot\left(\dot{\boldsymbol{x}}_{\nu}-\langle\dot{\boldsymbol{x}}_{\nu}\rangle\right)+\frac{1}{2}\ln\det\frac{2\pi}{\Omega}\boldsymbol{B}\right].\label{eq:logPost}
\end{equation}
This posterior distribution, albeit complicated, is analytical up
to integration of equation \eqref{MRE} and thus, can be easily traced
by automatic differentiation. 

Having obtained the maximum a posteriori (MAP) estimates $\boldsymbol{\theta}^{*}$
one can therefore easily obtain the Hessian $\boldsymbol{H}\equiv-\boldsymbol{\nabla\nabla}\ln P(\boldsymbol{\theta}|\dot{\boldsymbol{x}})|_{\boldsymbol{\theta}^{*}}$
of the posterior distribution and thus, calculate the Laplacian approximation
to the evidence for the model $\mathcal{M}_{i}$ \cite{mackay2003information}
\begin{equation}
P(\dot{\boldsymbol{x}}|\mathcal{M}_{i})\approx\underbrace{P(\dot{\boldsymbol{x}}|\boldsymbol{\theta}^{*},\mathcal{M}_{i})}_{Best\,fit\,likelihood}\underbrace{P(\boldsymbol{\theta}^{*}|\mathcal{M}_{i})(2\pi)^{k/2}\det{}^{-\frac{1}{2}}\boldsymbol{H}}_{Occam\,factor},\label{eq:Evd-1}
\end{equation}
and most importantly the \emph{Bayesian credible intervals}, or error
bars of the MAP estimates. Not relying on finite differences-methods
enables us to quantify the uncertainties of our predictions in a much
more reliable manner. 

Due to the \emph{just-in-time (JIT) compilation }of `JAX' \cite{jax2018github},
a python machine learning library, it is even possible to compute
the multi-dimensional integral for the model evidence 
\[
P(\dot{\boldsymbol{x}}|\mathcal{M}_{i})=\int P(\dot{\boldsymbol{x}}|\boldsymbol{\theta},\mathcal{M}_{i})P(\boldsymbol{\theta}|\mathcal{M}_{i})d\boldsymbol{\theta}
\]
exactly within seconds (nestle). With this, we can pursue the task
of Bayesian model averaging for epidemiological models. 

Operationally , the stochastic population density vector $\boldsymbol{x}$
is given by the data itself. Its derivative $\dot{\boldsymbol{x}}$
is simply obtained by finite differences (jax.numpy.gradient) \cite{jax2018github},
no fitting is applied. The mean $\bar{\boldsymbol{x}}$ satisfying
the MREs \eqref{MRE} can be found by solving this system of ODEs
for a given set of parameters $\boldsymbol{\theta}$ (jax.experimental.ode.odeint),
and therefore, is part of the optimization process of finding the
MAPs $\boldsymbol{\theta}^{*}$. The coefficients of the system size
expansion $\boldsymbol{A},\,\boldsymbol{J}$ and $\boldsymbol{B}$
depend on the mean. 

\section{Prediction\label{sec:Prediction}}

In the previous section, we described the inference of model parameters
$\boldsymbol{\theta}=(\theta_{1},\ldots,\theta_{k})$ given data $\boldsymbol{X}$
in the form of a time series, c.f. equation (\ref{eq:tseries}). We
now discuss how the results of this inference are used for prediction
in the \lstinline!pyross.forecast! module. An example where the forecasting
module is used is given in section \ref{sec:Inference-and-forecasting}.

\subsection{Posterior predictive distributions\label{subsec:Posterior-predictive-distributio}}

Once given data $\boldsymbol{X}$ has been used to infer parameters
$\boldsymbol{\theta}$ of a model $\mathcal{M}_{i}$, the probability
to make an observation $\boldsymbol{Y}$ is given by 

\begin{equation}
P(\boldsymbol{Y}|\boldsymbol{X},\mathcal{M}_{i})=\int P(\boldsymbol{Y}|\boldsymbol{X},\boldsymbol{\theta},\mathcal{M}_{i})P(\boldsymbol{\theta}|\boldsymbol{X},\mathcal{M}_{i})d\boldsymbol{\theta},\label{eq:PosPred}
\end{equation}
where $\boldsymbol{Y}$ is the observation whose probability one wants
to predict. The first term in the integral, called model uncertainty,
is the probability that, for given data $\boldsymbol{X}$ and parameters
$\boldsymbol{\theta}$, one observes $\boldsymbol{Y}$. The second
term, called parameter uncertainty, represents the probability of
the parameters $\boldsymbol{\theta}$ itself given the data $\boldsymbol{X}$.
As indicated in the notation, all three probabilities are conditional
on $\mathcal{M}_{i}$. In the present context, $\boldsymbol{X}$ is
typically a fully or partially observed time series up to the present,
and $\boldsymbol{Y}$ denotes future values of the time series, or
more generally any function of the time series. To use the right-hand
side of equation (\ref{eq:PosPred}) for calculating predictions,
explicit expressions for both factors need to be obtained. We first
approximate the equation as
\begin{equation}
P(\boldsymbol{Y}|\boldsymbol{X},\mathcal{M}_{i})\approx\frac{1}{N}\sum_{j=1}^{N}P(\boldsymbol{Y}|\boldsymbol{X},\boldsymbol{\theta}_{j},\mathcal{M}_{i}),\label{eq:PosPred-1}
\end{equation}
where $\boldsymbol{\theta}_{j}$ are $N$ independent samples drawn
from the distribution $P(\boldsymbol{\theta}|\boldsymbol{X},\mathcal{M}_{i})$.
In the following sections, we describe both this sampling process,
and how the corresponding terms in equation (\ref{eq:PosPred-1})
are evaluated in \lstinline!pyross.forecast!.

\subsection{Data uncertainty}

Uncertainty in the available data needs to be considered when fitting
model parameters. Sources of uncertainty are twofold. On the one hand,
authorities may be not able to report all the infected cases or deads.
On the other hand, uncertainty arises from the fact that tests can
never be perfect. The consequence of this uncertainty is that the
reported numbers of infectives do not match the actual numbers considered
in our models.

One way to reduce data uncertainty is to focus on refined models that
attribute specific compartments to the data available with the least
uncertainty. These compartments can be (see the bestiary \ref{chap:Bestiary})
compartments for hospitalised individual, people in ICU, deads, \emph{etc}.
Our inference methods in principle allow one to extract model parameters
from such incomplete information. However, as it can be expected that
these three classes cover only a small fraction of the overall infected
population (in particular older age groups), the suitability of these
data for extensive model fitting is questionable.

It would therefore be desirable to also use data from more universal
testing campaigns. The only way to handle uncertainty in these numbers
is to model the testing process explicitly. Such testing models typically
need to be informed by the number of tests being performed per day
and a guess for the specificity of the tests.

To do so, all individuals who are positively tested are moved in a
new class $T_{+}$ (confirmed cases)\footnote{Note that this class can be confused with the quarantine class $Q$
when confirmed people are quarantined (see SEAIRQ model in Sec. \ref{sec:SEAIIRQ}).
Otherwise, we interpret $T_{+}$ as a subclass of infected people
who can still infect susceptibles.}. Note that people in $T_{+}$ are not isolated by default and can
still infect others if they are not quarantined.

For the same reasons as every compartment models considered in PyRoss,
the testing process is intrinsically stochastic (although it becomes
deterministic in the law of large numbers limit). It is generically
defined by: 
\begin{enumerate}
\item the probability to actually report a confirmed case when a test is
conducted, 
\item the distribution of tests in time. 
\end{enumerate}

\subsubsection{Random testing in the population}

When a test is conducted, it can be performed completely at random
by selecting one individual in the whole population or it can be restricted
to some subpopulation.

By noting $I(t)$ the number of infected --- yet not tested ---
individuals, and $T_{+}(t)$ the number of positively tested individuals,
the simplest model is to consider a perfectly random testing procedure
among the $N_{\mathrm{testpop}}(t)=N-T_{+}(t)$ individuals in the
population who have not been tested positive yet. The probability
to detect \emph{one} infected individual at time $t$ thus reads as
\begin{equation}
p_{+}(t)=\frac{I(t)}{N_{\mathrm{testpop}}(t)}\,.\label{eq:3}
\end{equation}

Possible refinements of the testing process include more selective
processes like symptomatic testing, contact tracing, etc.

\subsubsection{Markovian testing process}

To go further, one needs to make some assumption about the distribution
of tests in time. The simplest assumption is to assume that the tests
are also performed randomly in time, without any memory of the tests
conducted earlier (Markov assumption). If one denotes by $\tau(t)$
the rate at which tests are performed, the number of tests conducted
between $t$ and $t+\Delta t$ is drawn from a Poisson distribution
of parameter $\int_{t}^{t+\Delta t}\tau(u)\mathrm{d}u$.

The rates at which infected people are tested positive thus reads
\begin{equation}
w_{\mathrm{test}}(t)=\tau(t)p_{+}(t)\,.\label{eq:8}
\end{equation}
We recall that when a positive case is detected, the global state
of the system $(I(t),T_{+},\dots{})$ is updated as $(I(t)-1,T_{+}(t)+1,\dots{})$.

In practice, one typically knows from data the average number of tests
per day. If one assumes $\tau(t)$ to be constant over one day, the
average number of tests conducted in one day is $\tau\times1\,\text{day}$.

At the expense of introducing non-Markovian features into the dynamics,
the distribution of tests in time is not limited to be Poissonian,
and one can imagine to conduct a strictly fixed number of tests per
day, or even more complex time distributions.

\subsection{Parameter uncertainty\label{subsec:Parameter-uncertainty}}

With equation (\ref{eq:GPost}) we have a Gaussian approximation for
the distribution of the parameters $\boldsymbol{\theta}$, which,
for data $\boldsymbol{X}$ and model $\mathcal{M}_{i}$ , is given
by

\begin{equation}
P(\boldsymbol{\theta}|\boldsymbol{X},\mathcal{M}_{i})\approx\mathcal{{N}}\exp\left(-\frac{1}{2}\Delta\boldsymbol{\theta}^{T}\boldsymbol{H}\Delta\boldsymbol{\theta}\right),\label{eq:GPost-1}
\end{equation}
where $\mathcal{{N}}$ is a normalisation constant, $\Delta\boldsymbol{\theta}=\boldsymbol{\theta}-\boldsymbol{\theta}^{*}$
with $\boldsymbol{\theta}^{*}$ the MAPs, and $\boldsymbol{H}$ the
Hessian of the log-likelihood function. To generate numerical samples
$\boldsymbol{\theta}_{j}$ for system parameters, we draw from this
Gaussian distribution. While in the parameter inference for compartment
models, the components of the parameters $\boldsymbol{\theta}$ typically
represent rates between compartments, which are positive, the Gaussian
distribution equation (\ref{eq:GPost-1}) in principle allows for
arbitrary real values. To avoid unphysical values for system parameters,
we redraw a sample $\boldsymbol{\theta_{j}}$ if any of its vector
component is negative, meaning we use a truncated version of the Gaussian
distribution.

\subsection{Model uncertainty\label{subsec:Model-uncertainty}}

For a deterministic model and a definite initial condition we have

\begin{equation}
P(\boldsymbol{Y}|\boldsymbol{X},\boldsymbol{\theta},\mathcal{M}_{i})=\delta(\boldsymbol{Y}-\boldsymbol{Y}(\boldsymbol{X},\boldsymbol{\theta},\mathcal{M}_{i}))\label{eq:DetPred}
\end{equation}
where $\boldsymbol{Y}(\boldsymbol{X},\boldsymbol{\theta},\mathcal{M}_{i})$
is the deterministic value for $\boldsymbol{Y}$, calculated from
model $\mathcal{{M}}_{i}$ using the parameters $\boldsymbol{\theta}$
and the initial condition determined by the data $\boldsymbol{X}$.
For a deterministic model, the approximate posterior predictive distribution
equation (\ref{eq:PosPred-1}) is thus given by

\begin{equation}
P(\boldsymbol{Y}|\boldsymbol{X},\mathcal{M}_{i})\approx\frac{1}{N}\sum_{j=1}^{N}\delta(\boldsymbol{Y}-\boldsymbol{Y}(\boldsymbol{X},\boldsymbol{\theta_{j}},\mathcal{M}_{i})).\label{eq:PosPred-2}
\end{equation}
Obtaining $\boldsymbol{Y}(\boldsymbol{X},\boldsymbol{\theta},\mathcal{M}_{i})$
in practice means one has to integrate the model equations, which
for the nonlinear ODEs corresponding to the compartment models is
achieved by numerical integration using \lstinline!pyross.deterministic!.
To estimate (\ref{eq:PosPred-1}) for a stochastic model, we also
use (\ref{eq:PosPred-2}), but now $\boldsymbol{Y}(\boldsymbol{X},\boldsymbol{\theta}_{j},\mathcal{M}_{i})$
represents a realisation of the stochastic dynamics, generated using
\lstinline!pyross.stochastic!. 

\subsection{Model averaging}

In section \ref{subsec:Posterior-predictive-distributio} we discussed
predicting the outcome of an observation $\boldsymbol{Y}$, assuming
a given observation $\boldsymbol{X}$ and a model $\mathcal{M}_{i}$.
However, typically there are multiple models that are, in principle,
compatible with the observation $\boldsymbol{X}$. A more refined
forecast is thus obtained by not only averaging over parameter uncertainties
and model uncertainties, but also over models, i.e.\,by considering

\begin{equation}
P(\boldsymbol{Y}|\boldsymbol{X})=\sum_{i}P(\boldsymbol{Y}|\boldsymbol{X},\mathcal{M}_{i})\,P(\mathcal{M}_{i}|\boldsymbol{X}),\label{eq:PosPred-3}
\end{equation}
where the summation is over the models $\mathcal{M}_{i}$ considered,
the first term in the sum was discussed in section \ref{subsec:Posterior-predictive-distributio},
and the second term in the sum is the probability that, of all the
models considered, $\mathcal{M}_{i}$ is the correct model, was discussed
in section \ref{subsec:Model-evidence}. In practice, incorporating
a model-average in the prediction formula \ref{eq:PosPred-1}, we
thus obtain

\begin{equation}
P(\boldsymbol{Y}|\boldsymbol{X})\approx\frac{1}{N}\sum_{j=1}^{N}P(\boldsymbol{Y}|\boldsymbol{X},\boldsymbol{\theta}_{j},\mathcal{M}_{j}),\label{eq:PosPred-1-1}
\end{equation}
where now for the $j$-th trajectory, first a model $\mathcal{M}_{j}$
is randomly drawn with probability given by equation (\ref{eq:Gevd})
(properly normalised so that the sum over all models considered is
unity), and then a random sample $\boldsymbol{\theta}_{j}$ for the
parameters of that model is drawn from the distribution $P(\boldsymbol{\theta}|\boldsymbol{X},\mathcal{M}_{j})$,
as discussed in section (\ref{subsec:Parameter-uncertainty}). Finally,
the probability appearing in the $j$-th term is obtained as described
in section (\ref{subsec:Model-uncertainty}).

\subsubsection{}

\section{Interventions}

To achieve a desired goal, such as reducing the number of infectives,
one would like to understand how a given intervention influences the
time evolution of a model. As will be discussed in this chapter, \lstinline!pyross!
offers several ways of implementing and optimising intervention parameters.
See examples \ref{sec:Prescribed-control} and \ref{sec:Prescribed-control}
for examples involving prescribed- and optimised intervention parameters.

\subsection{Non-pharmaceutical interventions (NPI) }

Non-pharmaceutical interventions (NPIs) are strategies that mitigate
the spread of a disease by suppressing its normal pathways for transmission.
These include social distancing, wearing masks, working from home,
and isolation of vulnerable populations. In contrast to pharmaceutical
interventions, which are slow to develop but effective in the long
term, NPIs can be rapidly implemented but are generally too costly
to maintain indefinitely. In the modelling framework of \lstinline!pyross!,
we represent NPIs as modifications to the contact matrix $\boldsymbol{C}$,
the elements $C_{ij}$ of which describe the number of contacts between
age groups $i$ and $j$ (see Eq.\ref{eq:CM}). Without any NPI, we
typically consider the contact matrix as a sum

\begin{equation}
C_{ij}(t)=C_{ij}^{H}+C_{ij}^{W}+C_{ij}^{S}+C_{ij}^{O}.\label{eq:full_contact_mat}
\end{equation}
where the four terms denote the number of contact at home, at work,
at school, and the other remaining contacts. We have written the above
by setting the constants $a^{H}=a^{S}=a^{W}=a^{0}=1$ in Eq.\ref{eq:CM}.
The class \lstinline!pyross.contactMatrix! provides an interface
to retrieve the individual contact matrices $\boldsymbol{C}_{H},\boldsymbol{C}_{W},\boldsymbol{C}_{S},\boldsymbol{C}_{O}$,
for several countries, obtained from Ref.\,\cite{prem2017projecting}.
In the presence of a NPI, typically in at least one of these spheres
contacts are reduced, and the corresponding contact matrix is to be
replaced by

\begin{equation}
\boldsymbol{C}_{ij}^{X}\rightarrow u_{i}\boldsymbol{C}_{ij}^{X}v_{j},\label{eq:full_contact_mat-1}
\end{equation}
where $X\in\{H,W,S,O\}$ labels the sphere, $u_{i}$ is the fraction
by which susceptible members of age group i reduce their contacts
and $v_{j}$ is the corresponding fraction for infective members.

\subsection{Intervention protocols\label{subsec:Control-strategies}}

An intervention typically consists of a time- and state-dependent
protocol for the contact matrix. The \lstinline!pyross! module currently
allows for two kinds of intervention strategies. Purely time-dependent
interventions, meaning a given time-dependent contact matrix $\mathbf{C}(t)$,
are available directly in the main simulation modules \lstinline!pyross.deterministic!
and \lstinline!pyross.stochastic!. An example for purely time-dependent
interventions is a full lockdown starting at time $t_{0}$ and released
at time $t_{1}$, for which the time-dependent contact matrix is

\begin{equation}
C_{ij}(t)=\begin{cases}
C_{ij}^{H}+C_{ij}^{W}+C_{ij}^{S}+C_{ij}^{O} & t<t_{0}\mathrm{\,\,or\,\,t}>t_{1}\\
C_{ij}^{H} & t_{0}<t<t_{1}.
\end{cases}\label{eq:full_lockdown_control}
\end{equation}
If one considers a system with uncertainty, either intrinsic from
the model or because of imperfect knowledge of system parameters,
then a purely time-dependent protocol could be insufficient. For example,
if at time $t_{1}$ in the protocol given by equation (\ref{eq:full_lockdown_control}),
the number of infectives has not decreased below some threshold, one
might not want to release the lockdown yet before that threshold is
met. To allow for interventions that dependent on time and state,
the submodule \lstinline!pyross.control! allows to consider event-driven
protocols. An event $E_{i}$ is a function, $E_{i}\equiv E_{i}(\mathbf{\boldsymbol{y}}(t),t)$,
where $\boldsymbol{y}(t)$ is the state of the dynamics, and it occurs
at a time $t^{*}$ if $E_{i}(\boldsymbol{\mathbf{\boldsymbol{y}}}(t^{*}),t^{*})=0$,
possibly with the additional requirement that the total derivative
$\mathrm{d/dt\,}E_{i}(\mathbf{\boldsymbol{y}}(t^{*}),t^{*})$ be either
positive or negative. In \lstinline!pyross.control!, the user provides
the program with a list of events $(E_{1},E_{2},...,E_{M})$, and
for each event supplies a contact matrix $\mathbf{C}_{i}$ which is
used in the further time evolution of the model once the event $E_{i}$
has occurred. An example for a protocol defined by events is given
by the two functions $E_{1}(y,t)=y_{i}-c_{1},E_{2}(y,t)=y_{i}-c_{2}$,
where $c_{1}>c_{2}$ and we require that the total derivative of $E_{1}$
be positive for an event, while the total derivative of $E_{2}$ be
negative for an event. For the contact matrix, we consider

\begin{equation}
C_{ij}(t)=\begin{cases}
C_{ij}^{H}+C_{ij}^{W}+C_{ij}^{S}+C_{ij}^{O} & \mathrm{{initially},}\\
C_{ij}^{H} & \mathrm{{if\,}\mathit{E}_{1}\,\mathrm{{occurs},}}\\
C_{ij}^{H}+C_{ij}^{W}+C_{ij}^{S}+C_{ij}^{O} & \mathrm{{if\,}\mathit{E}_{2}\,\mathrm{{occurs}.}}
\end{cases}\label{eq:full_lockdown_control-1}
\end{equation}
This protocol puts a lockdown into place once the population $\boldsymbol{\mathbf{y}}_{i}$
of compartment $i$ exceeds a threshold $c_{1}$, and releases this
lockdown once the population drops below $c_{2}$. In \lstinline!pyross.control!,
the user can define whether each event can only occur once or repeatedly;
if each event can only occur once, the user can furthermore set whether
the events can occur in arbitrary order, or only in the order they
are given.

For quarantine and contact tracing (Testing, contact-Tracing and Isolation
procedures), one can also imagine to model such interventions by adding
new compartments explicitly in the model (see for instance SEAIIRQ
in Sec. \ref{sec:SEAIIRQ}). For instance, when quarantine is decided
after a test has been performed, the rate at which the quarantine
compartment is filled is directly proportional to the average number
of tests currently conducted, which can be \emph{a priori} controlled. 
This other kind of intervention
is currently being implemented within PyRoss.

\subsection{Bayesian forecast of NPI}

Simulations can be used to determine whether a given control strategy
achieves a desired goal. If all system parameters are known exactly,
and if the dynamics of the model is deterministic, then a single simulation
using \lstinline!pyross.control! is sufficient to explore the consequences
of a given intervention strategy. For the case of uncertain parameters,
or a stochastic model, \lstinline!pyross.forecast! can be used to
generate an ensemble-forecasting for both purely time-dependent and
event-driven intervention strategies.

\subsection{Optimised intervention parameters}

Intervention protocols are typically considered to achieve a desired
goal, such as reducing the number of infectives. To frame this in
the language of an optimisation problem, we consider events $(E_{1},E_{2},...,E_{M})$
which depend on a parameter $\mathbf{\boldsymbol{c}}\in\mathbb{R}^{m}$.
An example for this are the two thresholds $\boldsymbol{\mathbf{c}}\equiv(c_{1},c_{2})\in\mathbb{R}^{2}$
from the protocol considered in section \ref{subsec:Control-strategies}.
We furthermore consider a cost functional $\mathcal{C}_{c}[\boldsymbol{y_{c}}]$,
which quantifies the cost of a given realisation $\boldsymbol{y_{c}}(t)$
of the dynamics. With the index $c$ we emphasise that the functional
can depend on $c$ both explicitly, and implicitly because the model
dynamics depends on the protocol. Taking into account both model-
and parameter uncertainty, the average cost is given by

\begin{equation}
\langle\mathcal{C}_{c}\rangle=\int\int\mathcal{C}_{c}[\boldsymbol{y}_{c}]P_{\boldsymbol{\theta}}[\boldsymbol{y}_{c}]\mathcal{D}[\boldsymbol{y}_{c}]P(\boldsymbol{\theta})d\boldsymbol{\theta},\label{eq:AvgCost}
\end{equation}
where $P_{\boldsymbol{\theta}}[\boldsymbol{y}_{c}]\mathcal{D}[\boldsymbol{y}_{c}]$
denotes the path-integral density of the trajectory $\boldsymbol{y}_{c}$
for given system parameters $\boldsymbol{\theta}$. The optimal protocol
$\boldsymbol{c}^{*}$ is defined as the protocol which minimises the
average cost,

\begin{equation}
c^{*}=\mathrm{{argmin}}_{c}\langle\mathcal{C}_{c}\rangle,\label{eq:cost_argmin}
\end{equation}
and to determine it numerically, a practical means of evaluating the
average cost of a protocol needs to be established. For deterministic
dynamics, there is only one path which occurs with probability one,
and the mean cost of a protocol is given by
\begin{equation}
\langle\mathcal{C}_{c}\rangle=\int\mathcal{C}_{c}[\boldsymbol{y}_{c}]P(\boldsymbol{\theta})d\boldsymbol{\theta},\label{eq:AvgCostDet}
\end{equation}
where $\boldsymbol{y}_{c}$ denotes the deterministic solution corresponding
to the parameters $\boldsymbol{\theta}$ and a given initial conditions.
An estimate for this average cost can be obtained numerically by approximating
$P(\boldsymbol{\theta})$ by a Gaussian, drawing $N$ samples $\boldsymbol{\theta}_{j}$
for this Gaussian, and averaging over the resulting cost, i.e.

\begin{equation}
\langle\mathcal{C}_{c}\rangle\approx\frac{1}{N}\sum_{j=1}^{N}\mathcal{C}_{c}[\boldsymbol{y}_{c}^{j}],\label{eq:AvgCostDetApproximate}
\end{equation}
where $\boldsymbol{y}_{c}^{j}$ is the deterministic solution of the
model dynamics subject to the protocol with parameters $\boldsymbol{c}$
and model parameters $\boldsymbol{\theta}_{j}$. If the model dynamics
is furthermore stochastic, we approximate the path-integral in equation
(\ref{eq:AvgCost}) by additionally averaging over realisations of
the model dynamics; then, equation (\ref{eq:AvgCostDetApproximate})
is still valid and $\boldsymbol{y}_{c}^{j}$ simply denotes a stochastic
realisation subject to the protocol with parameters $\boldsymbol{c}$
and model parameters $\boldsymbol{\theta}_{j}$. Equation (\ref{eq:AvgCostDetApproximate})
constitutes a numerically accessible means of evaluating the average
cost of a protocol, which can be used for numerical solution of the
minimisation problem (\ref{eq:cost_argmin}). While an optimisation
framework within \lstinline!pyross! is in the works, \lstinline!pyross.control!
can be conveniently used to generate the sample trajectories $\boldsymbol{y}_{c}^{j}$
which appear in equation (\ref{eq:AvgCostDetApproximate}). For an
example of a problem with optimised intervention parameters, see example
\ref{sec:Optimal-control}.

\section{Numerical methods\label{sec:Numerical-methods}}

\subsection{Exact sampling \label{subsec:Exact-sampling-of}}

The generate a realization of a discrete-state continuous time model
as described in section \ref{sec:Discrete-state-continuous-time-m},
\lstinline!pyross.stochastic! by default uses the Gillespie SSA algorithm
\cite{gillespie1977exact}. In this algorithm, an integration step

\begin{equation}
(t,\boldsymbol{n}(t))\longmapsto(t+\Delta t,\boldsymbol{n}(t+\Delta t))\label{eq:integration_step}
\end{equation}
consists of two parts, each of which involves a random choice:
\begin{enumerate}
\item Determine a waiting time $\Delta t$ until any population in the compartment
model changes (a ``reaction'' takes place).
\item Decide which population changes.
\end{enumerate}
For step 1, first the total reaction rate is calculated as $W=\sum_{\alpha}w_{\alpha}$,
and then an exponentially distributed random variable $\Delta t$
with mean $1/W$ is generated to determine the time at which the next
reaction occurs. For step 2, a random reaction $\alpha$ is then chosen
with probability $p_{\alpha}=w_{\alpha}/W$. The population vector
$\boldsymbol{n}$ is finally updated by adding the vector $\boldsymbol{r}_{\alpha}$,
as described in equation (\ref{eq:population_change}), i.e. 

\begin{equation}
\boldsymbol{n}(t+\Delta t)=\boldsymbol{n}(t)+\boldsymbol{r}_{\alpha}.
\end{equation}

\subsection{Acceleration by $\tau$-leaping}

For large populations the Gillespie algorithm described in section
\ref{subsec:Exact-sampling-of}, which simulates every event individually,
can be very slow. To accelerate stochastic simulations of large populations,
\lstinline!pyross.stochastic! also supports an integration method
called tau-leaping \cite{gillespie2001approximate}. Here, an integration
step (\ref{eq:integration_step}) consists of the following two parts:
\begin{enumerate}
\item A time interval $\tau$ is chosen deterministically, as discussed
further below.
\item For each possible reaction $\alpha$, a random sample $q_{\alpha}$
for the number of occurring reactions in the time interval $\tau$
is drawn from a Poisson distribution with mean $w_{\alpha}\cdot\tau$.
\end{enumerate}
The population vector $\boldsymbol{n}$ is then updated by adding
up all the population changes, i.e. 

\begin{equation}
\boldsymbol{n}(t+\tau)=\boldsymbol{n}(t)+\sum_{\alpha}q_{\alpha}\boldsymbol{r}_{\alpha},\label{eq:tau_leaping_update}
\end{equation}
where the vector $\boldsymbol{r}_{\alpha}$ was introduced in equation
(\ref{eq:population_change}). 

The time interval $\tau$ needs to be so small that the rates, which
depends on the current state, $w_{\alpha}\equiv w_{\alpha}(\boldsymbol{n})$,
does not change appreciably during all the reactions that take place
during $\tau$. To select $\tau$, \lstinline!pyross.stochastic!
uses the algorithm by Cao, Gillespie, and Petzold \cite{cao2006efficient}.
To avoid unphysical negative populations, which can be the consequence
of a tau-leaping step (\ref{eq:tau_leaping_update}), \lstinline!pyross.stochastic!
switches to the Gillespie SSA algorithm if any population is below
a threshold; identifying critical reaction channels and only propagating
those via the Gillespie SSA algorithm, as described in Ref.\,\cite{cao2006efficient},
will be added soon.

\subsection{Integration in the deterministic limit}

The deterministic dynamic of the models given in \ref{chap:Bestiary}
is obtained by numerical integration in \lstinline!pyross.deterministic!.
By default PyRoss uses \lstinline!scipy.integrate.odeint! for numerical
integrations. This is an adaptive time step integrator, which switches
between the backward differentiation formula for stiff problems \cite{wanner1996solving}
and Adams method for non-stiff problems \cite{hairer1993solving}.
Alternatively, we also allow to use integration methods from the package
Odespy \cite{langtangen2012tutorial} and other integrators from Scipy
\cite{virtanen2020scipy} such as \lstinline!scipy.integrate.solve_ivp!.
See chapter \ref{chap:Bestiary}, for deterministic dynamical systems. 

\subsubsection{}

\chapter{Bestiary\label{chap:Bestiary}}

In this chapter, we describe the implementations of age-structured
epidemiological compartment models in PyRoss. The basic variable in
this class of models is a metapopulation labeled by its epidemiological
state (susceptible, infectious, removed, etc) and additional attributes
like age, gender, geographic location and so on. The additional attributes
are what comprise the \textquotedbl structure\textquotedbl{} of the
model. Currently, PyRoss supports the models with susceptible ($S$),
infected ($I$), exposed ($E$), activated ($A$), quarantined ($Q$)
and removed ($R$) epidemiological states. Additionally, the infectious
class can be subdivided into $k$-stages. The progress of these variables
in time are described by chemical master equations and, when compartmental
fluctuations (CME) are small, by ordinary differential equations (ODE).
A hybrid method is also possible which switches from CME to ODE when
the population reaches a user defined threshold, at which point it
is assumed that random fluctuations are a negligible percentage of
the total population. These integration methods build the foundation
of PyRoss, upon which investigation into the effects of control such
as self-isolation or forecasting made from real world data can be
performed. 

PyRoss takes the inputs - age, contact structure and an epidemiological
compartment model - to simulate the deterministic and stochastic trajectories.
The age \cite{pyramid} and contact structures \cite{prem2017projecting}
can be obtained from published data. The demographic parameters which
determine contact matrices, together with their uncertainties, will
be discussed elsewhere. In this work, we assumed they are user-supplied.
In what follows, \ref{sec:SIR}-\ref{tab:SIRS}, we describe various
models available in PyRoss with increasing complexity. We also provide
a class to implement a generic user-defined compartment model in \ref{sec:Even-More-?}.

\section{SIR\label{sec:SIR}}

\begin{figure}
\centering\includegraphics[width=0.64\textwidth]{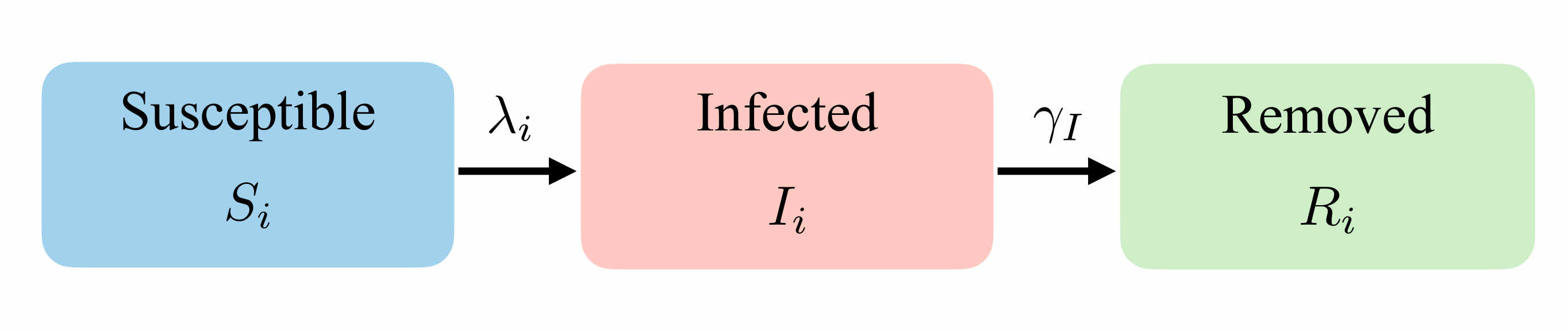}

\caption{\label{tab:SIR}\textbf{Schematic of the SIR model. }The parameters
for this model are: $\boldsymbol{\theta}=(\beta,\gamma_{I})$. The
class SIR can be instantiated in PyRoss using \lstinline!pyross.deterministic.SIR!.}
\end{figure}

We first present the well studied SIR model, where population within
age group $i$, is partitioned into susceptibles $S_{i}$, infectives
$I_{i}$, and removed individuals $R_{i}$. The sum of these is the
size of the population in age group $i$, $N_{i}=S_{i}+I_{i}+R_{i}$
\cite{anderson1992infectious,keeling2011modeling,towers2012social,ferguson2006strategies,hethcote2000mathematics}.
For this model, vital dynamics and the change in age structure on
the time scale of the epidemic in this model is ignored. Therefore
each $N_{i}$ and, consequently, the total population size
\begin{equation}
N=\sum_{i=1}^{M}N_{i}
\end{equation}
remain constant in time. With these assumptions the progress of the
epidemic is governed by the age-structured SIR model. Figure  \ref{tab:SIR}
shows the schematic. The deterministic limit of the SIR model is given
by the ODE:
\begin{align*}
\dot{S_{i}} & =-\lambda_{i}(t)S_{i},\\
\dot{I}_{i} & =\lambda_{i}(t)S_{i}-\gamma_{I}I_{i},\\
\dot{R}_{i} & =\gamma_{I}I_{i}.
\end{align*}
The rate of infection of a susceptible individual in age group $i$
is\emph{
\begin{equation}
\lambda_{i}(t)=\beta\sum_{j=1}^{M}\left(C_{ij}(t)\frac{I_{j}}{N_{j}}\right),\quad i,j=1,\ldots M\label{eq:li-1-1}
\end{equation}
}where $\beta$ is the probability of infection on contact (assumed
intrinsic to the pathogen) . We take the age-independent removal rate
$\gamma$ to be identical for both asymptomatic and symptomatic individuals
whose fractions are, respectively, $\alpha_{i}$ and $\bar{\alpha_{i}}=1-\alpha_{i}$.
The social contact matrix $C_{ij}$ denotes the average number of
contacts made per day by an individual in class $i$ with an individual
in class $j$. Clearly, the total number of contacts between group
$i$ to group $j$ must equal the total number of contacts from group
$j$ to group $i$, and thus, $N_{i}C_{ij}=N_{j}C_{ji}$.
\begin{figure}
\centering\includegraphics[width=0.99\textwidth]{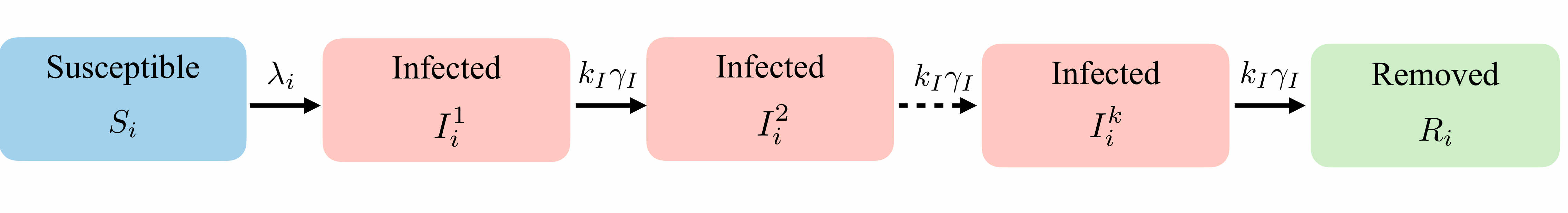}\caption{\label{tab:SIkR}\textbf{Schematic of the SIR with stages (SIkR) model.
}The parameters for this model are: $\boldsymbol{\theta}=(k_{I},\beta,\gamma_{I})$.
The class SIkR can be instantiated in PyRoss using \lstinline!pyross.deterministic.SIkR!.}
\end{figure}

The SIR model can be improved by adding more epidemiological states
as we describe below. Addition epidemiological states, like exposed
(E), where the individual has contracted the diseases but is not infectious,
or quarantined (Q), where the individual has contracted the disease,
is infectious, but cannot spread contagion because of confinement,
may be necessary for a better-resolved description. Despite these
limitations, the SIR model and its age-structured variant provide
the most parsimonious description of infectious disease and provide
a null model against which all others must be compared.

\section{SIR with stages (SIkR)\label{sec:SIR-with-stages}}

The SIR model considers only three mutually exclusive epidemiological
states: $S,I,R$. This leads to an exponentially distributed residence
time in the infectious state. Within the compartment framework, the
simplest way to make infectious period distributions more realistic
is to use stages ($k$ stages of infectious) \cite{lloyd2001realistic}.
The model SIR with stages (SIkR) is obtained by allowing $I$ class
is the SIR to have $k$-stages \cite{lloyd2001realistic}. The SIkR
model then has an infectious period with Erlang, Gamma distributions
with integer shape parameter, distribution \cite{anderson1980spread,wearing2005appropriate,krylova2013effects}.
The number of states $k$ can be adjusted to match empirically observed
infectious periods. Figure  \ref{tab:SIkR} shows the schematic. The
deterministic limit of the SIkR model is given as
\begin{align}
\dot{S_{i}} & =-\lambda_{i}(t)S_{i},\nonumber \\
\dot{I}_{i}^{1} & =\lambda_{i}(t)S_{i}-k_{I}\gamma_{I}I_{i}^{1},\label{eq:ageSIR-1-1}\\
\vdots\nonumber \\
\dot{I}_{i}^{k} & =k_{I}\gamma_{I}I_{i}^{k-1}-k_{I}\gamma_{I}I_{i}^{k},\nonumber \\
\dot{R}_{i} & =k_{I}\gamma_{I}I_{i}^{k}.\nonumber 
\end{align}
The rate of infection of a susceptible individual in age group $i$
is\emph{
\begin{equation}
\lambda_{i}(t)=\beta\sum_{j=1}^{M}\sum_{n=1}^{k}C_{ij}(t)\frac{I_{j}^{n}}{N_{j}},
\end{equation}
}

\section{SIIR}

\begin{figure}
\begin{centering}
\includegraphics[width=0.64\textwidth]{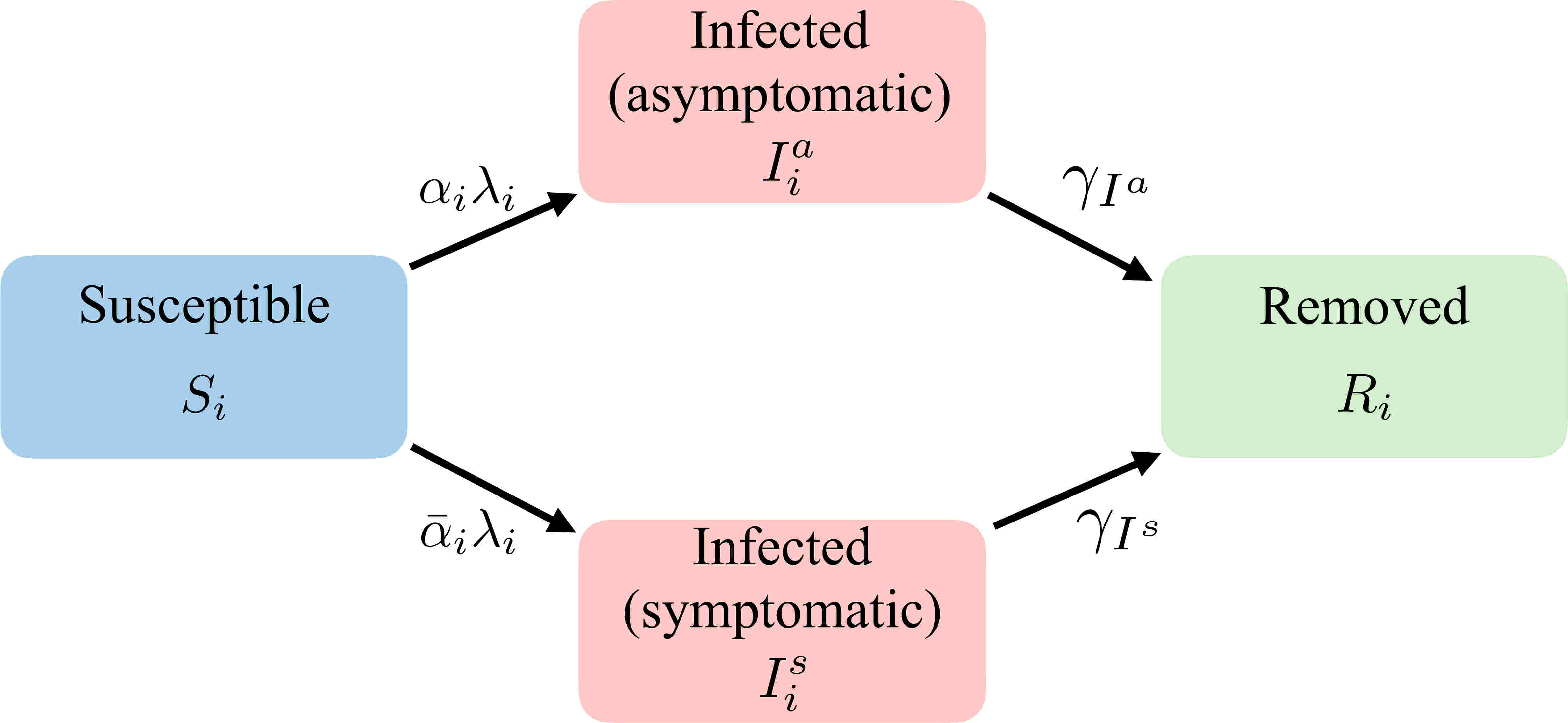}
\par\end{centering}
\caption{\label{tab:SIIR}\textbf{Schematic of the SIIR model. }The parameters
for this model are: $\boldsymbol{\theta}=(\alpha_{i},\beta,\gamma_{I^{a}},\gamma_{I^{s}})$.
The class SIIR can be instantiated in PyRoss using \lstinline!pyross.deterministic.SIR!.
Please note that both SIIR and SIR have been implemented as \lstinline!pyross.deterministic.SIR!
in PyRoss, as it possible to go from one to another by correct choice
of parameters.}
\end{figure}
We now extend the classic SIR model to an SIIR model, where the infective
class has been divided in asymptomatic $I_{i}^{a}$ and symptomatic
$I_{i}^{s}$. We assume that the rate of infection of a susceptible
individual in age group $i$ is \emph{
\begin{equation}
\lambda_{i}(t)=\beta\sum_{j=1}^{M}\left(C_{ij}^{a}(t)\frac{I_{j}^{a}}{N_{j}}+C_{ij}^{s}(t)\frac{I_{j}^{s}}{N_{j}}\right),\quad i,j=1,\ldots M\label{eq:li-1}
\end{equation}
}where $\beta$ is the probability of infection on contact (assumed
intrinsic to the pathogen) and $C_{ij}^{a}$ and $C_{ij}^{s}$ are,
respectively, the number of contacts between asymptomatic and symptomatic
infectives in age-group $j$ with susceptibles in age-group $i$ (reflecting
the structure of social contacts). We assume that symptomatic infectives
reduce their contacts compared to asymptomatic infectives and set
$C_{ij}^{s}=f^{s}C_{ij}^{a}\equiv f^{s}C_{ij}$, where $0\leq f^{s}\leq1$
is the proportion of contacts that are now avoided by these self-isolating
individuals (allowing also for compliance rates)

With these assumptions the progress of the epidemic is governed by
the age-structured SIIR model. Figure  \ref{tab:SIIR} shows the schematic.
The deterministic limit is given as,
\begin{align}
\dot{S_{i}} & =-\lambda_{i}(t)S_{i},\nonumber \\
\dot{I}_{i}^{a} & =\alpha_{i}\lambda_{i}(t)S_{i}-\gamma_{I^{a}}I_{i}^{a},\label{eq:ageSIR}\\
\dot{I}_{i}^{s} & =\bar{\alpha_{i}}\lambda_{i}(t)S_{i}-\gamma_{I^{s}}I_{i}^{s},\nonumber \\
\dot{R}_{i} & =\gamma_{I^{a}}I_{i}^{a}+\gamma_{I^{s}}I_{i}^{s}.\nonumber 
\end{align}
Here $\gamma_{I^{a}}$ is the removal rate for asymptomatic infectives,
$\gamma_{I^{s}}$ is the removal rate for symptomatic infectives,
$\alpha_{i}$ is the fraction of asymptomatic infectives.

\section{SEIR}

\begin{figure}
\centering\includegraphics[width=0.85\textwidth]{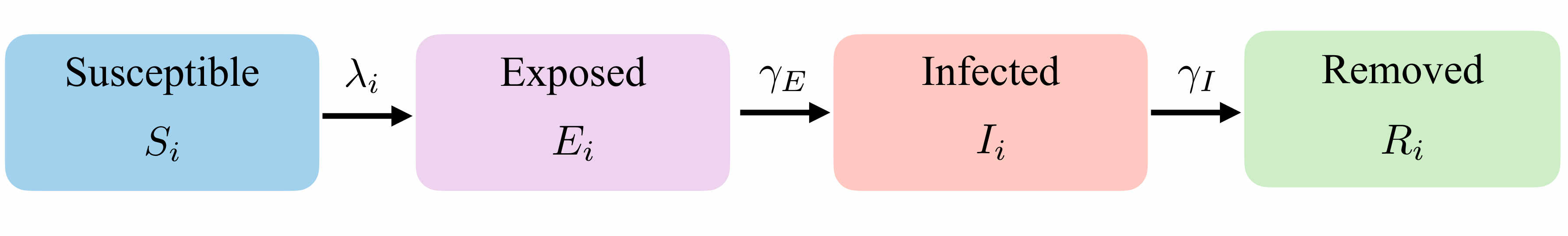}

\caption{\label{tab:SEIR}\textbf{Schematic of the SEIR model. }The parameters
for this model are: $\boldsymbol{\theta}=(\beta,\gamma_{I},\gamma_{E})$.
The class SEIR can be instantiated in PyRoss using \lstinline!pyross.deterministic.SEIR!.}
\end{figure}
The SIR model does not model the incubation period of a virus. This
can be included by adding to the SIR model an exposed E compartment
(to give an age-structured SEIR model) \cite{feng2000endemic,bailey1975mathematical,pastor2015epidemic,li1995global}.
Figure \ref{tab:SEIR} shows the schematic of the SEIR model. The
deterministic ODE giving its time-evolution is 
\begin{align}
\dot{S_{i}} & =-\lambda_{i}(t)S_{i},\nonumber \\
\dot{E}_{i} & =\lambda_{i}(t)S_{i}-\gamma_{E}E_{i}\nonumber \\
\dot{I}_{i} & =\gamma_{E}E_{i}-\gamma_{I}I_{i},\label{eq:ageSEIR-3-1}\\
\dot{R}_{i} & =\gamma_{I}I_{i}.\nonumber 
\end{align}
The rate of infection of a susceptible individual in age group $i$
is\emph{
\begin{equation}
\lambda_{i}(t)=\beta\sum_{j=1}^{M}\left(C_{ij}(t)\frac{I_{j}}{N_{j}}\right),\quad i,j=1,\ldots M\label{eq:li-1-1-1}
\end{equation}
}

\section{SEIR with stages (SEkIkR)}

\begin{figure}
\centering\includegraphics[width=1\textwidth]{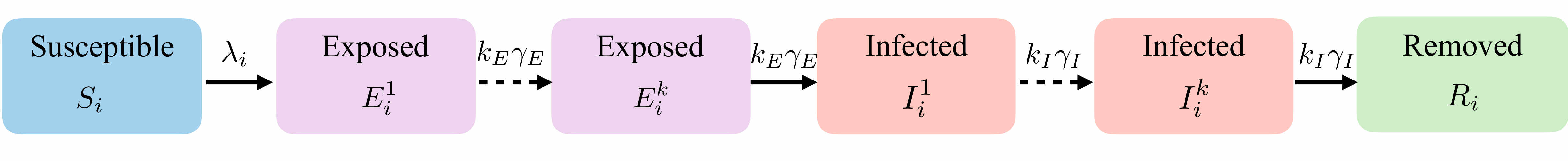}

\caption{\label{tab:SEkIkR}\textbf{Schematic of the SEIR with stages (SEkIkR)
model. }The parameters for this model are: $\boldsymbol{\theta}=(k_{I},k_{E},\beta,\gamma_{I},\gamma_{E})$.
The class SEkIkR can be instantiated in PyRoss using \lstinline!pyross.deterministic.SEkIkR!.}
\end{figure}
The SEIR model considers only four mutually exclusive epidemiological
states: $S,E,I,R$. This leads to an exponentially distributed residence
time in the incubating and infectious state. We use the same resolution
as in SIkR model, see \ref{sec:SIR-with-stages}, to obtain a more
realistic distribution of incubation and infectious times. The SEIR
model can be extended to an age-structured $k-$staged SEkIkR model.
Figure  \ref{tab:SEkIkR} shows the schematic. The ODE describing
SEIR is:
\begin{align}
\dot{S_{i}} & =-\lambda_{i}(t)S_{i},\nonumber \\
\dot{E}_{i}^{1} & =\lambda_{i}(t)S_{i}-k_{E}\gamma_{E}E_{i}^{1}\nonumber \\
\vdots\nonumber \\
\dot{E}_{i}^{k} & =k_{E}\gamma_{E}E_{i}^{k-1}-k_{E}\gamma_{E}E_{i}^{k}\\
\dot{I}_{i}^{1} & =k_{E}\gamma_{E}E_{i}^{k}-k_{I}\gamma_{I}I_{i}^{1},\nonumber \\
\vdots\nonumber \\
\dot{I}_{i}^{k} & =k_{I}\gamma_{I}I_{i}^{(k-1)}-k_{I}\gamma_{I}I_{i}^{k},\nonumber \\
\dot{R}_{i} & =k_{I}\gamma_{I}I_{i}^{k}.\nonumber 
\end{align}
The rate of infection of a susceptible individual in age group $i$
is\emph{
\begin{equation}
\lambda_{i}(t)=\beta\sum_{j=1}^{M}\sum_{n=1}^{k}C_{ij}(t)\frac{I_{j}^{n}}{N_{j}},
\end{equation}
}

\section{SEIIR}

\begin{figure}
\begin{centering}
\includegraphics[width=0.9\textwidth]{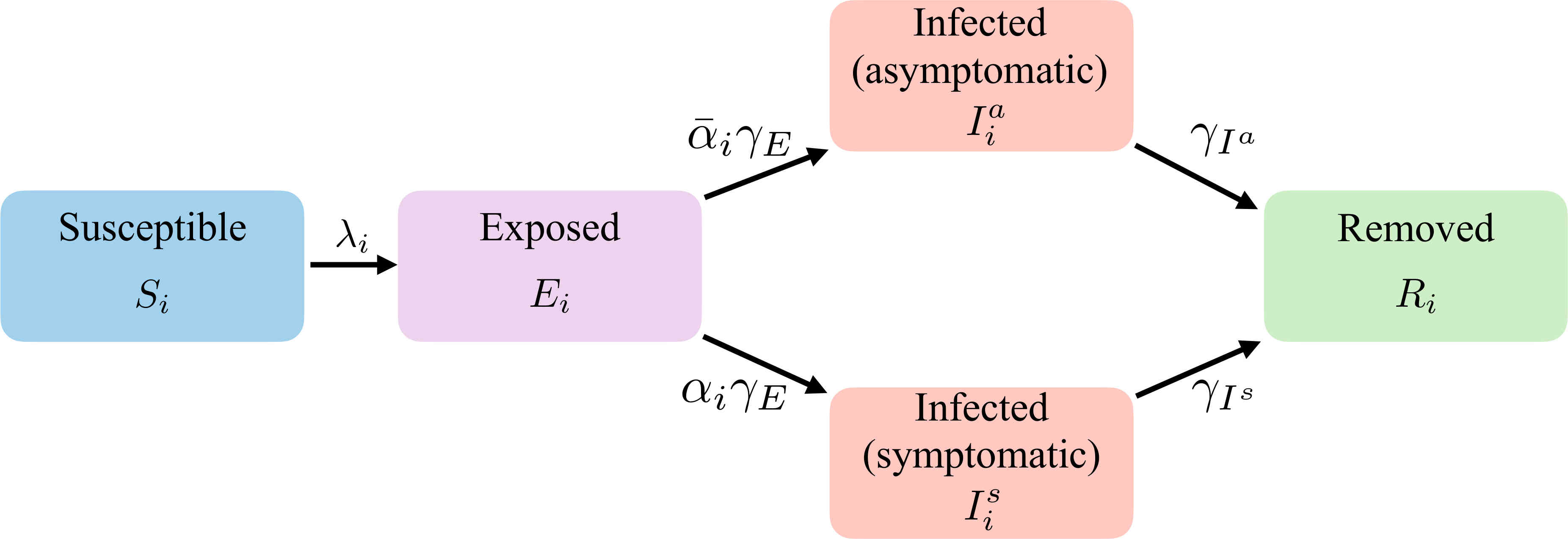}
\par\end{centering}
\caption{\textbf{Schematic of the SEIIR model. }The parameters for this model
are: $\boldsymbol{\theta}=(\alpha_{i},\beta,\gamma_{E},\gamma_{I^{a}},\gamma_{I^{s}})$.
The class SEIIR can be instantiated in PyRoss using \lstinline!pyross.deterministic.SEIR!.
Please note that both SEIIR and SEIR have been implemented as \lstinline!pyross.deterministic.SEIR!
in PyRoss as it possible to go from one to another by correct choice
of parameters.}
\end{figure}
We now extend the classic SIR model to an SIIR model, where the infective
class has been divided in asymptomatic $I_{i}^{a}$ and symptomatic
$I_{i}^{s}$. We assume that the rate of infection of a susceptible
individual in age group $i$ is \emph{
\begin{equation}
\lambda_{i}(t)=\beta\sum_{j=1}^{M}\left(C_{ij}^{a}\frac{I_{j}^{a}}{N_{j}}+C_{ij}^{s}\frac{I_{j}^{s}}{N_{j}}\right),
\end{equation}
}

The deterministic dynamics is given by the following ODE:

\begin{align}
\dot{S_{i}} & =-\lambda_{i}(t)S_{i},\nonumber \\
\dot{E}_{i} & =\lambda_{i}(t)S_{i}-\gamma_{E}E_{i}\nonumber \\
\dot{I}_{i}^{a} & =\alpha_{i}\gamma_{E}E_{i}-\gamma_{I^{a}}I_{i}^{a},\label{eq:ageSEIR-3}\\
\dot{I}_{i}^{s} & =\bar{\alpha_{i}}\gamma_{E}E_{i}-\gamma_{I^{a}}I_{i}^{s},\nonumber \\
\dot{R}_{i} & =\gamma_{I^{a}}I_{i}^{a}+\gamma_{I^{s}}I_{i}^{s}.\nonumber 
\end{align}

\section{SEIIR with stages (SEkIkIkR)}

\begin{figure}
\centering\includegraphics[width=0.99\textwidth]{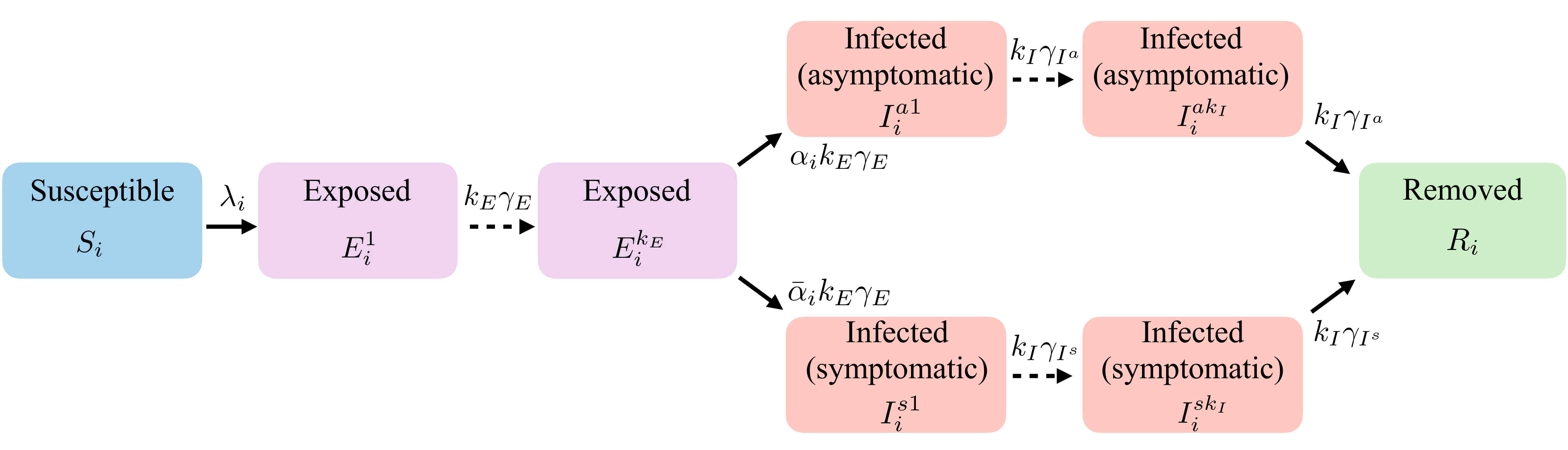}

\caption{\label{tab:SEkIkIkR}\textbf{Schematic of the SEIIR with stages (SEkIkIkR)
model. }The parameters for this model are: $\boldsymbol{\theta}=(k_{I},k_{E},\alpha_{i},\beta,\gamma_{I^{a}},\gamma_{I^{s}},\gamma_{E})$.
The class SEkIkIkR can be instantiated in PyRoss using \lstinline!pyross.deterministic.SEkIkIkR!.}
\end{figure}
We now extend the SEIIR model to have stages in exposed, asymptomatic
infectives, and symptomatic infectives classes. This is the the same
resolution as in SIkR model, see \ref{sec:SIR-with-stages}, to obtain
a more realistic distribution of incubation and infectious times.
Figure  \ref{tab:SEkIkIkR} shows the schematic. The deterministic
dynamics is given as

\begin{align}
\dot{S_{i}} & =-\lambda_{i}(t)S_{i},\nonumber \\
\dot{E}_{i}^{1} & =\lambda_{i}(t)S_{i}-k_{E}\gamma_{E}E_{i}^{1}\nonumber \\
\vdots\nonumber \\
\dot{E}_{i}^{k_{E}} & =k_{E}\gamma_{E}E_{i}^{k_{E}-1}-k_{E}\gamma_{E}E_{i}^{k_{E}}\\
\dot{I}_{i}^{a1} & =\alpha_{i}k_{E}\gamma_{E}E_{i}^{k}-k_{I}\gamma_{I^{a}}I_{i}^{a1},\nonumber \\
\vdots\nonumber \\
\dot{I}_{i}^{ak_{I}} & =k_{I^{a}}\gamma_{I^{a}}I_{i}^{a(k_{I}-1)}-k_{I}\gamma_{I^{a}}I_{i}^{ak_{I}},\nonumber \\
\dot{I}_{i}^{s1} & =\bar{\alpha_{i}}k_{E}\gamma_{E}E_{i}^{k_{E}}-k_{I}\gamma_{I^{s}}I_{i}^{a1},\\
\vdots\\
\dot{I}_{i}^{sk_{I}} & =k_{I}\gamma_{I^{s}}I_{i}^{s(k_{I}-1)}-k_{I}\gamma_{I^{s}}I_{i}^{sk_{I}},\\
\dot{R}_{i} & =k_{I}\gamma_{I^{a}}I_{i}^{ak_{I}}+k_{I}\gamma_{I^{s}}I_{i}^{sk_{I}}.\nonumber 
\end{align}

We assume that the rate of infection of a susceptible individual in
age group $i$ is \emph{
\begin{equation}
\lambda_{i}(t)=\beta\sum_{j=1}^{M}\sum_{n=1}^{k_{I}}\left(C_{ij}^{a}\frac{I_{j}^{an}}{N_{j}}+C_{ij}^{s}\frac{I_{j}^{sn}}{N_{j}}\right),
\end{equation}
}

\section{SEAIIR}

\begin{figure}
\begin{centering}
\includegraphics[width=0.99\textwidth]{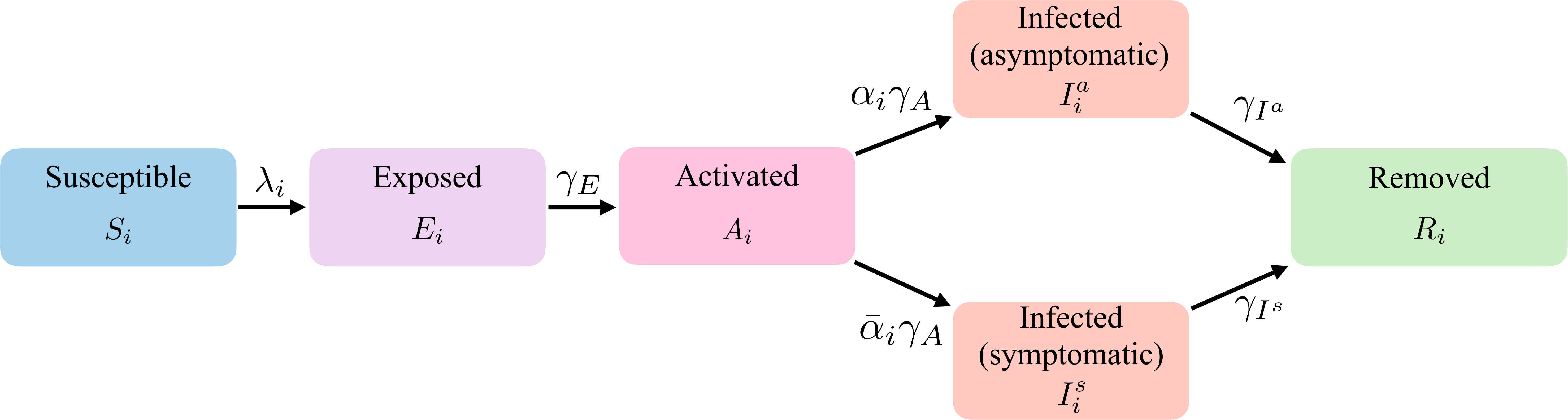}
\par\end{centering}
\caption{\textbf{Schematic of the SEAIIR model. }The parameters for this model
are: $\boldsymbol{\theta}=(\alpha_{i},\beta,\gamma_{E},\gamma_{A},\gamma_{I^{a}},\gamma_{I^{s}})$.
The class SEAIIR can be instantiated in PyRoss using \lstinline!pyross.deterministic.SEAIR!.
\label{tab:SEAIR}}
\end{figure}
This model is an extension of the SEIR model, introducing the additional
class $A$, which is both asymptomatic and infectious. In other words,
this models shows what ensues if \emph{everyone} who gets infected,
undergoes a latency period where they are both asymptomatic and infectious.
This class is potentially quite important, as there is some evidence
that people are infectious before they start showing symptoms. The
deterministic limit of this case

\begin{equation}
\begin{aligned}\dot{S}_{i} & =-\lambda_{i}(t)S_{i}\\
\dot{E}_{i} & =\lambda_{i}(t)S_{i}-\gamma_{E}E_{i}\\
\dot{A}_{i} & =\gamma_{E}E_{i}-\gamma_{A}A_{i}\\
\dot{I}_{i}^{a} & =\alpha_{i}\gamma_{A}A_{i}-\gamma_{I^{a}}I_{i}^{a}\\
\dot{I}_{i}^{s} & =\bar{\alpha_{i}}\gamma_{A}A_{i}-\gamma_{I^{s}}I_{i}^{s}\\
\dot{R}_{i} & =\gamma_{I^{a}}I_{i}^{a}+\gamma_{I^{s}}I_{i}^{s}
\end{aligned}
\end{equation}

The rate of infection of a susceptible individual in age group $i$
is\emph{
\begin{equation}
\lambda_{i}(t)=\beta\sum_{j=1}^{M}\left(C_{ij}^{a}\frac{I_{j}^{a}}{N_{j}}+C_{ij}^{a}\frac{A_{j}}{N_{j}}+C_{ij}^{s}\frac{I_{j}^{s}}{N_{j}}\right),
\end{equation}
}The $A$ and $I^{a}$ classes should behave virtually the same (so
their contact matrices should be equal). The two are kept distinct
to keep track of the fact that some people remain asymptomatic even
in the $I$ stage. Since it's difficult to find data on the ratio
of $I^{s}$ to $I^{a}$, it is possible to disregard the distinction
and simply use $I$ instead. 

\section{SEAI8R}

\begin{figure}
\begin{centering}
\includegraphics[width=0.99\textwidth]{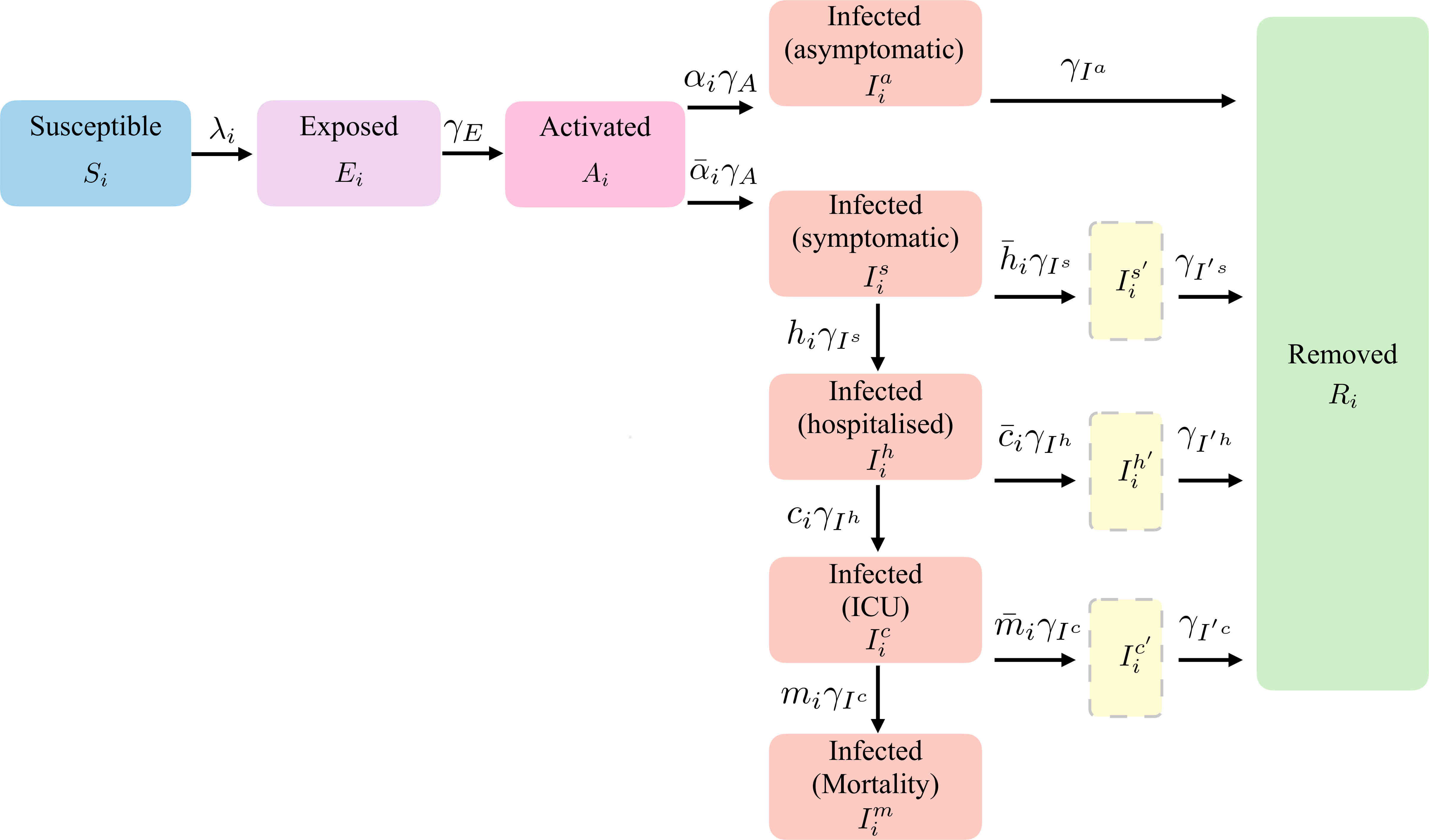}
\par\end{centering}
\caption{\textbf{Schematic of the SEAI8R model. }The class SEAI8R can be instantiated
in PyRoss using \lstinline!pyross.deterministic.SEAI8R!.\label{fig:SEAI8R}}
\end{figure}
This model is an extension of the SEAIIR model. There are now six
more types of infectives ($I_{i}^{h}$: infectives who are hospitalized,
$I_{i}^{c}$: infectives who are in ICU, $I_{i}^{m}$: mortality due
to the infection from ICU, $I_{i}^{s'}:$ intermediate stage between
symptomatic and removed, $I_{i}^{c'}:$ intermediate stage between
hospitalized and removed, and $I_{i}^{c'}:$ intermediate stage between
ICU and removed). The intermediate stages are needed to allow for
a fast progression of the disease while retaining the longer recovery
time and the ratios of people experiencing different levels of severity
of the disease. Figure \ref{fig:SEAI8R} shows the schematic. The
deterministic dynamics if given by the following ODE:

\begin{gather}
\dot{S_{i}}=-\lambda_{i}(t)S_{i}+\sigma_{i},\qquad\dot{E}_{i}=\lambda_{i}(t)S_{i}-\gamma_{E}E_{i},\qquad\dot{A}_{i}=\gamma_{E}E_{i}-\gamma_{A}A_{i}\nonumber \\
\dot{I}_{i}^{a}=\alpha_{i}\gamma_{A}A_{i}-\gamma_{I^{a}}I_{i}^{a},\qquad\dot{I}_{i}^{s}=\bar{\alpha_{i}}\gamma_{A}A_{i}-\gamma_{I^{s}}I_{i}^{s},\qquad\dot{I}_{i}^{s'}=\bar{h}_{i}\gamma_{I^{s}}I_{i}^{s}-\gamma_{I^{s'}}I_{i}^{s'}\nonumber \\
\dot{I}_{i}^{h}=h_{i}\gamma_{I^{s}}I_{i}^{s}-\gamma_{I^{h}}I_{i}^{h},\qquad\dot{I}_{i}^{h'}=\bar{c}_{i}\gamma_{I^{h}}I_{i}^{h}-\gamma_{I^{h'}}I_{i}^{h'},\qquad\dot{I}_{i}^{c}=c_{i}\gamma_{I^{h}}I_{i}^{h}-\gamma_{I^{c}}I_{i}^{c},\nonumber \\
\dot{I}_{i}^{c'}=\bar{m}_{i}\gamma_{I^{c}}I_{i}^{c}-\gamma_{I^{c'}}I_{i}^{c'},\qquad\dot{I}_{i}^{m}=m_{i}\gamma_{I^{c}}I_{i}^{c},\qquad\dot{N}_{i}=\sigma_{i}-m_{i}\gamma_{I^{c}}I_{i}^{c}\\
\dot{R}_{i}=\gamma_{I^{a}}I_{i}^{a}+\gamma_{I^{s'}}I_{i}^{s'}+\gamma_{I^{h'}}I_{i}^{h'}+\gamma_{I^{c'}}I_{i}^{c'}.\nonumber 
\end{gather}
The rate of infection of a susceptible individual in age group $i$
is\emph{
\begin{equation}
\lambda_{i}(t)=\beta\sum_{j=1}^{M}\left(C_{ij}^{a}\frac{I_{j}^{a}}{N_{j}}+C_{ij}^{a}\frac{A_{j}}{N_{j}}+C_{ij}^{s}\frac{I_{j}^{s}}{N_{j}}+C_{ij}^{h}\frac{I_{j}^{h}}{N_{j}}\right),
\end{equation}
}Here $\bar{h}_{i}=1-h_{i}$, $\bar{m}_{i}=1-m_{i}$, $C_{ij}^{s}=f^{s}C_{ij}^{a}\equiv f^{s}C_{ij}$
and $C_{ij}^{s}=f^{h}C_{ij}^{a}\equiv f^{h}C_{ij}$. We note the individuals
can be removed at any stage from either of the eight infection classes. 

\section{SEAIIRQ\label{sec:SEAIIRQ}}

\begin{figure}
\begin{centering}
\includegraphics[width=0.99\textwidth]{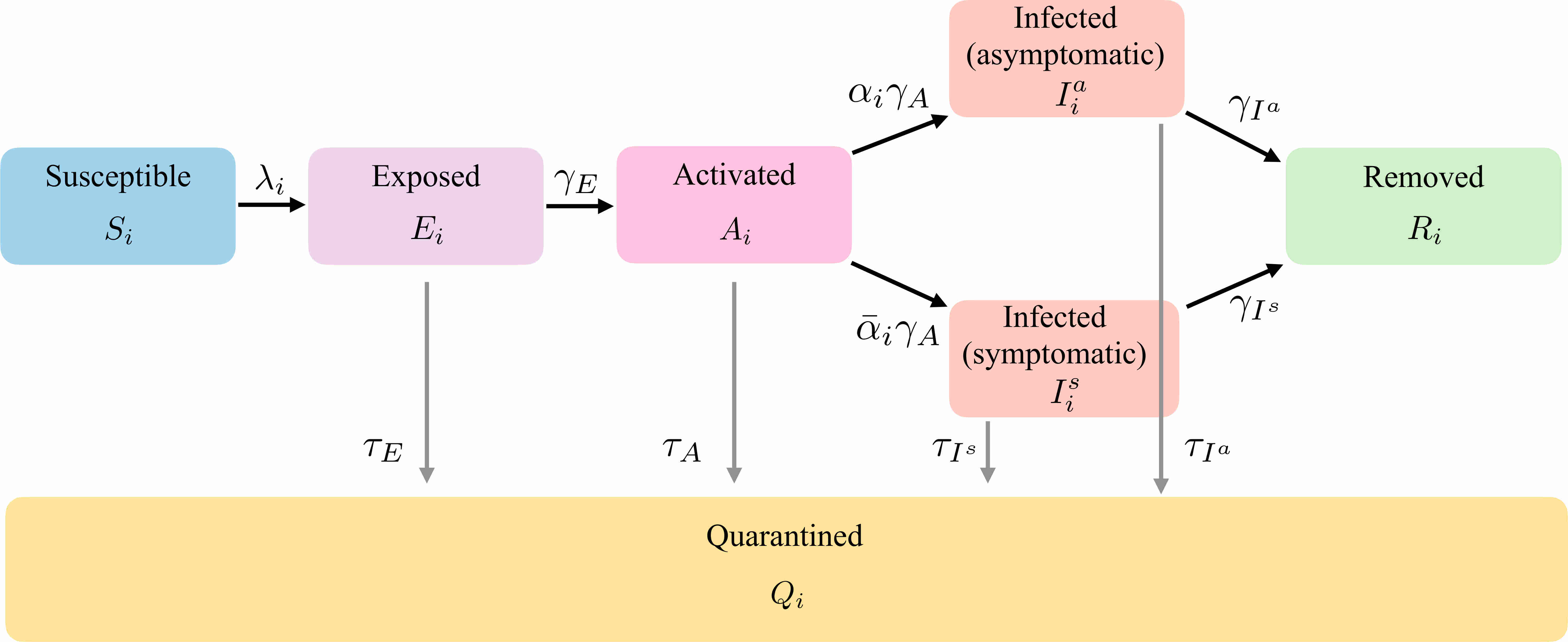}
\par\end{centering}
\caption{\textbf{Schematic of the SEAIIRQ model. }The parameters for this model
are: $\boldsymbol{\theta}=(\alpha_{i},\beta,\gamma_{E},\gamma_{A},\gamma_{I^{a}},\gamma_{I^{s}},\tau_{E},\tau_{A},\tau_{I^{a}},\tau_{I^{s}})$.
The class SEAIIRQ can be instantiated in PyRoss using \lstinline!pyross.deterministic.SEAIRQ!.\label{tab:SEAIRQ}}
\end{figure}
This model is an extension of the SEAIIR model. We introduce the $Q_{i}$
class, which may model individuals who have been tested and put into
quarantine (and can therefore not infect anyone else). This point
of $Q_{i}$ class is to model a possible an implementation of contact
tracing in PyRoss. Figure  \ref{tab:SEAIRQ} shows the schematic.
The deterministic dynamics of the SEAIRQ model is given as:

\begin{equation}
\begin{aligned}\dot{S}_{i} & =-\lambda_{i}(t)S_{i}\\
\dot{E}_{i} & =\lambda_{i}(t)S_{i}-(\gamma_{E}+\tau_{E})E_{i}\\
\dot{A}_{i} & =\gamma_{E}E_{i}-(\gamma_{A}+\tau_{A})A_{i}\\
\dot{I}_{i}^{a} & =\alpha_{i}\gamma_{A}A_{i}-(\gamma_{I^{a}}+\tau_{I^{a}})I_{i}^{a}\\
\dot{I}_{i}^{s} & =\bar{\alpha_{i}}\gamma_{A}A_{i}-(\gamma_{I^{s}}+\tau_{I^{s}})I_{i}^{s}\\
\dot{R}_{i} & =\gamma_{I^{a}}I_{i}^{a}+\gamma_{I^{s}}I_{i}^{s}\\
\dot{Q}_{i} & =\tau_{S}S_{i}+\tau_{E}E_{i}+\tau_{A}A_{i}+\tau_{I^{s}}I_{i}^{s}+\tau_{I^{a}}I_{i}^{a}
\end{aligned}
\end{equation}

The rate of infection of a susceptible individual in age group $i$
is\emph{
\begin{equation}
\lambda_{i}(t)=\beta\sum_{j=1}^{M}\left(C_{ij}^{a}\frac{I_{j}^{a}}{N_{j}}+C_{ij}^{a}\frac{A_{j}}{N_{j}}+C_{ij}^{s}\frac{I_{j}^{s}}{N_{j}}\right),
\end{equation}
}Here $\tau_{E,A,I^{s},I^{a}}$ is the testing rate in the population,
these are in general different for different classes. We have presumed
that people in the incubation stage $E$ can also be tested. 

\section{SIIRS\label{sec:SIIRS}}

\begin{figure}
\begin{centering}
\includegraphics[width=0.7\textwidth]{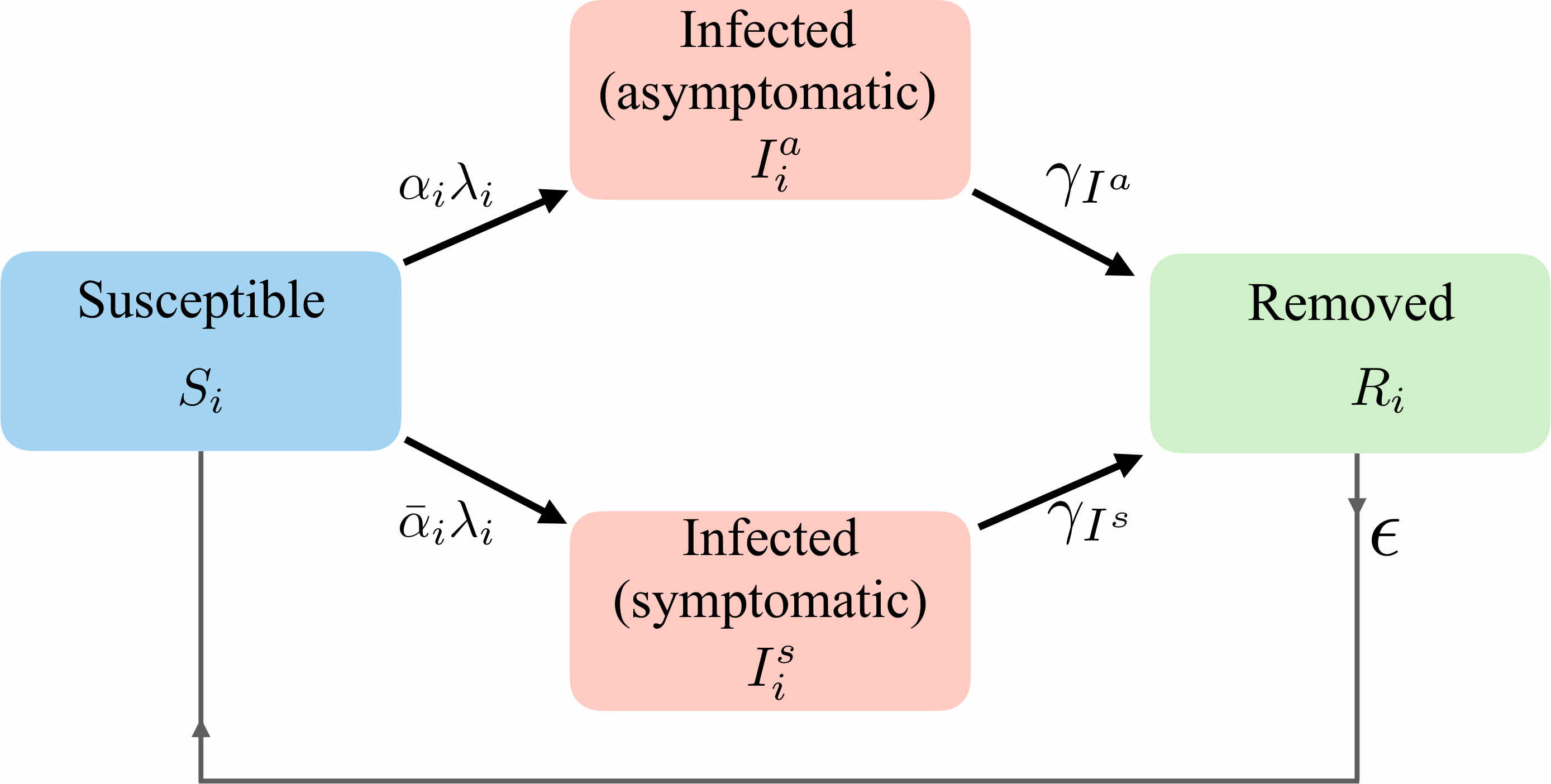}
\par\end{centering}
\caption{\textbf{Schematic of the SIIRS model. }The parameters for this model
are: $\boldsymbol{\theta}=(\alpha_{i},\beta,\gamma_{I^{a}},\gamma_{I^{s}},\epsilon)$.
The class SIIRS can be instantiated in PyRoss using \lstinline!pyross.deterministic.SIRS!.\label{tab:SIRS}}
\end{figure}
We now extend the age-structured SIR model to allow for removed persons
to be susceptible and for change in the population of each age group.
Figure  \ref{tab:SIRS} shows the schematic. The deterministic dynamics
of the resulting SIRS model is:
\begin{align}
\begin{aligned}\dot{S}_{i} & =-\lambda_{i}(t)S_{i}+\sigma_{i}+\epsilon(\gamma_{I^{a}}I_{i}^{a}+\gamma_{I^{s}}I_{i}^{s})\\
\dot{I}_{i}^{a} & =\alpha_{i}\lambda_{i}(t)S_{i}-\gamma_{I^{a}}I_{i}^{a}+l_{i}\\
\dot{I}_{i}^{s} & =\bar{\alpha_{i}}\lambda_{i}(t)S_{i}-\gamma_{I^{a}}I_{i}^{s}\\
\dot{R}_{i} & =\gamma_{I^{a}}I_{i}^{a}+\gamma_{I^{s}}I_{i}^{s}.\\
\dot{N}_{i} & =\sigma_{i}+l_{i}
\end{aligned}
\end{align}
Here $\epsilon$ is fraction of removed who is susceptible. $\sigma_{i}$
denotes of the arrival of new susceptibles, while $l_{i}$ are new
asymptomatic infectives. This means that $N_{i}$ is now dynamical.
The rate of infection of a susceptible individual in age group $i$
is\emph{ }same as in the SIIR model.

\section{Generic user-defined model\label{sec:Even-More-?}}

If the plethora of models described in the preceding sections are
not enough, then PyRoss provides the additional class \lstinline!pyross.deterministic.Spp!
(pronounced \emph{``S plus plus''}), which has the ability to simulate
any generic compartmental model. The model is specified by providing
a Python dictionary, and supports age-differentiated parameters. As
an example, the SIR model, defined in the Spp class, is given in Fig.\ref{fig:Spp}.
\begin{figure}
\begin{lstlisting}[language=Python,basicstyle={\scriptsize\ttfamily},breaklines=true]
model_spec = {     
		"classes" : ["S", "I", "R"],


  "S" : {         
		"linear"    : [],         
		"infection" : [ ["I", "-beta"] ]     
	},
    
	"I" : {         
		"linear"    : [ ["I", "-gamma"] ],         
		"infection" : [ ["I", "beta"] ]     
	},
    
	"R" : {         
		"linear"    : [ ["I", "gamma"] ],         
		"infection" : []     
	} 
} 
\end{lstlisting}

\caption{\label{fig:Spp}\textbf{Definition of the }\textbf{\emph{Spp}}\textbf{
class. }The \emph{Spp} class can be instantiated in PyRoss using \lstinline!pyross.deterministic.Spp!.}
\end{figure}

Currently, the \emph{Spp} class supports the two types of terms which
all the compartmental models above share: linear terms and infection
terms. The class could be used to simulate any generic age-structured
epidemiological compartment model, where the rates could be both time
and state dependent.

Note that \lstinline!pyross.deterministic.Spp! is designed with generality
rather than optimality in mind. A model implemented using \lstinline!pyross.deterministic.Spp!
will in general perform worse than any of the corresponding hard-coded
classes above.

\chapter{Applications}

In this chapter, we provide illustrative examples of usage of the
PyRoss library. 

\section{Basic reproductive ratio $\mathcal{R}_{0}$ from local rate of growth}

\begin{figure}[H]
\centering\includegraphics[width=0.8\textwidth]{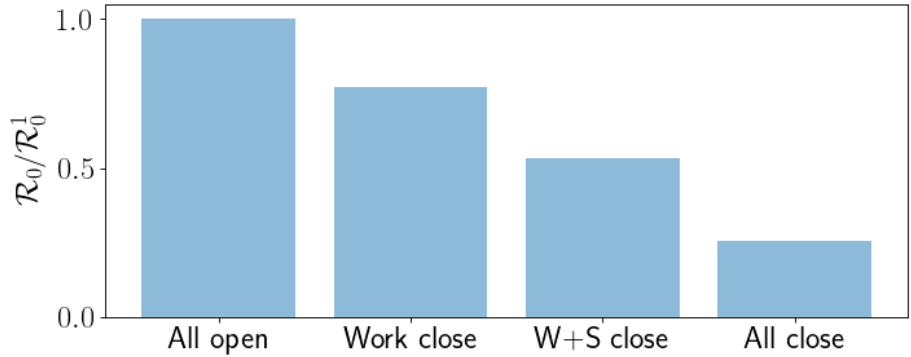}\caption{\textbf{The basic reproductive ratio as a function of intervention
measures.} The y-axis is normalized by the case without no intervention
measures. Here `W+S close' means both work and school are closed and
$\mathcal{R}_{0}^{1}$ is the basic reproductive ration without any
intervention measure. \label{fig:1-1}}
\end{figure}

Our first example is on computing the basic reproductive ratio $\mathcal{R}_{0}$
as a function of intervention measures. In a population of susceptibles
$S_{i}$, the expected number of secondary infections arising from
a single individual during the entire infectious period is defined
as $\mathcal{R}_{0}$ \cite{diekmann2010construction,heesterbeek2002brief}.
We obtain the basic reproductive ratio of the SIR model, defined in
section \ref{sec:SIR}, by linearising the dynamics about the disease-free
fixed point, where $S_{i}=N_{i}$. The time evolution of infectives
is governed by
\begin{equation}
\boldsymbol{J}=\gamma(\boldsymbol{L}-\boldsymbol{1}).
\end{equation}
Here $\boldsymbol{1}$ is the identity matrix and 
\[
\boldsymbol{L}=\frac{\alpha\beta}{\gamma}C_{ij}\frac{N_{i}}{N_{j}}.
\]
It is sufficient for the spectral radius of $\boldsymbol{L}$ to be
greater than unity for the epidemic to grow. The $\mathcal{R}_{0}$
is then obtained as the spectral radius of $\boldsymbol{L}$ \cite{diekmann2010construction,heesterbeek2002brief}:

\begin{equation}
\mathcal{R}_{0}\equiv\rho(\boldsymbol{L})=\text{max}\{|\Lambda_{1}|,\ldots,|\Lambda_{M}|\}.
\end{equation}
 We can now extend the linearisation at any point in time $t$ by
making the replacements $N_{i}$ $\longrightarrow$ $S_{i}(t)$ and
$C_{ij}\longrightarrow C_{ij}(t)$ in the expression for $\boldsymbol{L}$,
giving the time-dependent stability matrix $\boldsymbol{L}^{(t)}$
\cite{Singh2020}, and thus, define effective time-dependent basic
reproductive ratio 
\begin{equation}
\mathcal{R}_{\text{0}}^{\text{eff}}(t)\equiv\rho(\boldsymbol{L}^{(t)})=\text{max}\{|\Lambda_{1}^{(t)}|,\ldots,|\Lambda_{M}^{(t)}|\}
\end{equation}
We now provide illustrative examples of measuring $\mathcal{R}_{0}$
and $\mathcal{R^{\text{eff}}}_{0}(t)$. In Fig.(\ref{fig:1-1}), we
obtain the basic reproductive ration as intervention measures are
changed. 

\begin{lstlisting}[language=Python]
import pyross
import matplotlib.pyplot as plt, numpy as np

M=16                       ## number of age classes
my_data = np.genfromtxt('UK.csv', delimiter=',', skip_header=1)
Ni = (my_data[:,1]+my_data[:,2])[0:M]

# contact structure of the UK
ukCH, ukCW, ukCS, ukCO = pyross.contactMatrix.UK()
aH, aW, aS, aO = 1, 1, 1, 1
CH, CW, CS, CO = aH*ukCH, aW*ukCW, aS*ukCS, aO*ukCO
 
beta  = 1               # the probability of infection on contact
gIa   = 1./7            # removal rate of asymptomatic infectives 
gIs   = 1./7            # removal rate of symptomatic infectives 
alpha = 0.              # asymptomatic fraction
fsa   = 1               # suppresion of contact by symptomatics
parameters = {'alpha':alpha,'beta':beta, 'gIa':gIa,'gIs':gIs,'fsa':fsa, 'M':M, 'Ni':Ni}

contactMatrix = pyross.contactMatrix.SIR(CH, CW, CS, CO)
r0UK1 = contactMatrix.basicReproductiveRatio(parameters)

#switch off contacts at work
aH, aW, aS, aO = 1, 0, 1, 1
CH, CW, CS, CO = aH*ukCH, aW*ukCW, aS*ukCS, aO*ukCO
contactMatrix = pyross.contactMatrix.SIR(CH, CW, CS, CO)
r0UK2 = contactMatrix.basicReproductiveRatio(parameters)

#switch off contacts at work and school
aH, aW, aS, aO = 1, 0, 0, 1
CH, CW, CS, CO = aH*ukCH, aW*ukCW, aS*ukCS, aO*ukCO
contactMatrix = pyross.contactMatrix.SIR(CH, CW, CS, CO)
r0UK3 = contactMatrix.basicReproductiveRatio(parameters)

#switch off all contacts but at home
aH, aW, aS, aO = 1, 0, 0, 0
CH, CW, CS, CO = aH*ukCH, aW*ukCW, aS*ukCS, aO*ukCO
contactMatrix = pyross.contactMatrix.SIR(CH, CW, CS, CO)
r0UK4 = contactMatrix.basicReproductiveRatio(parameters)

r0 = np.array([r0UK1, r0UK2, r0UK3, r0UK4])/r0UK1
plt.bar(range(r0.size), r0, align='center', alpha=0.5); 
plt.ylabel(r'$\mathcal R_0/\mathcal R^{{1}}_0$')
labelY= ('All open', 'Work close', 'School close', 'All close' ); 
plt.xticks(range(r0.size),labelY);


\end{lstlisting}

\section{Beyond $\mathcal{R}_{0}$: Kreiss constant $\mathcal{K}(\boldsymbol{J})$\label{subsec:Linearised-growth-rates}}

In the previous section, we used linearisation to study growth of
epidemic. The prediction made by linearised theory is based on the
fact that the growth rate of a linear system is exponential in the
largest eigenvalue. This is well known in the epidemiology community,
where the value of $\mathcal{R}_{0}$ is used as a measure of how
dangerous an epidemic will become and on what timescale. On the other
hand, far less well known is the transient effect due to non-normality
of $\boldsymbol{J}$ on the initial dynamics. There have been some
papers \cite{Townley2007,Stott2012} in the ecology community on this
effect in the context of population dynamics. The best way to understand
why non-normality can be important is through an instructive example.
Let $\boldsymbol{x}=\bar{\boldsymbol{x}}+\boldsymbol{u}$, 

\[
\frac{d\boldsymbol{u}}{dt}=\boldsymbol{J}(t,\boldsymbol{\theta},\bar{\boldsymbol{x}})\cdot\boldsymbol{u}
\]

Consider, 
\[
\boldsymbol{J}=\begin{pmatrix}-1 & 500\\
0 & -2
\end{pmatrix}
\]
\begin{lstlisting}[language=Python]
import pyross
import numpy as np, matplotlib.pyplot as plt
import scipy.linalg as spl
from scipy.integrate import solve_ivp
from pyross.contactMatrix import characterise_transient

M = 2                   # the SIR model has no age structure
N = 100000              # the total population 

Ni = np.zeros((M))      # population in each group
fi = np.zeros((M))      # fraction of population in age age group
fi = np.array((0.25, 0.75))
for i in range(M):
    Ni[i] = fi[i]*N

beta  = 0.02            # infection rate
gamma = 0.007
gIa   = gamma           # removal rate of asymptomatic infectives
gIs   = gamma           # removal rate of symptomatic infectives
alpha = 0               # fraction of asymptomatic infectives
fsa   = 1               # the self-isolation parameter
Ia0 = np.array([0,0])   # the SIR model has only one kind of infective
Is0 = np.array([1,.1])  # we take these to be symptomatic
R0  = np.array([0,0])   # and assume there are no removed individuals initially
S0  = Ni-Ia0-Is0-R0             

# set the contact structure
C11, C22, C12 = 1,1,4
C = np.array(([C11, C12], [C12*fi[1]/fi[0], C22]))
def contactMatrix(t):
    return C

# duration of simulation and data file
Tf = 160;  Nt=160;

# instantiate model
parameters = {'alpha':alpha, 'beta':beta, 'gIa':gIa, 'gIs':gIs,'fsa':fsa}
model = pyross.deterministic.SIR(parameters, M, Ni)

# simulate model
data = model.simulate(S0, Ia0, Is0, contactMatrix, Tf, Nt)

# matrix for linearised dynamics
C=contactMatrix(0)
A=((beta*C-gamma*np.identity(len(C))).T*fi).T/fi
mcA=pyross.contactMatrix.characterise_transient(A, ord=1)
AP = A-np.max(np.linalg.eigvals(A))*np.identity(len(A))
mcAA = pyross.contactMatrix.characterise_transient(AP,ord=1)

# plot the data and obtain the epidemic curve
Sa  =data['X'][:,:1].flatten(); Sk=data['X'][:,1:M].flatten()
Isa =data['X'][:,2*M:2*M+1].flatten(); Isk=data['X'][:,2*M+1:3*M].flatten()
St=Sa+Sk;  It=Isa+Isk
# It = np.sqrt(Isa**2 + Isk**2)
t = data['t']

fig = plt.figure(num=None, figsize=(10, 8), dpi=80, facecolor='w', edgecolor='k')
plt.fill_between(t, 0, St/N, color="#348ABD", alpha=0.3)
plt.plot(t, St/N, '-', color="#348ABD", label='$S$', lw=4)
plt.fill_between(t, 0, It/N, color='#A60628', alpha=0.3)
plt.plot(t, It/N, '-', color='#A60628', label='$I$', lw=4)
Rt=N-St-It; plt.fill_between(t, 0, Rt/N, color="dimgrey", alpha=0.3)
plt.plot(t, Rt/N, '-', color="dimgrey", label='$R$', lw=4)
plt.autoscale(enable=True, axis='x', tight=True)

###Estimate from Kreiss constant
plt.plot(t,mcAA[2]*It[0]*np.exp(mcA[0]*t)/N,'-', color="green",
         label='$Estimate$', lw=4)
plt.yscale('log'); plt.xlabel("time"); plt.ylabel("% of population");  
plt.legend(fontsize=26); plt.grid();  plt.show()

def linear_system(t, x, A): return A@x
A2 = np.array([[3,2],[9,4]])
x0, tf  = [1,1], 1
ivp_exp = solve_ivp(linear_system, (0,tf), x0, args=[A2], t_eval=np.arange(0,tf,.1))
t=ivp_exp.t
Gamma = A2 - np.max(spl.eigvals(A2))*np.identity(len(A2))
mcA2 = characterise_transient(Gamma)
ivp_exp2 = solve_ivp(linear_system, (0,tf), x0, args=[Gamma], t_eval=np.arange(0,tf,.01))

f, ax = plt.subplots()
plt.plot(ivp_exp2.t,spl.norm(ivp_exp2.y.T, axis=1)/spl.norm(x0))
ax.set_xlabel("time");  ax.set_ylabel(r'$|u|/|u_0|$')
ax.plot(t, np.exp(mcA2[0]*t),"--",color="darkgreen")
ax.set_ylim((-.1,np.max(spl.norm(ivp_exp2.y.T, axis=1)/spl.norm(x0))*1.1))
t_trunc = t[np.where(t<mcA2[3])]
ax.plot(t_trunc,np.exp(mcA2[1]*t_trunc),"--",color="orange")
plt.axhline(y=mcA2[2],linestyle="dotted",color="black")    
plt.ylim([.98,1.4]); plt.annotate(r'Long time behaviour $\alpha (\Gamma)$', [.2,1.01])
plt.annotate(r'Initial growth rate $\omega (\Gamma)$',[.0,1.05], rotation=68)
plt.annotate(r'Kreiss constant $\mathcal{K} (\Gamma)$', [.4,1.3]); plt.show()

\end{lstlisting}
\begin{figure}
\begin{tabular}{cc}
\includegraphics[width=0.57\textwidth]{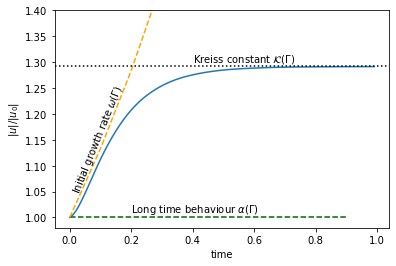} & \includegraphics[width=0.52\textwidth]{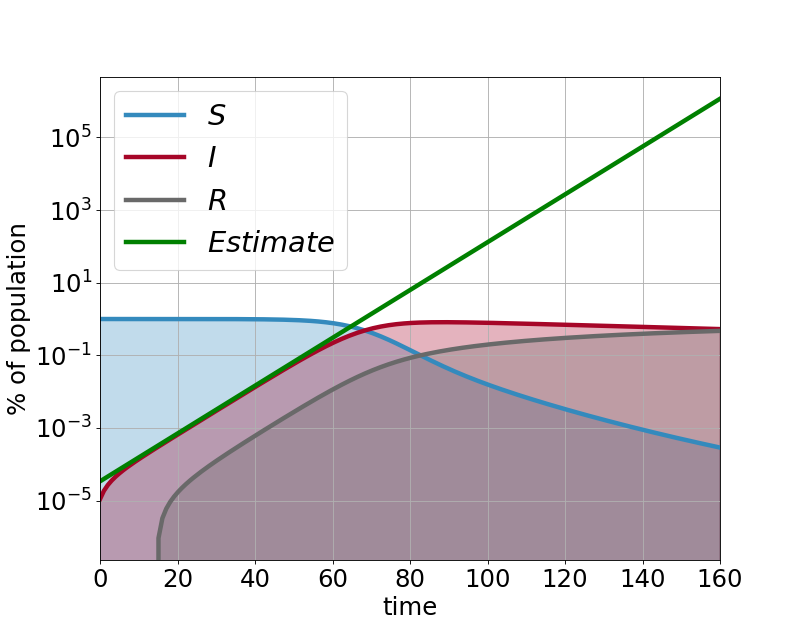}\tabularnewline
(a) & (b)\tabularnewline
\end{tabular}

\caption{\textbf{Effect of non-normality on the dynamics}. (a) The evolution
of the associated : (a) The evolution of the associated . The non-normality
of $\boldsymbol{C}$ results in a greater total number of infected,
plotted in green. . The system quickly saturates the Kreiss bound
and asymptotically tends towards normal evolution along it. (b) the
2-age structured SIR model with contact matrix\label{fig:Effect-of-non-normality}}
\end{figure}
Obviously the eigenvalues are $-1,-2$ so a simple eigenvalue criterion
would expect an exponential decay of any initial perturbations. However,
let $\boldsymbol{u}_{0}=\left(0,1\right)$ and it is obvious that
$|\boldsymbol{u}|$ is dramatically magnified in value. After a long
enough time, the system will obviously decay to $0$ but in a non-linear
system, transient amplification may have a dramatic effect on the
long term dynamics if a system is only locally stable around some
fixed point \cite{Asllani2018a,Asllani2018b}. The behaviour results
from the non-normality of $\boldsymbol{J}$. Normal matrices are defined
as $\boldsymbol{J}\boldsymbol{J}^{T}=\boldsymbol{J}^{T}\boldsymbol{J}$,
which, by the spectral theorem, is the criterion for diagonalisability.
Obviously eigenvalues alone cannot hope to encapsulate the full picture
here. Here we show these effects can be encapsulated by the introduction
of a single new multiplicative parameter and can be used to better
understand the initial growth dynamics. 

The transient is best characterized by the Kreiss constant $\mathcal{K}(\boldsymbol{J})$
which provides a lower bound to the maximum amplitude of a system
evolving under $\boldsymbol{J}$ \cite{trefethen2005spectra,Asllani2018a,Townley2007}.
For a system that grows with time, the ``extra increase'' can be
estimated by
\[
\mathcal{K}(\boldsymbol{J}-\lambda_{\text{Max}}\left(\boldsymbol{J}\right)),
\]
The associated system $\dot{\boldsymbol{u}}=\left(\boldsymbol{J}-\lambda_{Max}\boldsymbol{I}\right)\boldsymbol{u}=\boldsymbol{\Gamma}\boldsymbol{u}$
has the solution
\[
u(t)=\frac{e^{\boldsymbol{J}t}}{e^{\lambda_{\text{Max}}t}}u(0)\underset{t\rightarrow\infty}{\rightarrow}\mathcal{K}(\boldsymbol{\Gamma})u(0).
\]
In practice, the bound is well saturated over the time scale of the
transient $\tau$ which is usually much faster than any other dynamics
of the system (and can be estimated from pseudospectral methods \cite{trefethen2005spectra}
)
\[
u(t)\underset{t\rightarrow\tau}{\rightarrow}\mathcal{K}(\boldsymbol{\Gamma})u(0)
\]
Thus, the solution to our initial equation $\dot{u}=\boldsymbol{J}u$
can be estimated as 
\begin{equation}
u(t)=e^{\lambda_{\text{Max}}t}\mathcal{K}(\boldsymbol{\Gamma})u(0).\label{eq:est}
\end{equation}

In summary, the maximum eigenvalue and the Kreiss constant of the
associated system together characterise the initial growth of non-normal
evolution as seen in Eq.(\ref{eq:est}) and \ref{fig:Effect-of-non-normality}.
This new parameter acts as an amplification of the initial conditions
based on the degree of non-normality of our network and is general
for any type of network, for example age structure or geographical
information. 

\section{Stochastic sampling \label{sec:Stochastic-sampling}}

We now illustrate the usage of PyRoss for stochastic sampling of epidemics. 

\begin{lstlisting}[language=Python]
import numpy as np
import pyross
import matplotlib.pyplot as plt


M = 1                  # the SIR model has no age structure
Ni = 1000*np.ones(M)   # so there is only one age group
N = np.sum(Ni)         # and the total population is the size of this age group

beta  = 0.2            # infection rate
gIa   = 0.1            # removal rate of asymptomatic infectives
gIs   = 0.1            # removal rate of symptomatic infectives
alpha = 0              # fraction of asymptomatic infectives
fsa   = 1              # self-isolation of symtomatic infectives


Ia0 = np.array([0])     # the SIR model has only one kind of infective
Is0 = np.array([5])     # we take these to be symptomatic
R0  = np.array([0])     # and assume there are no removed individuals initially
S0  = N-(Ia0+Is0+R0)    # so that the initial susceptibles are obtained from S + Ia + Is + R = N

# there is no contact structure
def contactMatrix(t):
    return np.identity(M)

# duration of simulation and data file
Tf = 160;  Nt=160;

# instantiate model
parameters = {'alpha':alpha, 'beta':beta, 'gIa':gIa, 'gIs':gIs,'fsa':fsa}
model = pyross.stochastic.SIR(parameters, M, Ni)

# simulate model
data = model.simulate(S0, Ia0, Is0, contactMatrix, Tf, Nt)

# plot the compartments
pyross.utils.plotSIR(data)

# plot the basic reproductive ratio as a function of time
C=np.identity(M)
contactMatrix = pyross.contactMatrix.SIR(0.25*C, 0.25*C, 0.25*C, 0.25*C)
r0de = contactMatrix.basicReproductiveRatio(data, state='dynamic')
t = data['t']
plt.fill_between(t, 0, t*0+1, color="dimgrey", alpha=0.2)
plt.plot(r0de, '*', color='#A60628')
plt.xlabel('Days'); plt.grid()
plt.ylabel('Basic reproductive ratio')
plt.autoscale(enable=True, axis='x', tight=True);
plt.yticks(np.arange(0, 3, step=0.5)); plt.ylim(0,2.2);
plt.show()
\end{lstlisting}

\begin{figure}[H]
\centering\includegraphics[width=0.45\textwidth]{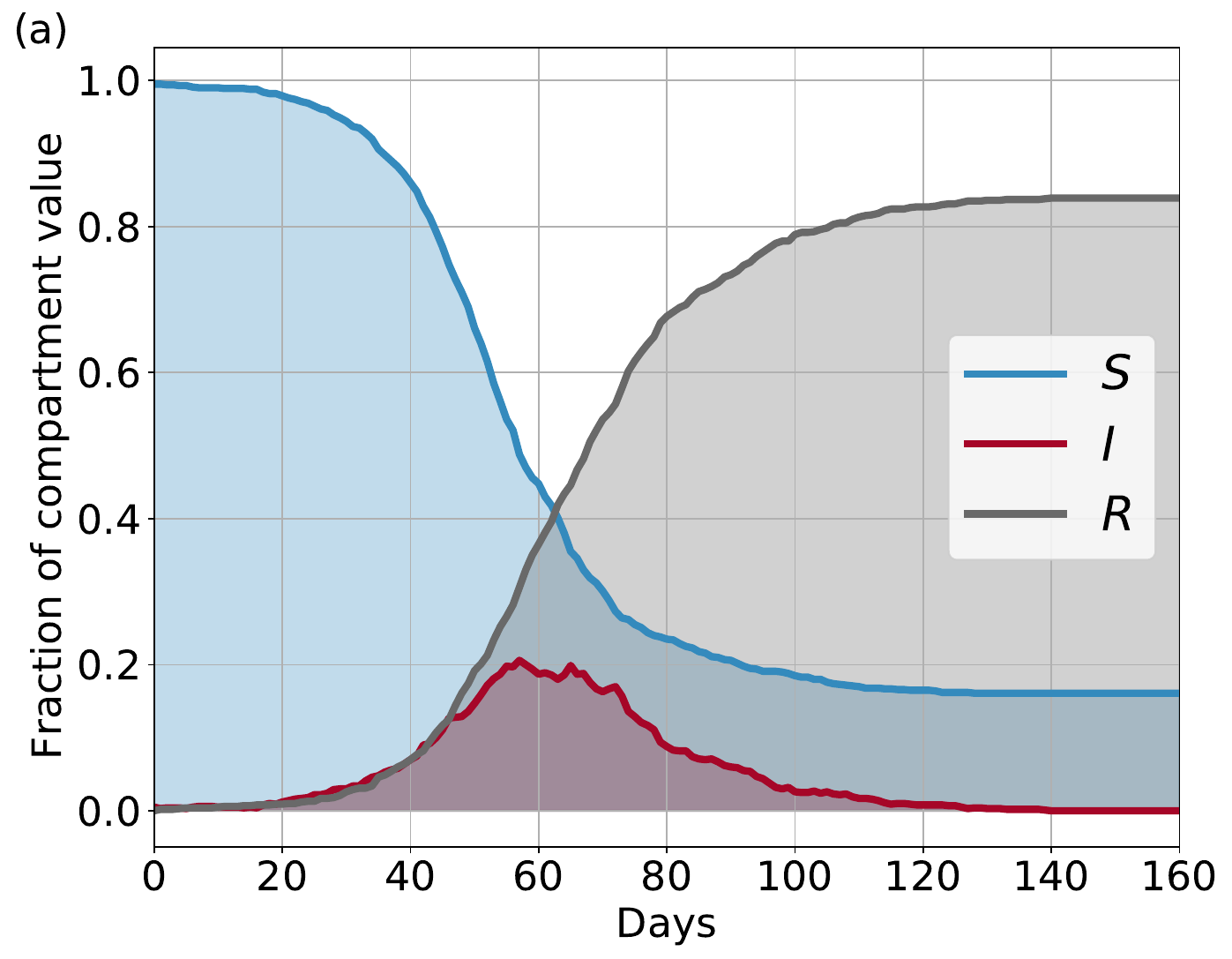}\qquad\includegraphics[width=0.45\textwidth]{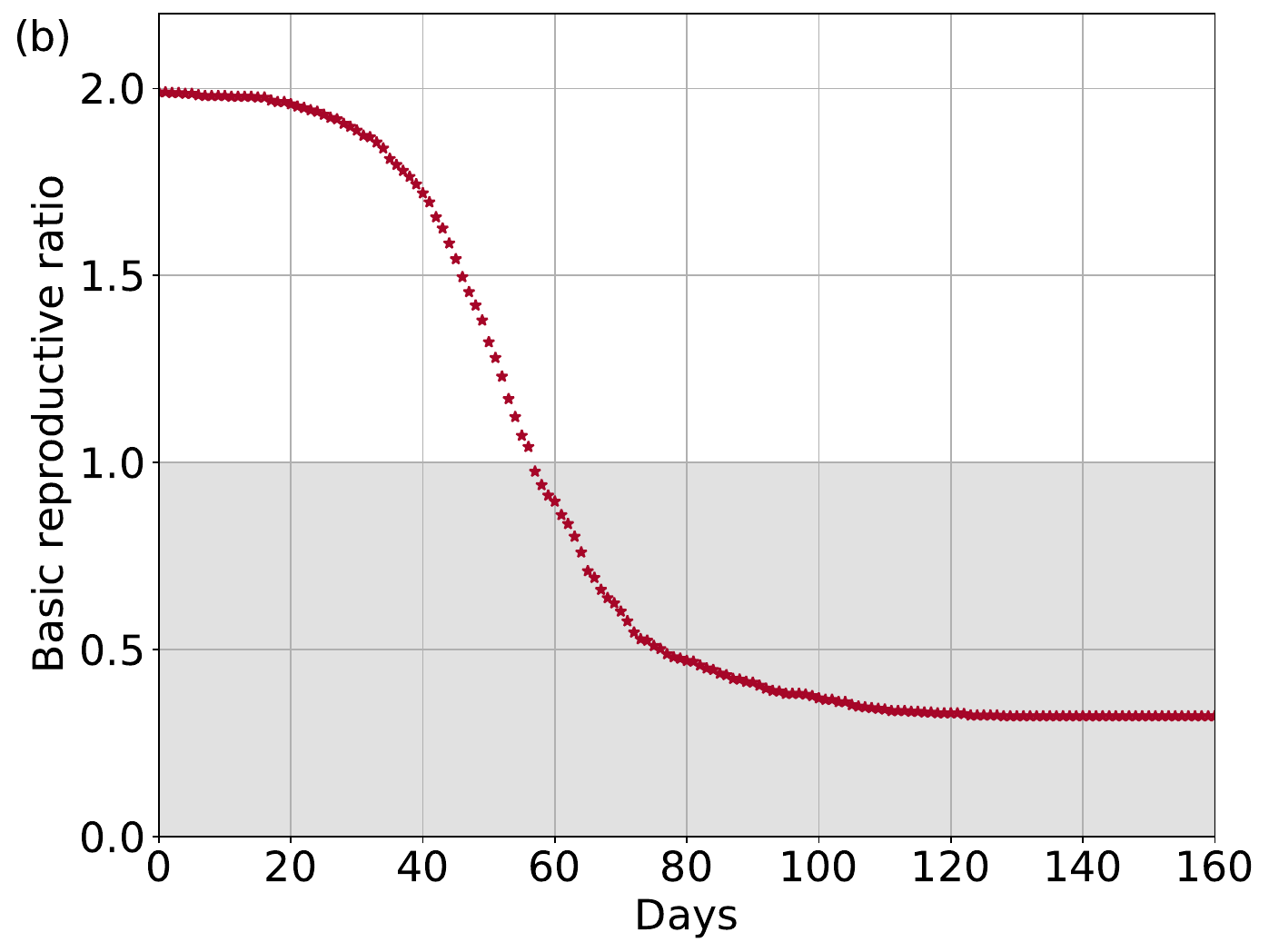}\caption{\textbf{A stochastic realisation of the SIR model, c.f.\,example
\ref{sec:Stochastic-sampling}.} Subplot (a) shows a resulting epidemic
curve, i.e.\,a plot of the number of susceptibles, infectives and
removed as a function of time. Subplot (b) depicts the corresponding
basic reproductive ratio. \label{fig:stochastic_sampling}}
\end{figure}

\section{Deterministic integration\label{sec:Deterministic-integration}}

We now illustrate the usage of PyRoss for deterministic sampling of
epidemics. 

\begin{lstlisting}[language=Python]
import numpy as np
import pyross
import matplotlib.pyplot as plt


M = 1                  # the SIR model has no age structure
Ni = 1000*np.ones(M)   # so there is only one age group
N = np.sum(Ni)         # and the total population is the size of this age group

beta  = 0.2            # infection rate
gIa   = 0.1            # recovery rate of asymptomatic infectives
gIs   = 0.1            # recovery rate of symptomatic infectives
alpha = 0              # fraction of asymptomatic infectives
fsa   = 1              # self-isolation of symtomatic infectives


Ia0 = np.array([0])     # the SIR model has only one kind of infective
Is0 = np.array([1])     # we take these to be symptomatic
R0  = np.array([0])     # and assume there are no removed individuals initially
S0  = N-(Ia0+Is0+R0)    # so that the initial susceptibles are obtained from S + Ia + Is + R = N

# there is no contact structure
def contactMatrix(t):
    return np.identity(M)

# duration of simulation and data file
Tf = 160;  Nt=160;

# instantiate model
parameters = {'alpha':alpha, 'beta':beta, 'gIa':gIa, 'gIs':gIs,'fsa':fsa}
model = pyross.deterministic.SIR(parameters, M, Ni)

# simulate model
data = model.simulate(S0, Ia0, Is0, contactMatrix, Tf, Nt)

# plot the compartments 
pyross.utils.plotSIR(data)

# plot the basic reproductive ratio as a function of time
C=np.identity(M); contactMatrix = pyross.contactMatrix.SIR(0.25*C, 0.25*C, 0.25*C, 0.25*C)
r0de = contactMatrix.basicReproductiveRatio(data, state='dynamic')
plt.plot(r0de, '*', color='#A60628')

\end{lstlisting}

\begin{figure}[H]
\centering\includegraphics[width=0.45\textwidth]{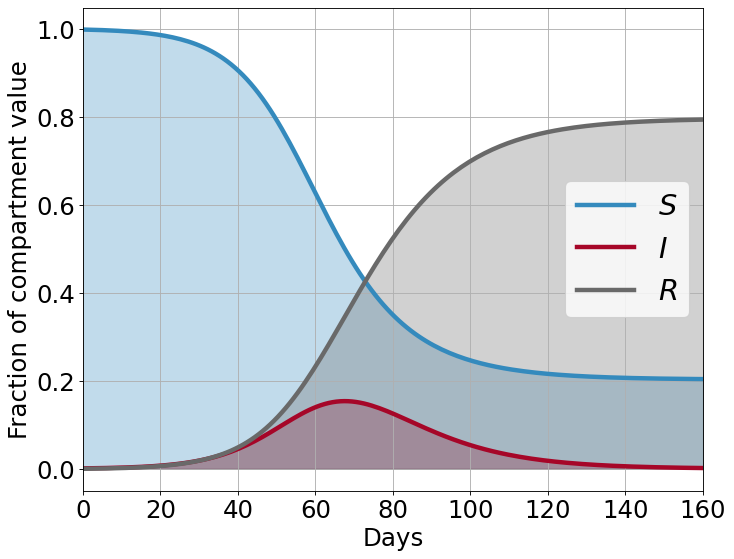}\qquad\includegraphics[width=0.45\textwidth]{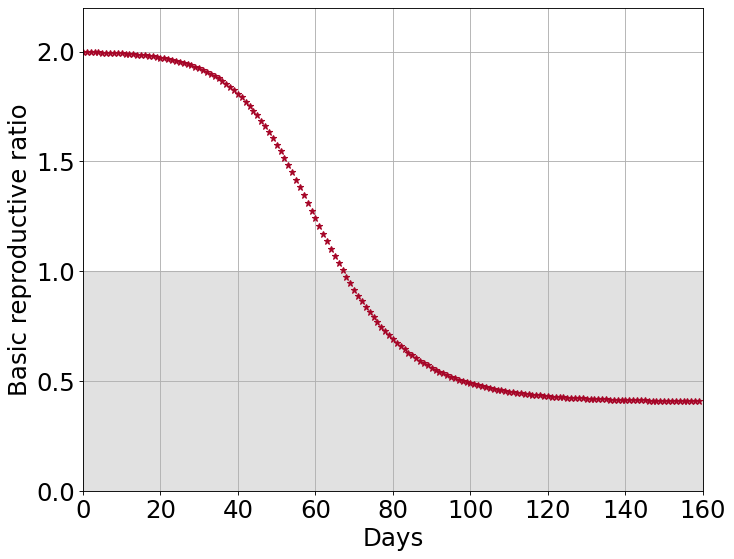}\caption{\textbf{A deterministic realisation of the SIR model, c.f.\,example
\ref{sec:Deterministic-integration}.} The left panel shows a resulting
epidemic curve, i.e.\,a plot of the number of susceptibles, infectives
and removed as a function of time. The right panel depicts the corresponding
basic reproductive ratio. \label{fig:1}}
\end{figure}

\section{Bayesian inference and forecasting\label{sec:Inference-and-forecasting}}

In this example, a trajectory of an SIIR model with two age groups
is generated using \lstinline!pyross.stochastic!. Based on the first
20 datapoints of the trajectory, \lstinline!pyross.inference! is
then used to infer the parameters of the time series. Finally, using
\lstinline!pyross.forecast!, predictions based the inferred parameters
are made and compared to the original SIIR trajectory.

\vspace{0.5cm}

\begin{lstlisting}[language=Python]
import numpy as np
import matplotlib.pyplot as plt
import pyross

# 1. Define model

# number of age groups and total population
M  = 2               # the population has two age groups
N  =  5e4            # and this is the total population

# exact parameters of reference trajectory
beta  = 0.02         # infection rate
gIa   = 1./7         # removal rate of asymptomatic infectives
gIs   = 1./7         # removal rate of asymptomatic infectives
alpha = 0.2          # fraction of asymptomatic infectives
fsa   = 0.8          # self-isolation of symtomatic infectives

# set the age structure
fi = np.array([0.25, 0.75])  # fraction of population in age age group
Ni = N*fi

# set up initial condition
Ia0 = np.array([10, 10])  # each age group has asymptomatic infectives
Is0 = np.array([10, 10])   # and also symptomatic infectives
R0  = np.array([0, 0])  # there are no removed individuals initially
S0  = Ni - (Ia0 + Is0 + R0) # initial number of susceptibles

# set the contact matrix
C = np.array([[18., 9.], [3., 12.]])
contactMatrix = lambda t: C


# 2. Use pyross stochastic to generate reference trajectory

Tf = 100; Nf = Tf+1 # reference trajectory is 100 days long
parameters = {'alpha':alpha, 'beta':beta, 'gIa':gIa, 'gIs':gIs,'fsa':fsa}
sto_model = pyross.stochastic.SIR(parameters, M, Ni)
data = sto_model.simulate(S0, Ia0, Is0, contactMatrix, Tf, Nf)
data_array = data['X']


# 3. Run inference on the first 20 days

# create array with first 20 days of data
Tf_inf = 20; Nf_inf = Tf_inf + 1
data_inf = (data_array/N)[:Nf_inf]

# parameters for inference
ftol = 1e-6
steps = 101
estimator = pyross.inference.SIR(parameters, M, fi, int(N), steps)

# initial guesses, bounds, and initial standard deviation for solver
alpha_g = 0.15
alpha_std = 0.2
alpha_bounds = (1e-3, 0.5)

beta_g = 0.05
beta_std = 0.1
beta_bounds = (1e-3, 1)

gIa_g = 0.13
gIa_std = 0.1
gIa_bounds = (1e-3, 1)

gIs_g = 0.15
gIs_std = 0.1
gIs_bounds = (1e-3, 1)

guess = np.array([alpha_g, beta_g, gIa_g, gIs_g])
stds = np.array([alpha_std, beta_std, gIa_std, gIs_std])
bounds = np.array([alpha_bounds, beta_bounds, gIa_bounds, gIs_bounds])
keys = ['alpha', 'beta', 'gIa', 'gIs']

params = estimator.infer_parameters(keys, guess, stds, bounds, data_inf, Tf_inf, Nf_inf, contactMatrix,
                                  global_max_iter=20, global_ftol_factor=10,
                                  verbose=True)
print(params)

# for forecasting we also need the covariance matrix
hess = estimator.compute_hessian(keys, params, guess, stds,
                        data_inf, Tf_inf, Nf_inf, contactMatrix)
cov = np.linalg.inv(hess)


# 4. Run forecast using inferred parameters and their covariance

# instantiate model
parameters = {'alpha': params[0], 'beta': params[1],
              'gIa': params[2], 'gIs': params[3],'fsa':fsa,'cov':cov}
model_forecast = pyross.forecast.SIR(parameters, M, Ni)

# initial condition for forecast is final state from inference
S0_fc = data_inf[-1,:M]*N
Ia0_fc = data_inf[-1,M:2*M]*N
Is0_fc = data_inf[-1,2*M:]*N

# run forecast
Tf_fc = Tf - Tf_inf; Nf_fc = Tf_fc+1 # simulation time for forecast
Ns = 500 # number of forecasting samples
result_fc = model_forecast.simulate(S0_fc, Ia0_fc, Is0_fc,
                                contactMatrix, Tf_fc, Nf_fc,Ns=Ns)
trajectories_fc = result_fc['X']
t_fc = result_fc['t'] + Tf_inf


# 5. Visualise result: plot fraction of symptomatic infectives

traj_ref = np.sum(data_array[:,2*M:],axis=-1) # reference trajectory
trajs_fc = np.sum(trajectories_fc[:,2*M:],axis=1) # predictions
mean_fc = np.mean( trajs_fc, axis=0) # mean prediction

fig, ax = plt.subplots(1,1,figsize=(7,5))
ax.axvspan(0, Tf_inf,label='Range used for inference',
           alpha=0.3,color='dodgerblue')
for i,e in enumerate(trajs_fc): # plot all forecasting trajectories
    ax.plot(t_fc,e/N,alpha=0.1)
ax.plot(traj_ref/N,lw=3,color='#A60628',label='Reference trajectory')
ax.plot(t_fc,mean_fc/N,
        ls='--',color='limegreen',label='Mean prediction',lw=3)
ax.set_xlim(0,np.max(t_fc))
ax.set_xlabel(r'Days')
ax.set_ylabel('Fraction of symptomatic infectives')
ax.legend(loc='upper right',fontsize=12)
plt.grid(); plt.show()
plt.close(fig)
\end{lstlisting}

\begin{figure}[H]
\centering\includegraphics[width=0.7\textwidth]{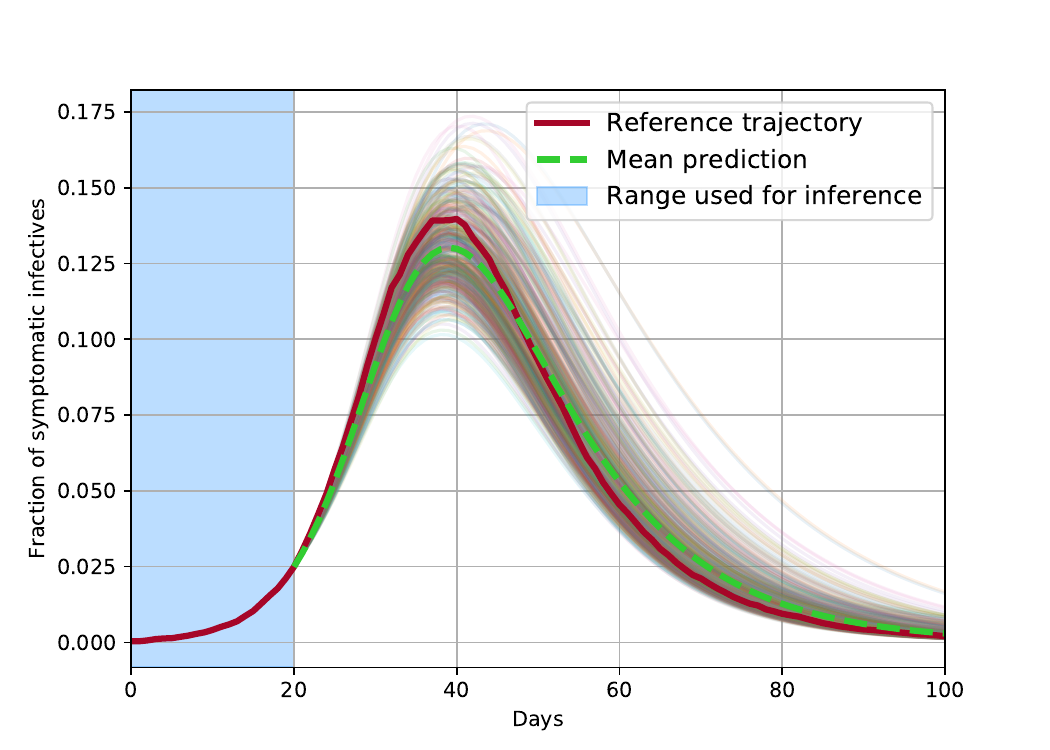}\caption{\textbf{Bayesian inference and forecasting.} Example \ref{sec:Inference-and-forecasting}
creates a trajectory (solid red line), estimates the model parameters
based on the beginning of the trajectory (blue shaded region), and
uses the estimated parameters to make predictions for the future time
evolution (thin colored lines, with mean given by a thick dashed line).}
\end{figure}

\section{Effect of event-driven NPI \label{sec:Prescribed-control}}

In this example, an event-driven protocol is defined: Once the number
of symptomatic infectives exceeds a threshold, the contact matrix
is reduced to the home contact; then, once the number of infectives
falls below another threshold, the original contact matrix is restored.
For an SEkIkIkR model with 16 age groups based on the UK age structure,
and UK contact matrices, we run this protocol in two variations: First,
we allow every event to happen at most one time; in the second run,
we allow events to repeat. Note that while this example employs UK
age groups and contact structure, the model parameters are not fitted
to real epidemiological data.

\vspace{0.5cm}

\begin{lstlisting}[language=Python]
import numpy as np
import matplotlib.pyplot as plt
import pyross


# 1. Load age structure and contact matrices of the UK

M=16  # number of age classes

# load age structure
my_data = np.genfromtxt('../data/age_structures/UK.csv',
                        delimiter=',', skip_header=1)
aM, aF = my_data[:, 1], my_data[:, 2]
Ni0=aM+aF
Ni = Ni0[:M] # consider first M age groups in data
N=np.sum(Ni) # total population is sum over all age groups

# get individual contact matrices
CH, CW, CS, CO = pyross.contactMatrix.UK()

# without interventions, the contact matrix is the sum of those
C = CH + CW + CS + CO


# 2. Define model

# parameters
alpha= 0.3         # fraction of symptomatics who self-isolate
beta = 0.0165      # probability of infection on contact
gE   = 1/2.72      # recovery rate of exposeds
kI   = 4;          # # of stages of I class
kE   = 4;          # # of stages of E class
gIa  = 1./7        # recovery rate of asymptomatic infectives
gIs  = 1./17.76    # recovery rate of symptomatic infectives
fsa  = 0.8         # the self-isolation parameter

# set up initial condition
S0 = np.zeros(M)
I0 = np.zeros((kI,M));
E0 = np.zeros((kE,M));
for i in range(kI):
    I0[i, 6:13]=14;  I0[i, 2:6]=13
for i in range(kE):
    E0[i, 0:16]=14
for i in range(M) :
    S0[i] = Ni[i] - np.sum(I0[:,i]) - np.sum(E0[:,i])
I0 = np.reshape(I0, kI*M)/kI;
E0 = np.reshape(E0, kE*M)/kE;


# 3. Define events and corresponding contact matrices

lockdown_threshold_1 = 1e6
lockdown_threshold_2 = 2000
#
events = []
contactMatrices = []
# Note that the event functions take a vector argument
# "xt" instead of a tuple (S, E0, Ia, Is).
# When defining the event functions for the SEkIkIkR
# model, the following correspondence has to be used:
# S  == xt[           :  M]
# E  == xt[1*M        :(1+kE)*M]
# Ia == xt[(1+kE)*M   :(1+kE+kI)*M]
# Is == xt[(1+kE+kI)*M:(1+kE+kI+kI)*M]
# (For any model, the order for the event functions is
#  exactly the same as in the arguments of model.simulation)

# This is a dummy event which will never occur;
# it is used to set the initial contact matrix
def event0(t,xt):
    return t + 1
event0.direction = +1
events.append(event0)
contactMatrices.append( C )

# Lockdown on
def event1(t,xt):
    return np.sum(xt[(1+kE+kI)*M:(1+kE+2*kI)*M]) - lockdown_threshold_1
event1.direction = +1 # need to pass threshold from below for event
events.append(event1)
contactMatrices.append( CH ) # only home contact matrix

# Lockdown off
def event2(t,xt):
    return np.sum(xt[(1+kE+kI)*M:(1+kE+2*kI)*M]) - lockdown_threshold_2
event2.direction = -1 # need to pass threshold from above for event
events.append(event2)
contactMatrices.append( C )


# 4. Run pyross.control simulation twice: Once with events only
#    occuring once, and once with events possibly several times

# instantiate model
parameters = {'beta':beta, 'gE':gE, 'gIa':gIa, 'gIs':gIs,
              'kI':kI, 'kE' : kE, 'fsa':fsa, 'alpha':alpha}
model = pyross.control.SEkIkIkR(parameters, M, Ni)

# Run 1: each event only occurs once
Tf = 2*365 # 2 years
Nt = (Tf +1)*10 # return 10 datapoints per day
result_1 = model.simulate(S0, E0, 0*I0, I0,
        events=events,contactMatrices=contactMatrices,
        Tf=Tf, Nf=Nt,events_repeat=False)

# Run 2: events can repeat
Tf = 8*365 # 8 years
Nt = (Tf +1)*10 # return 10 datapoints per day
result_2 = model.simulate(S0, E0, 0*I0, I0,
        events=events,contactMatrices=contactMatrices,
        Tf=Tf, Nf=Nt,events_repeat=True)


# 5. Visualise results

def plot_result(result,title=None):
    t_arr = result['t']
    traj = result['X']
    Is = model.Is(result)
    Is = np.sum(Is,axis=-1)
    events_occured = result['events_occured']
    #
    fig, ax = plt.subplots(1,1)
    if title != None:
        ax.set_title(title)
    ax.axhline(lockdown_threshold_1,label='Threshold event 1',
              ls='--',color='dodgerblue')
    ax.axhline(lockdown_threshold_2,label='Threshold event 2',
              ls='--',color='limegreen')
    for i,e in enumerate(events_occured[::2]):
        if 2*i + 1 < len(events_occured):
            if i == 0:
                label= 'NPI'
            else:
                label= ''
            ax.axvspan(e[0],events_occured[2*i+1][0],
                           label=label,
                          alpha=0.15,color='crimson')
    ax.plot(t_arr,Is,color='#A60628',label=r'$I_s$')
    ax.set_xlim(np.min(t_arr),np.max(t_arr))
    ax.set_ylabel('Compartment population'); ax.set_xlabel(r'Days')
    ax.legend(loc='upper right',fontsize=12)
    plt.grid(); plt.show(); plt.close(fig)

plot_result(result_1,title='Each event occurs once')
plot_result(result_2,title='Events can repeat')
\end{lstlisting}

\begin{figure}[H]
\centering\includegraphics[width=1\textwidth]{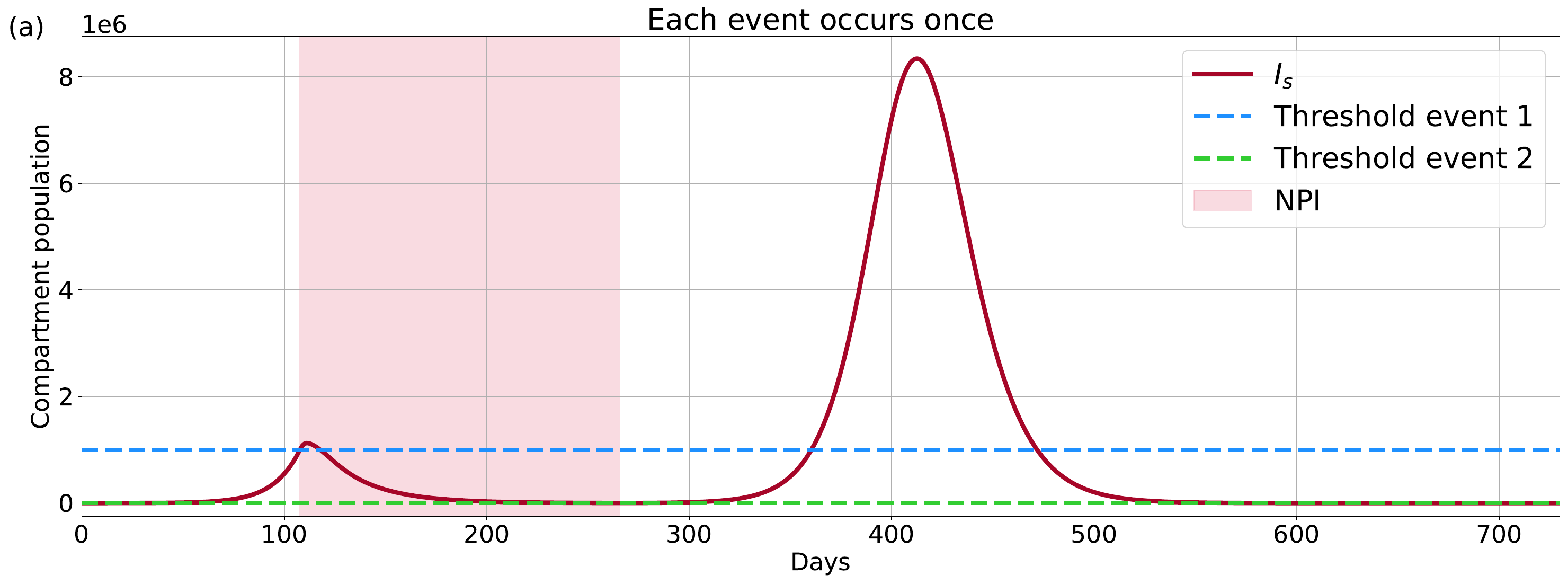}

\vspace{0.5cm}

\includegraphics[width=1\textwidth]{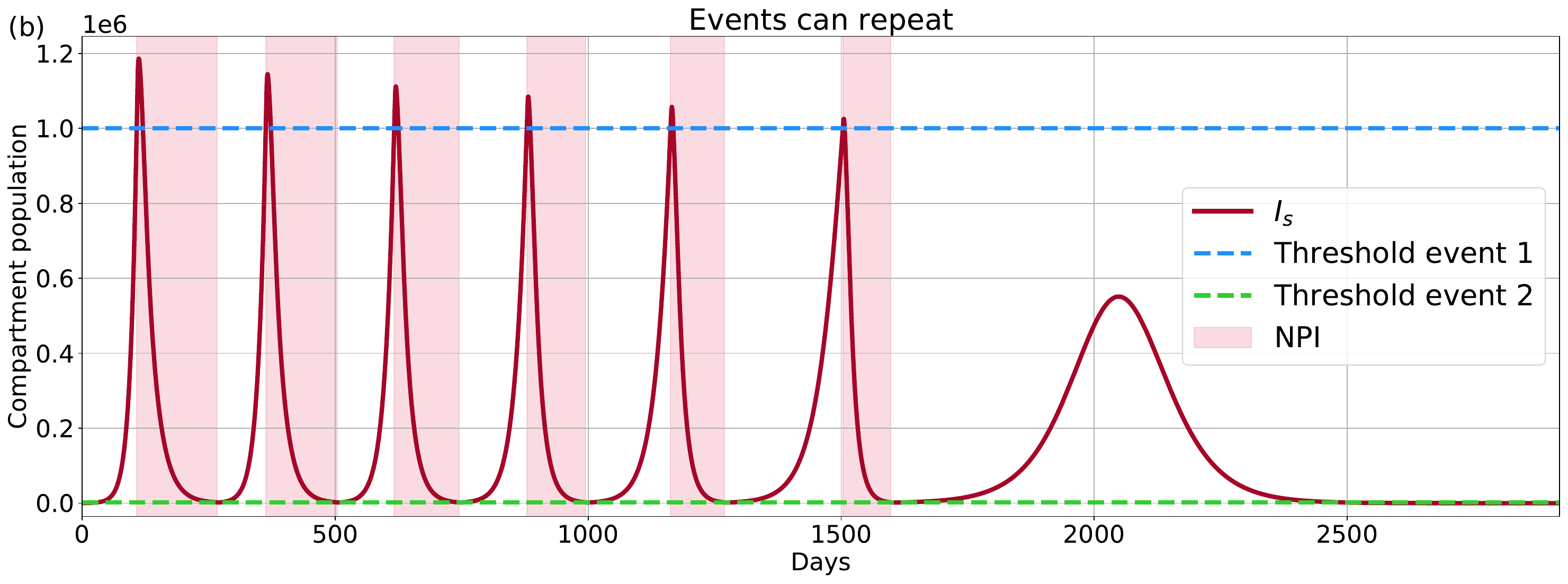}\caption{\textbf{Effect of event-driven NPI on model dynamics.} In example
\ref{sec:Prescribed-control}, an event-driven protocol is defined.
Once the number of symptomatic infectives exceeds a threshold, the
contact matrix is reduced to the home contacts; once the number of
infectives falls below another threshold, the original contact matrix
is restored. Subfigures (a) and (b) use the same events, but while
in (a) each event can only occur once, in (b) events can occur several
times.}
\end{figure}

\section{Optimising parameters of NPI\label{sec:Optimal-control}}

In the present example, an intervention protocol is optimised. We
consider the SIIR model with two age groups, with fictitious parameters
and a fictitious contact matrix. The protocol consists of two durations,
namely 
\begin{itemize}
\item the duration $T_{1}$ of a full lockdown (implemented in this toy
model by reducing the initial contact matrix via a prefactor $1/4$),
and
\item the duration $T_{2}$ a subsequent partial lockdown (implemented in
this toy model by reducing the initial contact matrix via a prefactor
$3/4$).
\end{itemize}
For illustrative purposes we choose the cost function

\begin{equation}
\mathcal{C}=\exp\left[\kappa\left(T_{1}+\frac{T_{2}}{5}\right)\right]+\frac{1}{\tau}\int_{0}^{t_{f}}\mathrm{d}t\,I_{s}(t),\label{eq:cost_in_example}
\end{equation}
where $t_{f}=1000$ days is the final time of the simulation, and
$I_{s}(t)$ is the total number of symptomatic infectives at time
$t$. The first term in equation (\ref{eq:cost_in_example}) can be
thought of as a cost for social distancing, while the second terms
models a cost of infection. The time scale $\tau$ determines the
relative importance of the two terms, and in our example we use $\tau=1$.
We additionally set $\mathcal{C}=\infty$ if at any time $I_{s}(t)>2000$,
which serves as a hard constraint that every protocol must keep the
number of infectives below a threshold. The rate $\kappa$ in the
first term of equation (\ref{eq:cost_in_example}) determines the
exponential increase in cost per time during lockdown, and in the
present example we minimise the total cost as a function of $(T_{1},T_{2})$,
considering three distinct scenarios: i) low cost of social distancing
($\kappa=0.01/$day), ii) intermediate cost of social distancing ($\kappa=0.1/$day),
and iii) high cost of social distancing ($\kappa=1/$day). Finally,
we compare the time series corresponding to these optimal protocols.

\vspace{0.5cm}

\begin{lstlisting}[language=Python]
import numpy as np
import matplotlib.pyplot as plt
import pyross
from multiprocessing import Pool
import cma
import time
from functools import partial


# 1. Define model

# number of age groups and total population
M  = 2               # the population has two age groups
N  =  5e4            # and this is the total population

# model parameters
beta  = 0.02         # infection rate
gIa   = 1./7         # removal rate of asymptomatic infectives
gIs   = 1./7         # removal rate of asymptomatic infectives
alpha = 0.2          # fraction of asymptomatic infectives
fsa   = 0.8          # the self-isolation parameter

# set the age structure
fi = np.array([0.25, 0.75])  # fraction of population in age age group
Ni = N*fi

# set up initial condition
Ia0 = np.array([10, 10])  # each age group has asymptomatic infectives
Is0 = np.array([10, 10])   # and also symptomatic infectives
R0  = np.array([0, 0])  # there are no removed individuals initially
S0  = Ni - (Ia0 + Is0 + R0) # initial number of susceptibles


# 2. Define function that generates two-step release protocol

def get_events(protocol= [50,100]):
    end_phase_1, end_phase_2 = protocol
    C = np.array([[18., 9.], [3., 12.]])
    events = []; contactMatrices = []
    # Note that the event functions take a vector argument
    # "xt" instead of a tuple (S, Ia, Is).
    # When defining the event functions for the SIR
    # model, the following correspondence has to be used:
    # S  == xt[ :M]
    # Ia == xt[1*M:2*M]
    # Is == xt[2*M:3*M]
    # (For any model, the order for the event functions is
    #  exactly the same as in the arguments of model.simulation)
    #
    # This is a dummy event which will never occur;
    # it is used to set the initial contact matrix
    events.append( lambda t, xt: 1. )
    contactMatrices.append( C )
    # Event 1: contact matrix reduced by a factor of 4 once
    # number of symptomatic infectives exceeds a threshold
    # ("start of full lockdown")
    lockdown_threshold_on = 1900
    def event1(t,xt):
        return np.sum(xt[2*M:3*M]) - lockdown_threshold_on
    event1.direction = +1 # need to pass threshold from below for event
    events.append(event1); contactMatrices.append( 0.25*C )
    # Event 2: contact matrix at 75% of original contact matrix after
    # a given time ("start of partial lockdown").
    def event2(t,xt):
        return t - end_phase_1
    events.append(event2); contactMatrices.append( 0.75*C )
    # Event 3: original contact matrix is restored after a given time
    # ("end of partial lockdown").
    def event3(t,xt):
        return t - end_phase_2
    events.append(event3); contactMatrices.append( C )
    return events, contactMatrices


# 3. Define cost function

maximal_value_for_Is = 2000
def evaluate_cost_function(t_arr,traj,protocol,events_occured,
                           sdc_prefactor=1.,
                           sdc_rate=1/7.):
    infinity = 1e300 # using np.inf would be more proper, but will
    # result in warnings throughout the minimization once a protocol
    # is probed that violates one of the hard constraints.
    #
    # hard constraints:
    # - times of protocol need to be non-negative
    if (np.array(protocol) < 0).any(): return infinity
    # - second stage of intervention needs to end after first stage ends
    if protocol[0] > protocol[1]: return infinity
    # - second stage of intervention needs to end before 700 days
    if protocol[1] > 700: return infinity
    # - number of symptomatic infectives should
    #   never exceed a given maximal value
    if np.max( np.sum( traj[2*M:3*M],axis=0 ) ) > maximal_value_for_Is:
        return infinity

    # Evaluate results of protocol.
    # Which phases have occured in the simulation?
    have_lockdown_start = False
    have_lockdown_end_0 = False
    have_lockdown_end_1 = False
    for i,e in enumerate(events_occured):
        if e[1] == 1:
            lockdown_start = e[0]
            have_lockdown_start = True
        elif e[1] == 2:
            lockdown_end_0 = e[0]
            have_lockdown_end_0 = True
        elif e[1] == 3:
            lockdown_end_1 = e[0]
            have_lockdown_end_1 = True

    # calculate cost function
    cost = 0.

    # cost of social distancing
    if have_lockdown_start:
        if have_lockdown_end_0:
            cost += sdc_prefactor* np.exp(sdc_rate* \
                                (lockdown_end_0-lockdown_start))
        else:
            cost += infinity
        #
        if have_lockdown_end_0 and have_lockdown_end_1:
            cost *= np.exp(0.2*sdc_rate*(lockdown_end_1-lockdown_end_0))

    # cost of infection
    cost += np.trapz( np.sum( traj[2*M:3*M],axis= 0 ),t_arr)
    return cost


# 4. Define function that evaluates the cost of a given protocol

def evaluate_cost_of_protocol(model,protocol,
                             verbose=False,return_trajectory=False,
                              sdc_prefactor=10.,
                              sdc_rate=1/7.):
    # define events corresponding to given protocol
    events, contactMatrices = get_events(protocol)
    # run simulation
    Tf = 1000; Nt = (Tf +1)*10
    result = model.simulate(S0,  Ia0, Is0,
                        events=events,contactMatrices=contactMatrices,
                        Tf=Tf, Nf=Nt)
    # evaluate cost
    cost = evaluate_cost_function(t_arr=result['t'],
                             traj=result['X'].T,
                             protocol=protocol,
                             events_occured=result['events_occured'],
                             sdc_prefactor=sdc_prefactor,
                             sdc_rate=sdc_rate)
    if return_trajectory:
        return cost, result['t'], result['X'], result['events_occured']
    else:
        return cost


# 5. Define minimiser

def minimizing_function(sdc_prefactor,sdc_rate,
                    protocol):
    #This is the function that will be minimised by the cma-es algorithm
    costs = evaluate_cost_of_protocol(model=model,
                                             protocol=protocol,
                                      sdc_prefactor=sdc_prefactor,
                                      sdc_rate=sdc_rate,
                                              return_trajectory=False)
    return costs

def find_optimal_protocol(initial_guess,sdc_prefactor,sdc_rate,
                          model,
                         verbose=True,initial_variance=5.):
    func = partial(minimizing_function, sdc_prefactor, sdc_rate)
    # set parameters for minimisation based on cma-es algorithm
    number_of_threads = 4
    number_iterations = 1000
    p = Pool(number_of_threads)
    options = cma.CMAOptions()
    options['popsize'] = 12
    # run minimisation
    start_time = time.time()
    optim = cma.CMAEvolutionStrategy(initial_guess, initial_variance,
                                     options)
    iteration = 0
    while not optim.stop() and iteration < number_iterations:
        positions = optim.ask()
        values = p.map(func, positions)
        optim.tell(positions, values)
        optim.disp()
        iteration += 1
    end_time = time.time()
    if verbose:
        print("Time: ", end_time - start_time)
        print("Value: ", optim.best.f)
        print("Parameters: ",optim.best.x)
    return optim.best.x


# 6. Find optimal protocol for three different scenarios

# Initialise model
parameters = {'alpha':alpha,'beta':beta, 'gIa':gIa,'gIs':gIs,'fsa':fsa}
model = pyross.control.SIR(parameters, M, Ni)
initial_guess=[70,200] # use same initial guess for all minimisations
sdc_prefactor = 1. # means that social distancing cost and
# cost of infection have the same weight in our model

# Low, intermediate, and high cost rate for social distancing:
sdc_rates = [1e-2,1e-1,1.]
optimal_protocols = []
for i,sdc_rate in enumerate(sdc_rates):
    optimal_protocol = find_optimal_protocol(initial_guess=initial_guess,
                                           sdc_prefactor=sdc_prefactor,
                                           sdc_rate=sdc_rate,
                                              model=model)
    optimal_protocols.append(optimal_protocol)


# 7. Visualise results

def plot_result(protocol,model,
                title='Optimised intervention parameters',
                t_max=None):
    #
    cost, t_arr, traj, events_occured = evaluate_cost_of_protocol(
                                            model=model,
                                             protocol=protocol,
                                              return_trajectory=True)
    Is = np.sum(traj[:,2*M:],axis=-1)

    fig, ax = plt.subplots(1,1,figsize=(7,5))
    ax.set_title(title)
    ax.axhline(maximal_value_for_Is,color='red',ls='--',
              label=r'Constraint')

    ax.axvspan(events_occured[0][0],events_occured[1][0],
              color='crimson',label='Intervention 1',alpha=0.2)
    ax.axvspan(events_occured[1][0],events_occured[2][0],
              color='orange',label='Intervention 2',alpha=0.2)
    ax.plot(t_arr,Is)
    if t_max == None:
        ax.set_xlim(np.min(t_arr),np.max(t_arr))
    else:
        ax.set_xlim(np.min(t_arr),t_max)
    ax.set_ylabel('Symptomatic infectives')
    ax.set_xlabel(r'time [days]')
    ax.legend(loc='upper right',fontsize=12,framealpha=1.)
    plt.show(); plt.close(fig)

optimal_protocol_labels = ['Low cost of social distancing',
                          'Intermediate cost of social distancing',
                          'High cost of social distancing']
t_max = [700,250,250]

for i,optimal_protocol in enumerate(optimal_protocols):
    plot_result(protocol=optimal_protocol,model=model,t_max=t_max[i],
           title=optimal_protocol_labels[i])
\end{lstlisting}
\begin{figure}[H]
\centering\includegraphics[width=1\textwidth]{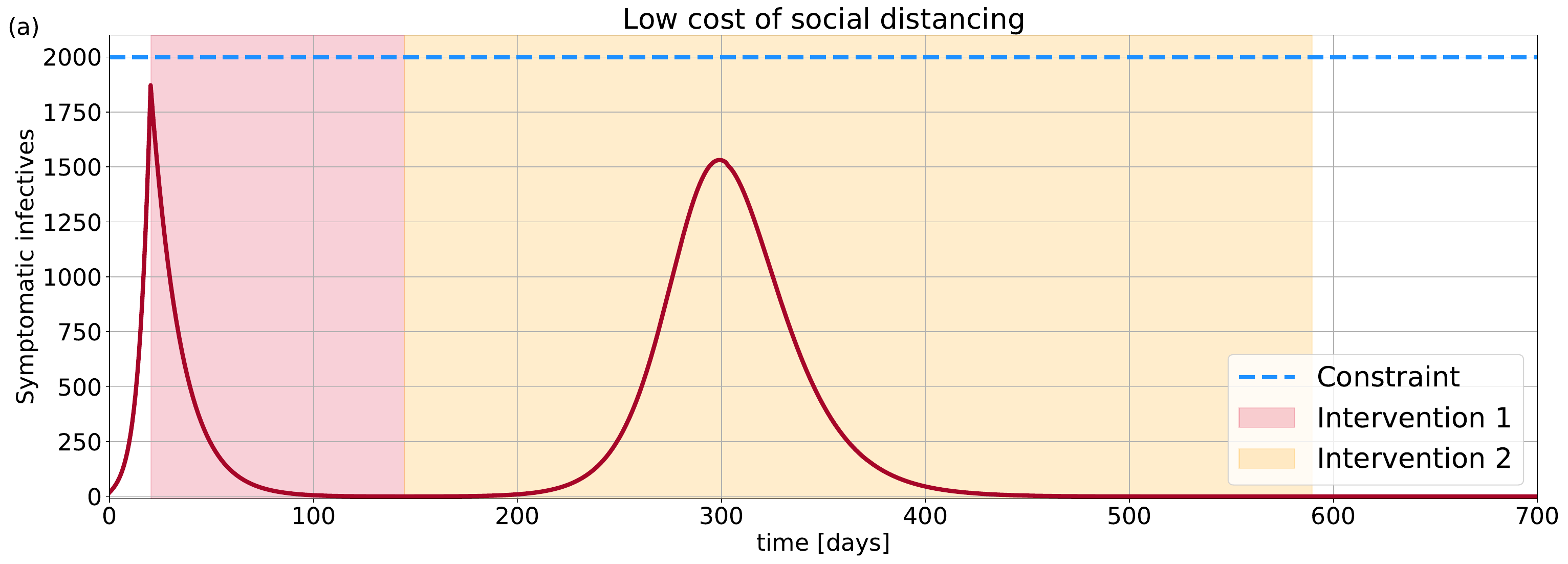}

\vspace{0.5cm}

\includegraphics[width=1\textwidth]{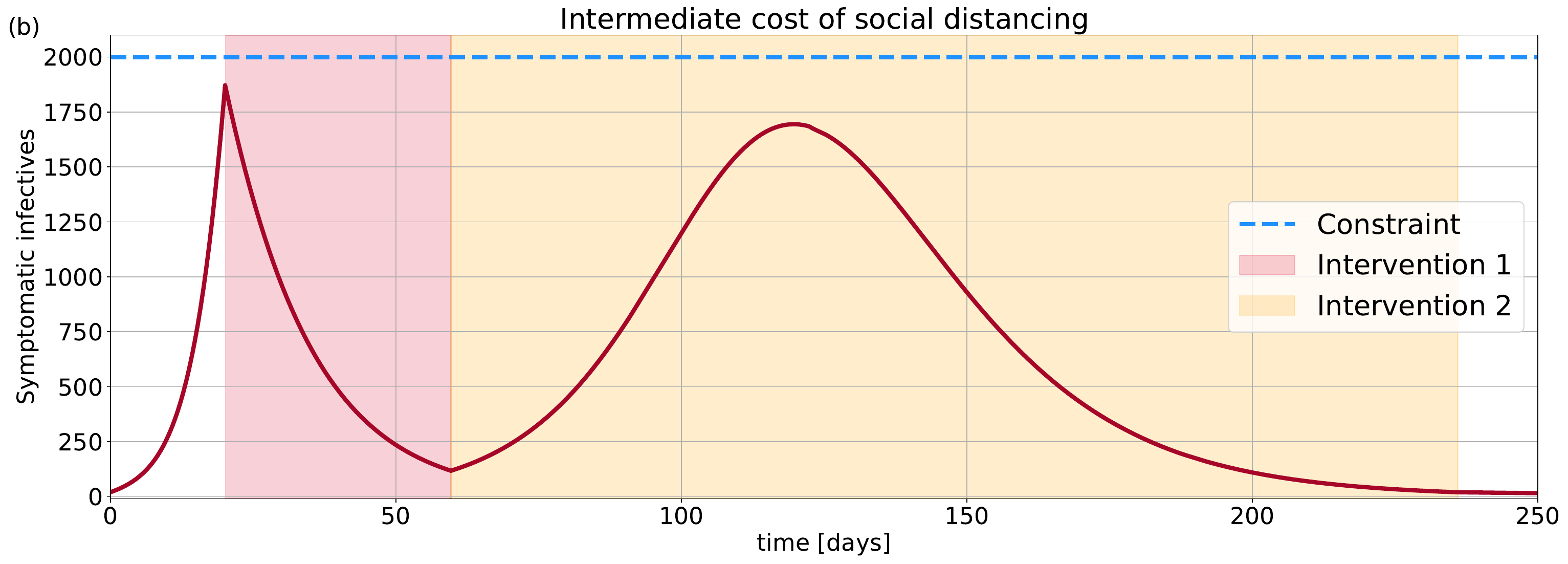}

\vspace{0.5cm}

\includegraphics[width=1\textwidth]{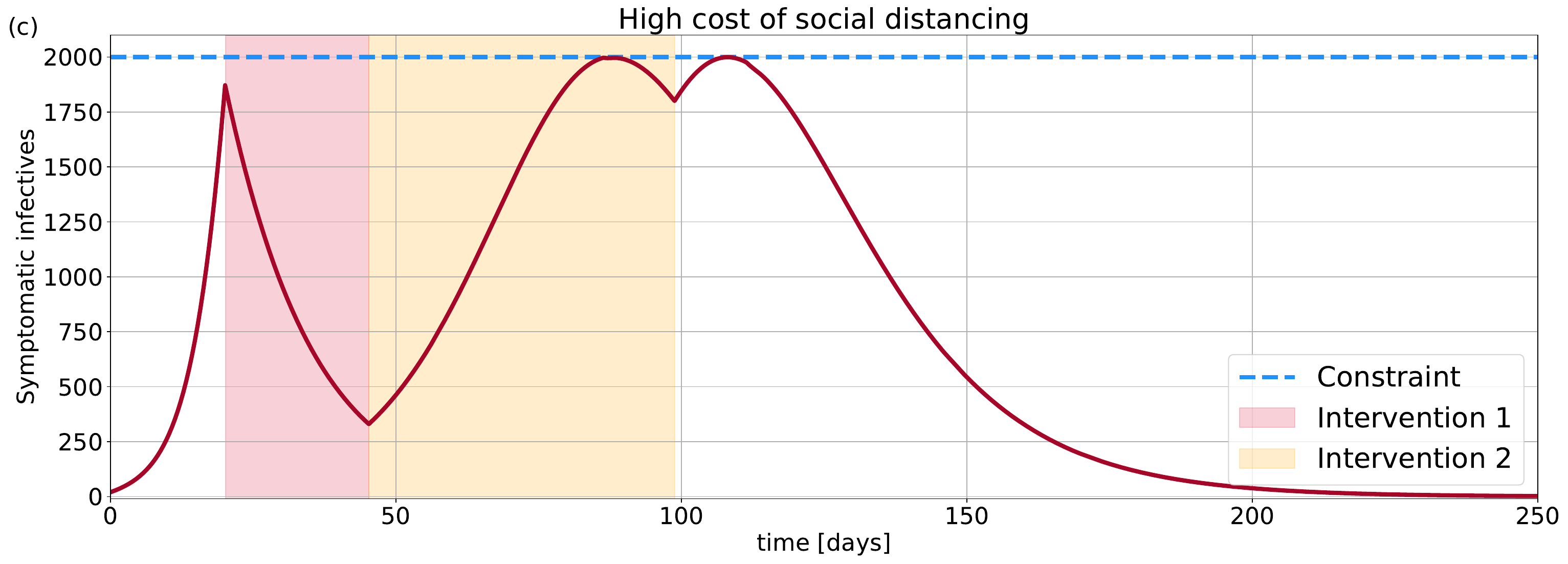}\caption{\textbf{Optimisation of NPI protocol parameters}, c.f. example \ref{sec:Optimal-control}.
The protocol consists of two durations, namely the time of a full
lockdown (initial contact matrix reduced via a prefactor 1/4), and
a subsequent partial lockdown (initial contact matrix reduced via
a prefactor 3/4). The cost of a protocol is comprised of a i) cost
for social distancing (= reducing the entries of the contact matrix),
and ii) a cost of infection. As a hard constraint, the protocol must
keep the number of infectives below a defined threshold, shown as
horizontal dashed line. The optimal protocol minimises the total cost.
Subplots (a), (b), (c), show three results of optimisation for the
same system, depending on the cost of social distancing. }
\end{figure}

\section{Model sensitivity\label{sec:Optimal-control-1}}

Here we illustrate the usage of PyRoss to study model sensitivity.
We use four different models, SIR, SEIR, SIR with stages, and SEIR
with stages. The models predict similar epidemic curve before the
lockdown, while the they show distinct behavior after lockdown. We
show that the addition of an exposed $E_{i}$ compartment (to give
an age-structured SEIR model) makes the infectious number grows beyond
the lockdown for the time scale of the incubation. Note that while
this example employs UK age groups and contact structure, the model
parameters are not fitted to real epidemiological data.

\begin{lstlisting}[language=Python]
import numpy as np
import pyross
import matplotlib.pyplot as plt

## population and age classes
M=16  ## number of age classes
my_data = np.genfromtxt('India-2019.csv', delimiter=',', skip_header=1)
Ni = (my_data[:, 1]+ my_data[:, 2])[0:M]

# Get individual contact matrices
CH, CW, CS, CO = pyross.contactMatrix.India()
# Generate contact matrix 
generator = pyross.contactMatrix.ContactMatrixFunction(CH, CW, CS, CO)


Tf=42;  Nf=600 
times= [21, Tf] # temporal boundaries between different contact-behaviour
aW, aS, aO = 0.0, 0.0, 0.0

# prefactors for CW, CS, CO:
interventions = [[1.0,1.0,1.0],      # before first time
                 [aW, aS, aO],       # between first and second time
                ]         
# generate corresponding contact matrix function
contactMatrix = generator.interventions_temporal(times=times,interventions=interventions)


# ## SIR
beta  = 0.01646692      # probability of infection on contact 
gIa   = 1./7            # removal rate of asymptomatic infectives 
gIs   = 1./7            # removal rate of symptomatic infectives 
alpha = 0.              # asymptomatic fraction
fsa   = 1               # suppresion of contact by symptomatics
# initial conditions    
Is_0 = np.zeros((M));  Is_0[6:13]=14;  Is_0[2:6]=13
Ia_0 = np.zeros((M))
R_0  = np.zeros((M))
S_0  = Ni - (Ia_0 + Is_0 + R_0)
parameters = {'alpha':alpha,'beta':beta, 'gIa':gIa,'gIs':gIs,'fsa':fsa}
model = pyross.deterministic.SIR(parameters, M, Ni)
dataSIR =model.simulate(S_0, Ia_0, Is_0, contactMatrix, Tf, Nf)
I1 = model.Is(dataSIR)


# ## SEIR
beta  = 0.027           # probability of infection on contact 
gIa   = 1./7            # removal rate of asymptomatic infectives 
gE    = 1/2.72          # removal rate of exposeds
gIs   = 1./7            # removal rate of symptomatic infectives 
alpha = 0.              # asymptomatic fraction
fsa   = 1               # suppresion of contact by symptomatics
Is_0 = np.zeros((M));  Is_0[6:13]=14;  Is_0[2:6]=13
Ia_0 = np.zeros((M))
E_0 = np.zeros((M)); 
R_0  = np.zeros((M))
S_0  = Ni - (Ia_0 + Is_0 + R_0)
param={'alpha':alpha,'beta':beta,'gIa':gIa,'gIs':gIs,'gE':gE,'fsa':fsa}
model = pyross.deterministic.SEIR(param, M, Ni)
dataSEIR =model.simulate(S_0, E_0, Ia_0, Is_0, contactMatrix, Tf, Nf)
I2 = model.Is(dataSEIR)


# SIR with stages (SIkR)
beta = 0.01324           # probability of infection on contact 
gI   = 1./7              # removal rate of  infectives 
gE   = 1/2.72            # removal rate of exposeds
kI   = 32;               # # of stages of I class
I0 = np.zeros((kI,M));
for i in range(kI):
    I0[i, 6:13]=14;  I0[i, 2:6]=13
S0 = np.zeros(M)  
for i in range(M) :
    S0[i] = Ni[i] - np.sum(I0[:,i])
I0 = np.reshape(I0, kI*M)/kI
parameters = {'beta':beta, 'gI':gI, 'kI':kI}
model = pyross.deterministic.SIkR(parameters, M, Ni)
dataSIkR=model.simulate(S0, I0, contactMatrix, Tf, Nf)
I3 = model.I(dataSIkR)


# SEIR with stages (SEkIkR)
beta = 0.0188           # probability of infection on contact 
gI   = 1./7             # removal rate of  infectives 
gE   = 1/2.72           # removal rate of exposeds
kI   = 32;              # # of stages of I class
kE   = 2;               # # of stages of E class 
S0 = np.zeros(M)  
I0 = np.zeros((kI,M));
E0 = np.zeros((kE,M));
for i in range(kI):
    I0[i, 6:13]=14;  I0[i, 2:6]=13
for i in range(kE):
    E0[i, 0:16]=14
for i in range(M) :
    S0[i] = Ni[i] - np.sum(I0[:,i]) - np.sum(E0[:,i])
I0 = np.reshape(I0, kI*M)/kI;
E0 = np.reshape(E0, kE*M)/kE;
parameters = {'beta':beta, 'gE':gE, 'gI':gI, 'kI':kI, 'kE' : kE}
model = pyross.deterministic.SEkIkR(parameters, M, Ni)
data=model.simulate(S0,E0,I0,contactMatrix,Tf,Nf); I4 = model.I(data)

asI1,asI2,asI3,asI4 = I1.sum(axis=1),I2.sum(axis=1),I3.sum(axis=1),I4.sum(axis=1) #sum over all ages


plt.plot(t, asI1, '-', lw=lwd, color='#A60628', label='SIR')
plt.plot(t, asI3, '-', lw=lwd, color='#348ABD', label='SIkR')
plt.plot(t, asI2, '-', lw=lwd, color='goldenrod', label='SEIR')
plt.plot(t, asI4, '-', lw=lwd, color='forestgreen', label='SEkIkR')
plt.xlabel('Days'); plt.ylabel('Infected individuals'); 
\end{lstlisting}
\begin{figure}[H]
\centering\includegraphics[width=1\textwidth]{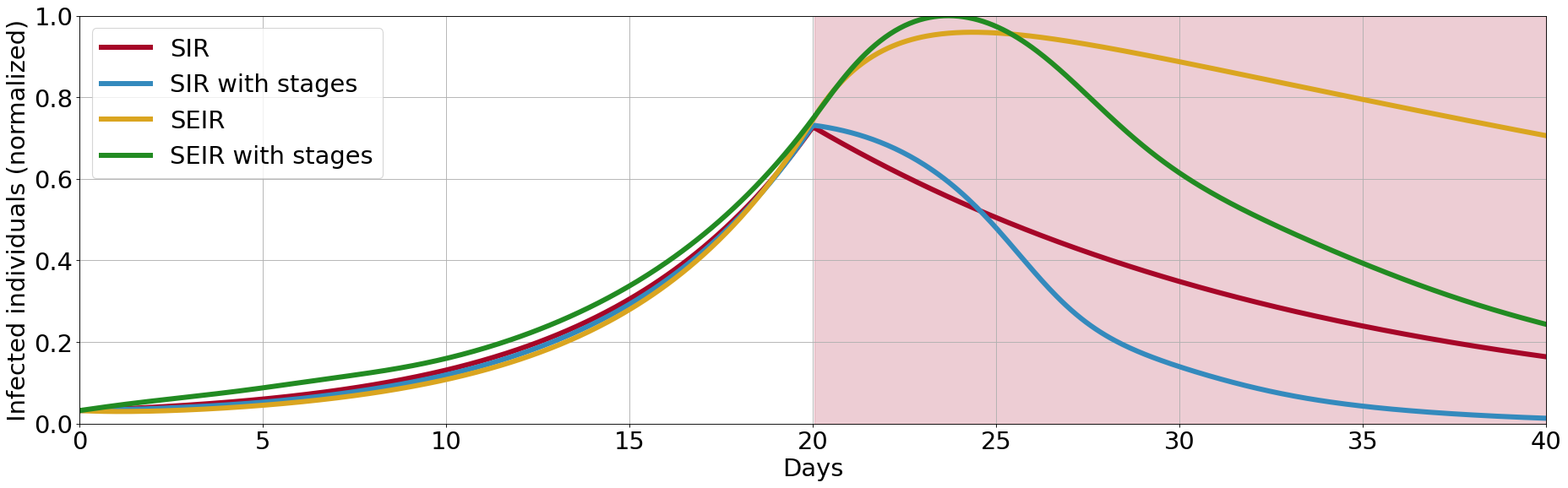}

\caption{\textbf{Model-dependence of dynamics after lockdown.} In example \ref{sec:Optimal-control-1},
the models SIR, SEIR, SIR with stages, and SEIR with stages, are considered.
Parameters are chosen such that the initial grow rates are comparable.
Initially, all models use the full contact matrix, which is the sum
of home, work, school, and other contacts. At a fixed time, a lockdown
is modeled by retaining only the home contacts. This plot illustrates
the resulting different model dynamics at the initial phase of the
lockdown.}
\end{figure}

\chapter{Conclusion}

This report has presented an overview of PyRoss, a Python library
that integrates compartment models of epidemics with Bayesian inference
and optimization tools.

The current focus of PyRoss, and hence of this report, is on well-mixed,
age-structured models with a modest number of contact channels (home,
school, workplace, other). However, PyRoss allows these channels to
be expanded indefinitely in principle, so that more complex contact
information relating to particular working or social environments
can be incorporated as available.

Likewise PyRoss allows the disease stages themselves to be subdivided
at will. This enables the statistics of objective medical states,
such as hospitalization and intensive care or ventilator use, to be
modelled. 
By also allowing an indefinitely expandable number of compartments
per disease stage, the residence time distribution in such stages
can be varied from the exponential decay of simple compartment models
to the sharply-peaked distribution that underlies time-since-infection
models \cite{kermack1927contribution,reddingius1971notes,sattenspiel1995structured,hoppensteadt1974age,diekmann2010construction,thieme1993may}.
Certain other mathematical limitations can be overcome similarly.

NPIs are viewed within PyRoss as strategies to influence contact matrices,
generally in a time-dependent fashion, through variations in contact
rate, or transmission rate at contact, or both. These interventions
can be represented in as much social detail as is resolved by the
contact matrices themselves.

A Bayesian system of inference for model parameters, and for forecasting
with error estimation, is fully integrated into PyRoss. The former
is a crucial feature which allows comparison between models based
on evidence. This can guide a principled expansion in the number
of compartments up to a level of granularity justified by the data,
and can warn when this has gone too far. The approach guards
against over-fitting of historic data, leading to precise but inaccurate
forecasting -- a risk faced by parameter-rich, data-poor models
in any field of science.

One can expect robust inference from historic data to prove crucial for forecasting
the future course of an epidemic, particularly where sequenced or nuanced
NPIs are involved. Accordingly, PyRoss's Bayesian parameter estimation tools allow
the fitting of time-dependent contact-matrix parameters representing
NPIs. Prior estimates (supplied perhaps by expert judgement) of the
effect of an NPI on contact statistics can therefore be continuously
improved, after the NPI is implemented, by feeding the observed results
back into the system (Fig.~\ref{IPIfig}). Observed outcomes
for one NPI (such as sudden lockdown) can inform the prediction
of others (such as a sequential, age-stratified unlock).

Within PyRoss, given a user-defined cost function that encodes their
harm, NPIs can be optimized and ranked by minimization of the chosen
cost function. The actual choice of cost functions is, of course,
morally and politically problematic. However, there are many other
areas of planning where saving life is weighed against economic cost,
including decisions about where and whether to build a new hospital,
or indeed a new road. 

The principles of compartment modelling embraced by PyRoss, when constrained by a disciplined approach to parameter and uncertainty estimation, are, we
believe, more powerful than is widely assumed, but of course not all-powerful.
Known challenges include situations in which the well-mixed approximation
does not hold at small scales, such as the fact that isolation by
household can allow rapid disease spread within each household while
preventing it at societal level. Redesign of the compartment structure
to reflect a different `unit of infection', such as a household, may
help in some such cases.

In contrast, there is no problem of principle in extending compartment
models to societies that are locally well mixed but heterogeneous
at larger scales. Different contact matrices can be assigned
at regional level and/or to micro-locations such as specific workplaces
or social venues. Once such geo-social compartments are resolved,
the transition rates between them are controlled by human mobility
which can itself be modelled by jumps into and out of compartments
representing journeys (within which infection can of course occur).
NPIs can then separately target either the locationally specific contact
matrices, or the transfer of people between locations. In a forthcoming
report we will describe PyRossGeo, an extension of PyRoss along these
lines.

This report has focussed on the design principles and capabilities
of PyRoss, illustrated by very simple examples of its use. We
have not deployed it directly here to address aspects of the current
COVID-19 crisis, but for an example of its use in doing so, see \cite{Singh2020}.
(Note that since that paper was written, the capabilities of PyRoss
have advanced dramatically.)

The PyRoss library is open-source. We positively encourage its use
by other scientists, whether seasoned epidemic modellers or those
new to the field. We hope it can contribute
to a stronger scientific platform for evidence-based decision making
as the current pandemic continues, and also before the next one.

\section*{Acknowledgement}

The work described in this report was completed in the six weeks
spanning April to mid-May 2020 as a contribution to the Rapid Assistance
in Modelling the Pandemic (RAMP) initiative, coordinated by the Royal
Society. The GitLab platform enabled seamless collaboration under
lockdown conditions. We are grateful to the numerous developers who
contribute to the open-source platform on which PyRoss is built and
without which it would not exist. We acknowledge advice from Graeme
Ackland, Daan Frenkel, Julia Gog, Chris Rogers, and Richard Wilkinson.
We thank the code review team of RAMP's Rapid Review Group at Oxford
for their scrutiny of the PyRoss library and for their suggestions
for improvement; RAMP's Red Team at Edinburgh further code review
and stress testing; and those who opened issues and offered suggestions
on GitHub. Remaining defects in the PyRoss library and in this report
are the responsibility of the authors. This work was funded in part
by the European Research Council under the EU\textquoteright s Horizon
2020 Program, Grant No.~740269; by the Royal Society through a Research
Professorship held by MEC, and by an Early Career Grant to RA from
the Isaac Newton Trust.

\chapter*{Appendix: Simple versus Complicated Models\label{appendixCovidSim}}

\begin{figure*}[t]
\centering\includegraphics[scale=0.4]{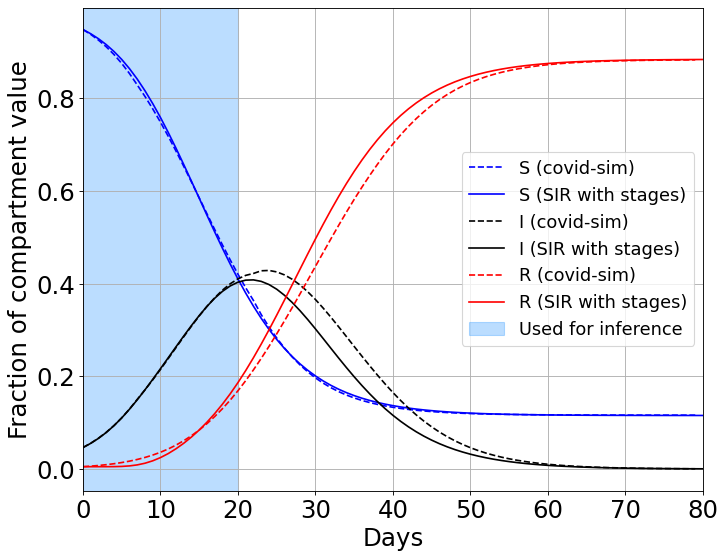}\caption{\textbf{Predictions from an unstructured SI5R model in Simulana}.
Simulana is an imagined country where epidemics are correctly described
by the Covid-Sim algorithm of \cite{ferguson2020impact} with some particular parameter set. \label{fig:age-contact-1-1}}
\end{figure*}

Let us imagine an alternative universe which contains a country resembling
the UK, called Simulana, for which the CovidSim model, with a specific
parameter set as selected in \cite{ferguson2020impact}, describes
the disease dynamics as well as one could wish. The government of Simulana allow the
epidemic to run its course without NPIs. Suppose that the only data
available to modellers in Simulana are time-resolved national aggregate
statistics $(S(t),I(t),R(t))$ for the number of susceptibles, infecteds,
and removeds. To forecast the epidemic, the modellers adopt an age-unstructured
SIkR model, with $k$, the number of infected sub-compartments discussed
above, set to $k=5$. We emphasise that this is an utterly basic model.
Nonetheless, by fitting to the available data up to (say) the point where $S=I$,
Simulana's modellers are able to predict its remaining course reasonably
well, as shown in Figure \ref{fig:age-contact-1-1}. Note that only
the maximum a posteriori (MAP) prediction is shown although PyRoss
can deliver the full posterior distribution if required.

These predictions raise the following question: {\em
Unless the specific CovidSim parameters that correctly describe epidemics in Simulana
are disclosed upfront by revelation, is CovidSim in general much
better at predicting epidemics there than, say, SI5R?}

Interpreting this question in a strictly Bayesian sense of `which
model has more evidence', we suspect the answer is `no'
-- although to confirm this quantitatively would require a more precise
delineation of CovidSim's free parameters than we currently have.
This is because the Bayesian `Occam factors' penalize models with
parameters that are unconstrained by the data available: the evidence for these models is small, whereas parsimonious models always have larger evidence if equally capable of fitting the data. 

But
even if the question is interpreted less formally, we might also suspect a negative answer. Given
the limited medical data available in Simulana, it is not clear that this can constrain
CovidSim's parameters sufficiently to give forecasts that are any
more reliable than those of much simpler models. This is despite the fact that using CovidSim the epidemic in Simulana's could be forecast
perfectly in principle if the correct parameters were known.

In the above example, the modellers of Simulana received only aggregate
patient data; alternatively they might receive data stratified by
age, geography etc.. In this case the shortcomings of SI5R for Simulana
would soon be revealed by the Bayesian approach. However, a generalized
compartment model might well still beat CovidSim at `predicting its own future'. Of particular
promise are models whose compartments broadly aligned with the granularity
of the data available, creating inbuilt parsimony.

It is very important to recognize that the above paragraphs contain
no specific criticism of CovidSim itself, or any other particular
model. Instead they express a more general
concern about basing predictions on {\em any} model whose parameters
have proliferated beyond the ability of data to estimate them. Such
concerns apply not only for epidemics but across many other disciplines.
They are one of the reasons so many scientists today prefer a Bayesian
approach to model and parameter selection, as offered by PyRoss for compartment models.

\end{document}